\documentclass[preprint,12pt,times,authoryear]{elsarticle}

\usepackage{amssymb}
\usepackage{amsmath}
\usepackage{amsthm}
\newtheorem{remark}{Remark}
\newtheorem{assumption}{Assumption}

\usepackage{booktabs}
\usepackage{subcaption}
\usepackage{float}
\usepackage{enumitem}
\usepackage{verbatim}
\usepackage{mathtools}
\usepackage{multirow}
\usepackage{xcolor}
\usepackage{algorithm}

\DeclareMathOperator*{\E}{\mathbb{E}}

\newcommand{\mD}{\mbox{\textbf{D}}}
\newcommand{\mI}{\mbox{\textbf{I}}}
\newcommand{\mM}{\mbox{\textbf{M}}}
\newcommand{\mC}{\mbox{\textbf{C}}}
\newcommand{\mW}{\mbox{\textbf{W}}}

\newcommand{\vX}{\mbox{\boldmath{$X$}}}
\newcommand{\vx}{\mbox{\boldmath{$x$}}}
\newcommand{\vy}{\mbox{\boldmath{$y$}}}

\newcommand{\vz}{\mbox{\boldmath{$z$}}}
\newcommand{\vmu}{\mbox{\boldmath{$\mu$}}}
\newcommand{\vs}{\mbox{\boldmath{$s$}}}

\usepackage{ulem}  
\newcommand{\purple}[1]{\textcolor{purple}{\sout{#1}}} 

\journal{Spatial Statistics}

\begin{document}

\begin{frontmatter}



\title{A Robust Nonparametric Framework for Detecting Repeated Spatial Patterns} 


\author[label1]{Rajitha Senanayake} 
\author[label1,label2]{Pratheepa Jeganathan}

\affiliation[label1]{organization={Department of Mathematics and Statistics, McMaster University},
            city={Hamilton},
            state={Ontario},
            country={Canada}}

\affiliation[label2]{organization={School of Computational Science and Engineering, McMaster University},
            city={Hamilton},
            state={Ontario},
            country={Canada}}

\begin{abstract}
Identifying spatially contiguous clusters and repeated spatial patterns (RSP) characterized by similar underlying distributions that are spatially apart is a key challenge in modern spatial statistics. Existing constrained clustering methods enforce spatial contiguity but are limited in their ability to identify RSP. We propose a novel nonparametric framework that addresses this limitation by combining constrained clustering with a post-clustering reassigment step based on the maximum mean discrepancy (MMD) statistic. We employ a block permutation strategy within each cluster that preserves local attribute structure when approximating the null distribution of the MMD. We also show that the MMD$^2$ statistic is asymptotically consistent under model-free spatial mixing conditions. This two-stage approach enables the detection of clusters that are both spatially distant and similar in distribution. Through simulation studies that vary spatial dependence, cluster sizes, shapes, and multivariate dimensionality, we demonstrate the robustness of our proposed framework in detecting RSP. We further illustrate its applicability through an analysis of spatial proteomics data from patients with triple-negative breast cancer. Overall, our framework presents a methodological advancement in spatial clustering, offering a flexible and robust solution for spatial datasets that exhibit repeated patterns.
\end{abstract}




\end{frontmatter}



\section{Introduction}
\label{sec1}

Spatial clustering refers to the identification of regions in a spatial domain where observations exhibit similar attribute values or statistical properties, resulting in groups of spatially contiguous and homogeneous units.  Typically, spatial clusters are contiguous, as locations in close proximity tend to share similar attribute values. However, similar spatial patterns can also emerge repeatedly in distant locations, forming discrete patches referred to as repeated spatial pattern (RSP). These may reflect shared underlying processes such as climate patterns \citep{benevento2024correlation}, socioeconomic trends \citep{murtagh_survey_1985}, and disease dynamics \citep{kulldorff1997spatial}. Despite the relevance of RSP detection in practical applications, many existing spatial clustering methods are limited in their capacity to identify RSP.

One domain that increasingly relies on spatial clustering is spatial omics  \citep{Hartman_2024, keren_structured_2018}.  Recent advances in spatial technologies have greatly expanded the ability to study biological complexity within tissues. Reflecting this impact, spatial proteomics was named the 2024 Method of the Year by Nature Methods \citep{spatial-proteomics-2024}. These developments have increased the need for robust spatial clustering methods that can accommodate the heterogeneous nature of spatial omics data. Such data may consist of continuous measurements (e.g., protein marker intensities, gene expression levels), categorical annotations (e.g., cell type classifications), or binary indicators (e.g., presence or absence of specific markers). Consequently, spatial clustering methods must be flexible enough to handle any of these data types, making them one of the tools to uncover spatio-temporal patterns by integrating multimodal data from the Human Tumor Atlas Network (HTAN) \citep{de2025sharing}.

Several methods have been developed for clustering spatial omics data. For example, \texttt{Giotto} identifies spatially homogeneous regions by modeling gene expression patterns using a hidden Markov random field (HMRF) \citep{dries_giotto_2021}. Moreover, \texttt{Banksy} uses k-means clustering, Leiden community detection, or many other algorithms, with spatial information encoded as node attributes derived from azimuthal Gabor filters (AGFs) that capture local spatial structure \citep{singhal_banksy_2024}. \texttt{Proust} employs a graph-based autoencoder to learn hybrid feature representations that integrate gene expression and spatial proximity, followed by nonparametric or model-based clustering on the learned embeddings to identify spatial domains \citep{yao_spatial_2024}.

While methods such as \texttt{Giotto}, \texttt{Banksy}, and \texttt{Proust} leverage graph-based spatial representations or enhanced feature embeddings for spatial domain detection, they exhibit limitations in handling certain data modalities and RSP. For instance, the HMRF framework in \texttt{Giotto} models gene expression patterns and can identify spatial patches with similar expression distributions that may align with RSP. However, it is not directly applicable to categorical or binary data. \texttt{Banksy} and \texttt{Proust}, on the other hand, may fail to capture RSP, particularly in the presence of varying spatial dependence or irregular spatial shapes.

Spatial clustering methods can be broadly categorized into three types: (1) attribute-based methods, (2) integrated methods that consider both attribute similarity and spatial proximity, and (3) methods that explicitly balance these two components to detect RSP. Attribute-based methods, such as k-means and hierarchical clustering, primarily focus on clustering based on attribute similarity. These methods can incorporate spatial information by treating coordinates as additional attributes or by enhancing spatial representation, such as splines or Gabor filters. While they are typically nonparametric and flexible across diverse data types, they often fail to ensure spatial contiguity. Integrated methods, including constrained hierarchical clustering (CHC) and scan statistic-based clustering, combine spatial proximity with attribute similarity to identify spatially contiguous clusters \citep{guenard_hierarchical_2022,miranda_regk-means_2017,ester_density-based_1996}. The CHC explicitly incorporates spatial constraints to ensure contiguity, whereas scan statistics detect regions with significantly higher attribute densities than expected under a null model, assessing statistical significance \citep{abolhassani2021up}. Despite their utility, these integrated methods may struggle to identify RSP because they weigh spatial proximity and attribute similarity differently. 

To address the above limitation, the methods that counterbalances spatial and attribute contributions have emerged. The \texttt{STICC} method constructs subregions using Gaussian Markov random fields and employs a mixture modeling framework to assign these subregions to clusters \citep{kang_sticc_2022}. \texttt{STICC} can detect RSP by clustering subregions that are spatially contiguous and RSP. However, its reliance on Gaussian assumptions and the combinatorial complexity of the subregion-to-cluster assignment procedure may limit its applicability to non-Gaussian data types and large-scale spatial omics datasets.

RSPs have also been explored in the spatio-temporal clustering literature. For example, the Spatial-CHC algorithm constructs a temporal correlation matrix using Gaussian rank correlation and a spatial correlation matrix based on a Matérn covariance function. These matrices are combined into a single dissimilarity matrix by computing a geodesic average on the Riemannian manifold of correlation matrices. Hierarchical clustering is then applied to this combined dissimilarity matrix to identify spatial patterns that recur over time \citep{benevento2024correlation}. However, this approach is specifically designed to capture temporal repetition of spatial patterns, rather than RSP within a spatial domain. 

While parametric approaches such as \texttt{STICC} offer a data-generative process well-suited for detecting RSP, their limitations motivate our development of a more flexible, nonparametric framework. The method proceeds in two stages: it first applies constrained clustering to ensure spatial contiguity, followed by a post-clustering comparison to identify regions that share similar spatial distribution. The framework is broadly applicable and can be paired with any constrained clustering method to detect RSP within the spatial domain.

Our main contribution is a novel post-clustering framework, \texttt{repSpat}, designed to identify RSP by addressing the common over-segmentation produced by constrained clustering methods. Specifically, we introduce multiple hypotheses testing approach that
(1) quantifies similarity in distributions between clusters using the maximum mean discrepancy squared (MMD$^2$) statistic, which is well suited for detecting general differences in multivariate distributions \citep{gretton2006kernel}; and
(2) employs a block permutation strategy that preserves local spatial structure within clusters, enabling valid inference under these unknowns \citep{guillot2013dismantling}.
This combination allows \texttt{repSpat} to reassign labels to spatially distant clusters that arise from similar underlying spatial processes, extending the utility of existing constrained clustering methods to settings with RSP.

The remainder of the paper is organized as follows. Section 2 describes the proposed methodology, \texttt{repSpat}, in detail. Section 3 presents a simulation study evaluating the performance of \texttt{repSpat} under varying conditions and comparing these results with those obtained from \texttt{Banksy}. Section 4 demonstrates the application of our approach to spatial proteomics data. Finally, Section 5 concludes with a discussion and future directions.

\section{Methodology}

Let $\vX(\vs)$ denote a multivariate spatial process, where $\vs \in \mathcal{S} \subseteq \mathbb{R}^2$ is a spatial location in a continuous domain $\mathcal{S}$, and $\vX(\vs) \in \mathcal{X} \subseteq \mathbb{R}^p$ is a $p$-dimensional attribute vector observed at $\vs$.  Let $\lbrace \vs_1, \ldots, \vs_n\rbrace$ denote a set of $n$ spatial sampling locations. 

We now introduce the \texttt{repSpat} framework, which is designed to detect repeated spatial patterns (RSP) in $\vX(\vs)$. The proposed \texttt{repSpat} framework consists of four main steps:

\begin{enumerate}[label=(\arabic*)]
	\item Constrained clustering: Apply constrained agglomerative hierarchical clustering (CAHC) to obtain spatially contiguous clusters.
	\item Testing spatial invariance: Perform pairwise comparisons of cluster distributions using the maximum mean discrepancy squared (MMD$^2$) statistic.
	\item Block permutation: Approximate the null distribution of the MMD$^2$ statistic while preserving local dependence.
	\item Cluster reassignment: Reassign cluster labels based on pairwise similarity.
\end{enumerate}
A detailed algorithm is provided in Algorithm~\ref{algo1}.

\begin{algorithm}[h]
	
	\caption{\texttt{repSpat} Framework}
	\label{algo1}
	\begin{enumerate}[label=Step \arabic*:, leftmargin=*]
		\item Construct the dissimilarity matrix $\mathbf{D}$ and spatial links $\ell_{ij}$ using $m$-nearest neighbors.
		\item Apply constrained agglomerative hierarchical clustering (CAHC) to obtain $G$ clusters.
		\item For each pair of clusters $(g,h)$:
		\begin{enumerate}[label=\alph*), leftmargin=*]
			\item Compute the MMD$^2$ statistic between $X^{(g)}$ and $X^{(h)}$.
			\item Approximate the null distribution of the MMD$^2$ statistic using block permutation.
			\item Perform hypothesis testing and adjust for multiple comparisons.
		\end{enumerate}
		\item Construct a graph where nodes represent clusters and edges indicate similarity.
		
		\begin{enumerate}[label=\alph*), leftmargin=*]
			\item Drop the edges where the hypothesis testing adjusted for multiple comparison rejected the null hypothesis. 
			\item Reassign cluster labels based on connected components.
		
		\end{enumerate}
		
	\end{enumerate}

\end{algorithm}

Although we use CAHC as the initial clustering step in this study, the proposed \texttt{repSpat} framework is not restricted to this choice and can be applied to any constrained clustering method that produces spatially contiguous clusters.

To illustrate the steps of the \texttt{repSpat} framework, we consider a simulated example shown in Figure~\ref{fig_sim_setup}, where RSP are generated across spatially distant regions. In the following subsections, we elaborate on Algorithm~\ref{algo1} and describe how each step of the \texttt{repSpat} method applies in this illustrative example.

\subsection{Constrained Clustering}

 In Step 1-2 of Algorithm~\ref{algo1}, we  employ constrained agglomerative hierarchical clustering (CAHC) to identify spatially contiguous clusters \citep{guenard_hierarchical_2022}.

Based on observations $\lbrace \vX(\vs_1), \ldots, \vX(\vs_n)\rbrace$, we define an attribute-based dissimilarity matrix $\mD$, where each entry $d_{ij}$ represents the dissimilarity between observations at locations $\vs_i$ and $\vs_j$.  Following \citet[Section 14.3.1]{hastie2009elements}, we define the dissimilarity as an aggregation across attributes, $d_{ij} = \sum_{k=1}^{p} d_k(i,j)$, where $d_k(i,j)$ denotes the dissimilarity between observations at $\vs_i$ and $\vs_j$ for the $k$-th attribute. The choice of $d_k(i,j)$ depends on the data type. For example, when $\vX(\vs_i) \in \mathbb{R}^p$ consists of continuous attributes, we use the Euclidean distance $d_{ij} = \|\vX(\vs_i) - \vX(\vs_j)\|_2$. For binary or multinomial attributes, we use the Jaccard dissimilarity $d_{ij} = 1 - \frac{\vX(\vs_i) \cdot \vX(\vs_j)}{\|\vX(\vs_i)\|_1 + \|\vX(\vs_j)\|_1 - \vX(\vs_i) \cdot \vX(\vs_j)}.$

 To incorporate spatial contiguity constraints in clustering, we define spatial links $\ell_{ij} \in \{0,1\}$ based on the neighborhood relationships among spatial locations $\{\vs_1, \ldots, \vs_n\}$, which together form an adjacency matrix \citep[Section 13.3.1]{legendre2012numerical}. For a continuous domain, we construct an $m$-nearest neighbor structure. That is, for each location $\vs_i$, we set $\ell_{ij} = 1$ if $\vs_j$ is among the $m$ closest locations to $\vs_i$ in terms of spatial distance, and $\ell_{ij} = 0$ otherwise. Under this construction, the contiguity constraint requires that clusters are merged only when there exists at least one spatial link between them. Alternative definitions of spatial links, such as distance thresholds or adjacency structures, can also be used depending on the application.

 Following \citet{guenard_hierarchical_2022}, the dissimilarity matrix $\mD$ is constrained by the spatial links $\ell_{ij}$, such that $d_{ij}$ is considered only when $\ell_{ij}=1$.  CAHC then proceeds  by initially assigning each observation $\vX(\vs_i)$ to its own cluster. At each step, it merges the pair of clusters with the smallest attribute-based dissimilarity $d_{ij}$, subject to a spatial contiguity constraint.   This process continues until either all observations are grouped into a single connected cluster or constraints are met.  At each step, the inter-cluster distances are updated via the Lance--Williams formula \citep{lance_general_1967}.

 The Lance--Williams formula provides a recursive scheme for updating inter-cluster dissimilarities after merging two clusters $i$ and $j$ to form a new cluster $h$. The dissimilarity between the new cluster $h$ and another cluster $k$ is updated as
\begin{equation}
\label{eq:lw-update}
d_{k,h} = \alpha_i d_{i,k} + \alpha_j d_{j,k} + \beta d_{i,j} + \gamma \lvert d_{i,k} - d_{j,k} \rvert,
\end{equation}
where $d_{i,k}$, $d_{j,k}$, and $d_{i,j}$ denote the dissimilarities between clusters $i$ and $k$, $j$ and $k$, and $i$ and $j$, respectively, and $\alpha_i, \alpha_j, \beta,$ and $\gamma$ depend on the chosen linkage method (e.g., single, complete, average, or Ward linkage) and the values for common linkage methods are provided in Table~\ref{tab_param}.

 CAHC requires two tuning parameters: (1) the neighborhood size $m$, defined as the number of nearest neighbors used to construct spatial links, and (2) the number of clusters $G$. These parameters control the strength of spatial contiguity and the granularity of the resulting clusters.   In practice, we select both parameters using the silhouette score as described below.

The silhouette score is a metric for evaluating clustering quality, based on the trade-off between within-cluster cohesion and between-cluster separation \citep{rousseeuw_silhouettes_1987}. A higher silhouette score indicates that clusters are both compact and well-separated. We adapt this metric to incorporate spatial links $\ell_{ij}$, yielding a spatially-informed silhouette score.  Without incorporating spatial links, the silhouette score compares all clusters, including those that are not spatially connected. Such comparisons are not relevant under the contiguity constraint in CAHC. 

Suppose CAHC produces $G$ clusters, denoted by ${\mathcal{C}^{(1)}, \ldots, \mathcal{C}^{(G)}}$. For a spatial location $i \in \mathcal{C}^{(g)}$, the average within-cluster dissimilarity is defined as
\begin{equation}\label{a_i}
a(i) = \frac{1}{n_g - 1} \sum_{\substack{j \in \mathcal{C}^{(g)} \ j \neq i}} d_{ij},
\end{equation}
where $n_g = |\mathcal{C}^{(g)}|$ is the number of points in cluster $\mathcal{C}^{(g)}$.

To compute the average between-cluster dissimilarity, we restrict comparisons to clusters that are spatially adjacent to $\mathcal{C}^{(g)}$.  That is, clusters sharing at least one spatial link with $\mathcal{C}^{(g)}$.  The modified between-cluster dissimilarity for point $i$ is given by
\begin{equation}\label{b_i}
b(i) = \min_{\mathcal{C}^{(\ell)} \neq \mathcal{C}^{(g)}: \ell_{g\ell} = 1} \left( \frac{1}{|\mathcal{C}^{(\ell)}|} \sum_{j \in \mathcal{C}^{(\ell)}} d_{ij} \right),
\end{equation}
where $\ell_{g\ell} = 1$ indicates at least one spatial link exists between any point in $\mathcal{C}^{(g)}$ and any point in $\mathcal{C}^{(\ell)}$. The spatially-informed silhouette score for point $i$ is then
\begin{equation}\label{silhouette}
sh(i) = \frac{b(i) - a(i)}{\max\lbrace a(i), b(i) \rbrace}.
\end{equation}

This modified silhouette score provides a more accurate assessment of clustering quality in the spatial setting by jointly evaluating cluster compactness and separation, while respecting spatial constraints. 

\subsubsection{Illustrative Example (Algorithm~\ref{algo1}: Step 1--2)}

In the illustrative example with $n = 2{,}500$ spatial locations, we consider candidate neighbourhood sizes $m = 2, \ldots, 10$ and numbers of clusters $G = 2, \ldots, 10$. The spatially-informed silhouette score is used to select these parameters, and the average silhouette score is maximized at $m = 8$ and $G = 7$, as shown in Figure~\ref{fig_sil}(a). The resulting cluster labels for all points are presented in Figure~\ref{fig_sim_const_results}.

We now turn to assessing whether the clusters identified by CAHC exhibit RSP. To this end, we define a multiple hypotheses testing procedure based on the maximum mean discrepancy squared (MMD$^2$) statistic.

\subsection{Testing for Repeated Spatial Patterns (RSP) Across Clusters}
\label{test-repSpat}

Given $G$ clusters obtained from CAHC using a selected neighborhood size $m$, we assess whether any pair of clusters shares a similar spatial distribution to identify RSP.

Let $\mathcal{C}^{(g)} = \{\vs_1^{(g)}, \ldots, \vs_{n_g}^{(g)}\}$ and $\mathcal{C}^{(h)} = \{\vs_1^{(h)}, \ldots, \vs_{n_h}^{(h)}\}$ denote two such clusters, each corresponding to a spatial subregion. The associated multivariate attributes are given by
$\vX^{(g)} = \{ X(\vs_i^{(g)}) \in \mathbb{R}^p : i = 1, \ldots, n_g \}$ and
$\vX^{(h)} = \{ X(\vs_j^{(h)}) \in \mathbb{R}^p : j = 1, \ldots, n_h \}$. Let $\text{P}^{(g)}$ and $\text{P}^{(h)}$ denote the joint distributions of $\vX^{(g)}$ and $\vX^{(h)}$, respectively. We test the null hypothesis $\text{H}_0: \text{P}^{(g)} = \text{P}^{(h)}$ and the alternative is that $\text{H}_A: \text{P}^{(g)} \neq \text{P}^{(h)}$.

To test this hypothesis, we employ a nonparametric test statistic, the maximum mean discrepancy squared (MMD$^2$), which measures the difference between $P^{(g)}$ and $P^{(h)}$ based on their embeddings in a reproducing kernel Hilbert space (RKHS) \citep{gretton2006kernel}. We now formally define the RKHS and MMD$^2$.

We follow the definition of RKHS given in \citep[pp. 6–9]{berlinet2011reproducing}. Let $k: \mathcal{X} \times \mathcal{X} \to \mathbb{R}$ be a positive definite kernel, where $\mathcal{X} \subseteq \mathbb{R}^p$. The RKHS, $\mathcal{H}_k$ associated with $k$ is a Hilbert space of functions $f: \mathcal{X} \to \mathbb{R}$ such that, for each $\vx \in \mathcal{X}$, the evaluation functional $\mathcal{F}_{\vx}: \mathcal{H}_k \to \mathbb{R}$, defined by $\mathcal{F}_{\vx}(f) = f(\vx)$, is bounded and linear. By Riesz's representation theorem, there exists a unique representer $k(\cdot, \vx) \in \mathcal{H}_k$ such that the reproducing property holds: $ f(\vx) = \langle f, k(\cdot, \vx) \rangle_{\mathcal{H}_k}$. This property also implies that for any $\vx, \vy \in \mathcal{X}$, the kernel function satisfies $k(\vx, \vy) = \langle k(\cdot, \vx), k(\cdot, \vy) \rangle_{\mathcal{H}_k}$.

We now define the MMD$^2$ between the distributions $P^{(g)}$ and $P^{(h)}$ in $\mathcal{H}_k$, following \cite{gretton2006kernel}:
\begin{equation}
\operatorname{MMD}^{2}\left(P^{(g)}, P^{(h)}; \mathcal{H}_k\right)
= \lVert \mu_{P^{(g)}} - \mu_{P^{(h)}} \rVert^2_{\mathcal{H}_k},
\end{equation}
where $\mu_{\text{P}^{(g)}} := \mathbb{E}_{\vx \sim \text{P}^{(g)}}\left[k(\cdot, \vx)\right]$ and $\mu_{\text{P}^{(h)}} := \mathbb{E}_{\vx \sim \text{P}^{(h)}}\left[k(\cdot, \vx)\right]$ are the kernel mean embeddings of $\text{P}^{(g)}$ and $\text{P}^{(h)}$ in $\mathcal{H}_k$, respectively. The properties of the mean embeddings are studied in \cite{muandet_kernel_2017}.

Given observed samples $\vx^{(g)} = \lbrace \vx_i^{(g)}\rbrace_{i=1}^{n_g}$ and $\vx^{(h)} = \{\vx_j^{(h)}\}_{j=1}^{n_h}$ drawn from $\text{P}^{(g)}$ and $\text{P}^{(h)}$, respectively, the empirical estimate of MMD$^2$ is given by
\begin{align}
\widehat{\operatorname{MMD}}^{2}(X^{(g)}, X^{(h)}; \mathcal{H}_k) =\; &
\frac{1}{n_g^2} \sum_{i=1}^{n_g} \sum_{j=1}^{n_g} k\left(\vx_i^{(g)}, \vx_j^{(g)}\right) 
+ \frac{1}{n_h^2} \sum_{i=1}^{n_h} \sum_{j=1}^{n_h} k\left(\vx_i^{(h)}, \vx_j^{(h)}\right) \notag \\
&- \frac{2}{n_g n_h} \sum_{i=1}^{n_g} \sum_{j=1}^{n_h} k\left(\vx_i^{(g)}, \vx_j^{(h)}\right). \label{eq:mmd}
\end{align}

The kernel function computes a measure of similarity between two vectors by implicitly evaluating their inner product in a high-dimensional feature space. For $k$ to be a valid reproducing kernel, it must satisfy Mercer’s conditions: it must be symmetric, continuous, and yield a positive semi-definite kernel matrix \citep[Theorem 2.1]{muandet_kernel_2017}. Many radial basis function (RBF) kernels, such as the Gaussian kernel,  define similarity based on a distance measure $\lVert\vx - \vy\rVert$, which enters directly into the kernel evaluation.  For multivariate binary attributes, Euclidean distance may be replaced with the Jaccard distance.  Table~\ref{tab_kernel} summarizes the three kernels investigated in this study, where the choice of distance is adapted to the data type. 

\begin{table}[t]
	\centering
	\caption{Three kernel functions based on the Euclidean distance $\|\vx - \vy\|$. The parameters $\sigma$ and $c$ are kernel-specific bandwidth or shape parameters.}
	\renewcommand{\arraystretch}{1.4}
	\begin{tabular}{@{} l l @{}} 
		\toprule
		Kernel & Formula \\  
		\midrule
		Gaussian (RBF) kernel        & $\exp\left(-\|\vx - \vy\|^{2} / \sigma^2\right)$  \\
		Laplacian kernel             & $\exp\left(-\|\vx - \vy\|_{1} / \sigma\right)$ \\
		Inverse multiquadratic (IMQ) kernel & $\left(\|\vx - \vy\|^{2}_{2} + c^2\right)^{-1/2}$  \\
		\bottomrule
	\end{tabular}
	\label{tab_kernel}
\end{table}

In \ref{app1}, we conduct simulations by varying the number of spatial points $n$, attribute dimension $p$, spatial autocorrelation $\eta$, kernel type, and kernel-specific parameters to investigate the sampling distribution of the $\widehat{\operatorname{MMD}}^{2}$ statistic under scenarios where the underlying distributions of two clusters are either identical or different. In Figures \ref{fig:mmd-prop1}-\ref{fig:mmd-prop3}, we find that the inverse multiquadratic (IMQ) kernel with shape parameter $c \geq 1$ shifts the distribution of $\widehat{\operatorname{MMD}}^{2}$ further away from zero when the cluster distributions differ, thereby enhancing sensitivity. 

The kernel parameter (e.g., bandwidth in Gaussian and Laplacian kernels, shape parameter in the IMQ kernel) influences the power of the $\widehat{\operatorname{MMD}}^{2}$ test under the alternative hypothesis by controlling the kernel's sensitivity to differences in distributions. In high-dimensional settings with multivariate attributes, larger kernel parameters are often necessary. This is because, as dimensionality increases, pairwise distances between observations tend to grow, causing kernel similarity values to shrink toward zero and reducing the ability to distinguish between distributions.

To address this, the kernel parameter should be scaled relative to the distribution of pairwise distances. For instance, in Figures \ref{fig:mmd-prop4} - \ref{fig:mmd-prop5} and Figures \ref{fig:mmd-prop6}-\ref{fig:mmd-prop7}, the Gaussian and Laplacian kernels requires a bandwidth $\sigma$ that is at least 2 to avoid near-zero kernel values for the simulated data in Section \ref{sim-sett}. Similarly, for the IMQ kernel, the parameter $c$ should be set to balance attenuation of long-range distances while preserving sensitivity to meaningful local variation. Thus, appropriate kernel parameter selection is critical for maintaining test power, especially in high-dimensional or sparse-feature regimes, or when the distance measure is not normalized to lie within $[0,1]$.

Note that the $\widehat{\operatorname{MMD}}^{2}$ statistic in Eq.~\eqref{eq:mmd} is biased, but consistent when the observations in $\vx^{(g)}$ and  $\vx^{(h)}$  are independent samples from $P^{(g)}$ and $P^{(h)}$, respectively \citep[Theorem 4]{gretton2006kernel}. Next, we show that $\widehat{\operatorname{MMD}}^{2}$ is consistent under spatial dependence.

 Under the null hypothesis, we assume that the spatial processes $\vX^{(g)}$ and $\vX^{(h)}$ within clusters $\mathcal{C}^{(g)}$ and $\mathcal{C}^{(h)}$ are disjoint. In Section~\ref{proof-mmd-con}, we show that the unbiased version of Eq.~\eqref{eq:mmd} is asymptotically consistent under a model-free mixing condition on $\vX^{(g)}$ and $\vX^{(h)}$, together with a summability condition on the dependence coefficients $Q_{\ell}$ in Eq.~\eqref{eq:ql}. Consequently, under the same conditions, the biased version in Eq.~\eqref{eq:mmd} is also asymptotically consistent.

As demonstrated in our simulation study, the variance of the $\widehat{\operatorname{MMD}}^{2}$ statistic remains finite even as the strength of spatial autocorrelation increases from weak to strong levels (see Figures~\ref{fig:mmd-prop2} and~\ref{fig:mmd-prop3}). These results highlight the suitability of $\widehat{\operatorname{MMD}}^{2}$ for analyzing the cluster results from the CAHC.

\subsubsection{Illustrative Example (Algorithm~\ref{algo1}: Step 3(a))}

In Step~3(a) of Algorithm~\ref{algo1}, for each pair of clusters $(\mathcal{C}^{(g)}, \mathcal{C}^{(h)})$, we compute the empirical $\widehat{\operatorname{MMD}}^{2}$ using Eq.~\eqref{eq:mmd}. 

In the illustrative example, the CAHC step yields $G = 7$ clusters, resulting in $\binom{7}{2} = 21$ pairwise comparisons. For each pair, we compute $\widehat{\mathrm{MMD}}^2(X^{(g)}, X^{(h)})$ using the inverse multiquadratic (IMQ) kernel with parameter $c = 1$. The resulting values quantify the dissimilarity between the distributions of cluster-specific attributes. The observed $\widehat{\operatorname{MMD}}^{2}_{\text{obs}}$ values for all cluster pairs are reported in Table~\ref{tab_mmd_example}. Pairs with smaller $\widehat{\operatorname{MMD}}^{2}_{\text{obs}}$ values indicate clusters with more similar distributions, motivating the subsequent hypothesis testing and graph construction steps.

\subsection{Block Permutation}\label{block}

Motivated by Mantel’s test for spatial clustering \citep[pp. 14–16]{schabenberger2017statistical}, recommendations for structure-preserving resampling in spatial data \citep{guillot2013dismantling}, and subsampling over the  subregions to compute variance of a statistic \citep{sherman1996variance},  we propose a block permutation approach to approximate the null distribution of $\widehat{\operatorname{MMD}}^{2}$.

Under the null hypothesis that $\vX^{(g)} \sim \vX^{(h)}$, $\widehat{\operatorname{MMD}}^{2}$ in Eq.~\eqref{eq:mmd} is expected to be small. Standard permutation tests rely on the assumption that observations are exchangeable under the null. However, in our context, observations $\vX^{(g)}$ and $\vX^{(h)}$ may exhibit spatial or structural dependencies, such as local RSP within clusters. In particular, the kernel similarity $k(\vx_i^{(g)}, \vx_j^{(g)})$ may capture local attribute similarity that would be disrupted if individual observations were randomly permuted. For instance, if $\vx_j^{(g)}$ is locally similar to $\vx_i^{(g)}$ but is replaced with a randomly selected $\vx_k^{(g)}$, the similarity $k(\vx_i^{(g)}, \vx_k^{(g)})$ may not preserve the same structure,  leading to invalid approximations of the null distribution.

To address this, we partition each cluster $\mathcal{C}^{(g)}$ and $\mathcal{C}^{(h)}$ into blocks based on attribute similarity using the $k$-means clustering algorithm.  Because the clusters identified by constrained clustering are already assumed to be approximately spatially proximal, we do not impose additional spatial constraints when forming blocks. If the CAHC procedure retains the dendrogram, block definitions can alternatively be obtained by cutting the dendrogram at a fixed height within each cluster, without requiring additional clustering. 

Once blocks are defined, entire blocks are permuted between $\mathcal{C}^{(g)}$ and $\mathcal{C}^{(h)}$ to approximate the sampling distribution of $\widehat{\operatorname{MMD}}^2$ under the null. Our method is conceptually related to structured resampling methods such as shift permutations \citep{guillot2013dismantling}, which aim to preserve spatial dependence when generating null distributions. The detailed block permutation procedure for testing between clusters $\mathcal{C}^{(g)}$ and $\mathcal{C}^{(h)}$ is outlined in \ref{blk-perm}.

For each hypothesis test, the observed statistic $\widehat{\operatorname{MMD}}^{2}_{\text{obs}}$ is compared against the null distribution obtained via block permutation. To account for multiple testing across all distinct cluster pairs identified by CAHC, we apply the Benjamini–Hochberg procedure to control the false discovery rate (FDR) at a threshold of 0.05 \citep{benjamini1995controlling}. Finally, \texttt{repSpat} reassigns cluster labels based on the pairwise hypothesis tests.

\subsubsection{Illustrative Example (Algorithm~\ref{algo1}: Steps 3(b)--3(c))}

In the illustrative example, Steps~3(b)--3(c) of Algorithm~\ref{algo1} are used to approximate the null distribution of the $\widehat{\operatorname{MMD}}^{2}$ statistic via block permutation and to perform hypothesis testing for each pair of clusters.

Figure~\ref{blk_perm_illus_example} shows the sampling distributions of $\widehat{\operatorname{MMD}}^{2}$ obtained from the block permutation procedure. The first two rows correspond to comparisons between clusters generated from the same underlying distribution (within-region comparisons), where the $\widehat{\operatorname{MMD}}^{2}$ values are concentrated near zero. The remaining rows correspond to comparisons between clusters generated from different distributions (between-region comparisons), where the $\widehat{\operatorname{MMD}}^{2}$ values are shifted away from zero.

These results illustrate that the block permutation procedure provides a clear separation between the null and alternative distributions, enabling reliable hypothesis testing for detecting distributional differences between clusters. Table~\ref{tab:pvalues} shows the hypothesis testing results for all cluster pairs.

\subsection{Reassigning the Cluster Labels}\label{cl-repart}

To guide the final reassignment step, we adopt a graph-based approach in which each node represents a cluster obtained from CAHC,  and an edge represents a pair of clusters for which there is insufficient evidence to conclude that their distributions differ. Each edge is labeled with the corresponding observed $\widehat{\operatorname{MMD}}^{2}_{\text{obs}}$ value. 

As a rule of thumb, clusters that form fully connected subgraphs (cliques) are reassigned to a common label. When fully connected subgraphs do not form, the reassignment decision can be guided by the magnitude of the $\widehat{\operatorname{MMD}}^{2}_{\text{obs}}$ values, with smaller values indicating higher similarity between clusters.

\subsubsection{Illustrative Example (Algorithm~\ref{algo1}: Step 4(a)--4(b))}

Figure~\ref{sample_pwr_graph} illustrates how the graph is constructed, how cliques are identified, and how cluster labels are reassigned.

\section{Simulation Study}
\label{sim-study}

 We evaluate the performance of the proposed \texttt{repSpat} framework through simulation studies designed to assess its ability to detect RSP under varying spatial dependence structures and cluster configurations. 
\subsection{Simulation Settings}
\label{sim-sett}

 We conduct simulation studies to evaluate the performance of \texttt{repSpat} in detecting repeated spatial patterns (RSP). In addition to assessing its ability to identify RSP, we examine the false discovery rate (FDR) control of $\widehat{\operatorname{MMD}}^{2}$ test when paired with block permutation. We also compare the clustering accuracy of \texttt{repSpat} with that of \texttt{Banksy}, which can detect RSP when appropriately tuned for spatial neighborhood features.  We consider \texttt{Banksy} as a benchmark because it is a recently proposed method for spatial domain detection in spatial omics data, incorporates spatial neighborhood information, and is implemented in an accessible R/Bioconductor package with documented tuning procedures. In contrast, methods such as \texttt{STICC} are developed under Gaussian model assumptions and are tailored to settings with smaller sample sizes, which differ from the large-scale, nonparametric spatial setting considered in this study. 

\begin{figure}[t]
	\centering
	\includegraphics[width=0.8\textwidth]{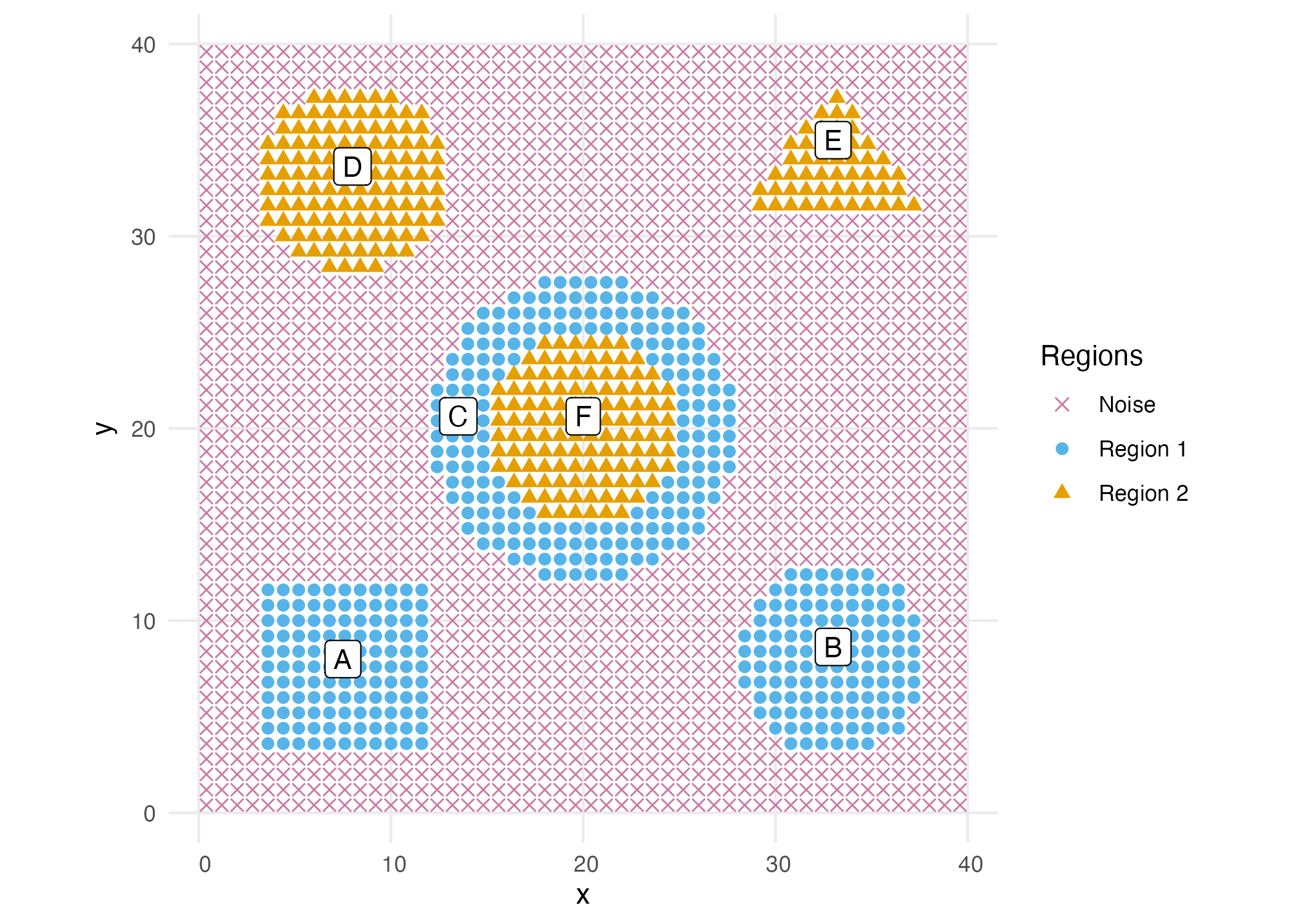}
	\caption[Simulation domain setting with 2500 coordinates]{Simulation setup with 2500 spatial locations. Repeated spatial patterns are generated as Region 1 and Region 2, each comprising three patches. The remaining area is a noise region.}
	\label{fig_sim_setup}
\end{figure}

Each simulated dataset consists of three regions: Region 1, Region 2, and a background noise region. Region 1 and Region 2 each contain three spatially disjoint patches with with distinct spatial shapes as shown in Figure~\ref{fig_sim_setup}. The three patches in Region 1 are generated from the same underlying distribution, and similarly the three patches in Region 2 are generated from another common distribution. Thus, Regions 1 and 2 represent RSP. The remaining area is treated as a noise region and is generated independently from a standard Gaussian distribution with mean 0 and variance 1.

The simulations vary the total number of observations ($n$), the number of features ($p$), and the strength of spatial dependence ($\eta$). We consider the combinations $n \in \{2500, 4900\}$, $p \in \{5, 10\}$, and $\eta \in \{0.3, 0.8\}$, resulting in eight simulation settings. Here, a simulation setting refers to one fixed combination of $(n,p,\eta)$.   We simulate data from a Gaussian Markov random field (GMRF) to control simulation settings through model parameters.  In each setting, $p$ multivariate attributes are simulated independently using conditional autoregressive (CAR) models.

Following \cite{besag1974spatial}, for a finite number of locations $n$, the joint distribution for one attribute $\vX(\vs)= \left(X(\vs_1), X(\vs_2), \ldots, X(\vs_n)\right)^{T}$ is
\begin{equation}\label{Joint_CAR}
    \vX(\vs) \sim \text{Normal}_n\left( \vmu,\, (\mI_n - \mC)^{-1} \mM \right),
\end{equation}
where $\vmu$ is the $n \times 1$ mean vector, $\mI_n$ is the $n \times n$ identity matrix, and $\mM = \text{diag}(\tau_1^2, \ldots, \tau_n^2)$ is a diagonal matrix of variances assuming $\tau_i^2 = \tau^2 , \quad \forall i$. Moreover, we assume isotropic dependence such that $\mC = \eta \mW$, where $\eta$ is the spatial autocorrelation parameter and $\mW$ is the spatial weight matrix.  Here, $\mW = (w_{ij})$ is a row-standardized spatial weight matrix constructed from the $m$-nearest neighbor structure, where $w_{ij} > 0$ if $\vs_j$ is among the $m$ nearest neighbors of $\vs_i$ and $w_{ij} = 0$ otherwise, with each row normalized to sum to one. This normalization ensures that $(\mI_n - \eta \mW)$ is invertible for $\eta$ in an appropriate range.

Spatial coordinates are simulated over a square domain $[0, 40] \times [0, 40]$, allowing infill asymptotics,  where the spatial domain is fixed while the number of sampling locations increases, leading to decreasing distances between neighboring points \citep[Section 5.8]{cressie2015statistics}.   Figure~\ref{fig_sim_setup} shows an example configuration with $n=2500$, containing six distinct regions (labeled A through F), which include circles, a triangle, a rectangle, and a donut-shaped region (comprising inner and outer circles). Within each region, spatial attributes $\vX(\vs)$ are independently generated for $p$ attributes using the CAR model in \eqref{Joint_CAR}, with region-specific parameters designed to produce RSP.

In Figure~\ref{fig_sim_setup}, patches A, B, and C share the same parameters and form Region 1; patches D, E, and F form Region 2 with a different set of parameters. For Region 1, we use mean $\vmu_1 = 4\mathbf{1}_{n_{1}}$ and variance $\tau_1^2 = 1$; for Region 2, we use mean $\vmu_2 = 10\mathbf{1}_{n_{2}}$ and variance $\tau_2^2 = 1$. All six patches share an 8-nearest neighbor spatial weight matrix and the same autocorrelation parameter $\eta$. To account for background noise, we include an additional noise region, where the realizations are drawn from a standard Gaussian distribution with mean 0 and variance 1.

 In the baseline simulation design, all simulated features are signal features because each feature contributes to the difference in distributions between Region 1 and Region 2. To better reflect real-world data settings, we additionally consider versions of each simulation setting with irrelevant features by adding one, two, and three independent noise variables to all regions. These noise variables are generated independently from a standard Gaussian distribution and do not contribute to the differences between Region 1 and Region 2.

\subsection{Simulation Results}

 For each simulation setting, defined by a fixed combination of $(n, p, \eta)$, we generate 100 independent datasets, referred to as 100 simulation runs. Each dataset is generated using the same data-generating process but with different random realizations of the spatial field and noise variables.

Both \texttt{repSpat} and \texttt{Banksy} are applied to the same simulated datasets within each setting. For each simulation run, we apply constrained agglomerative hierarchical clustering (CAHC) as the initial clustering step for \texttt{repSpat}.  The tuning parameters, including the neighbourhood size and the number of clusters, are selected by maximizing the spatially-informed silhouette score over candidate values based on a representative run under the setting where all features are signal.  These selected parameters are then held fixed across the remaining runs and are also used for the corresponding settings with additional noise variables. We use the inverse multiquadric (IMQ) kernel with $c = 1$ to compute $\widehat{\operatorname{MMD}}^{2}$. Larger values of $c$ may increase the power of the hypothesis test, as illustrated in Figures~\ref{fig:mmd-prop1}--\ref{fig:mmd-prop3}.

\begin{figure}[t]
     \centering
     \begin{subfigure}[b]{0.46\textwidth}
         \centering
         \includegraphics[width=\textwidth]{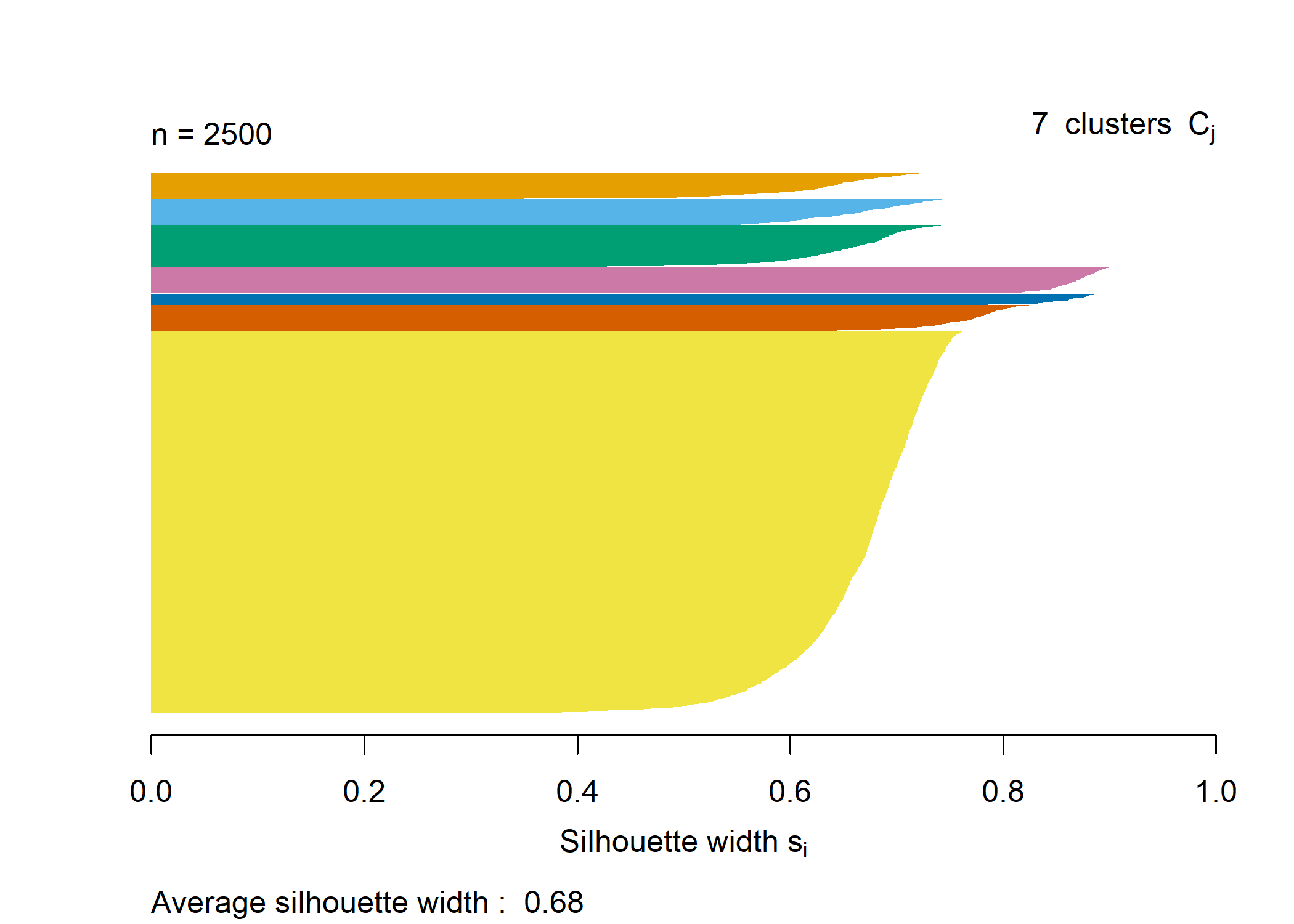}
         \caption{}
         \label{adj_sil1}
     \end{subfigure}
     \hspace{0.05\textwidth}
     \begin{subfigure}[b]{0.46\textwidth}
         \centering
         \includegraphics[width=\textwidth]{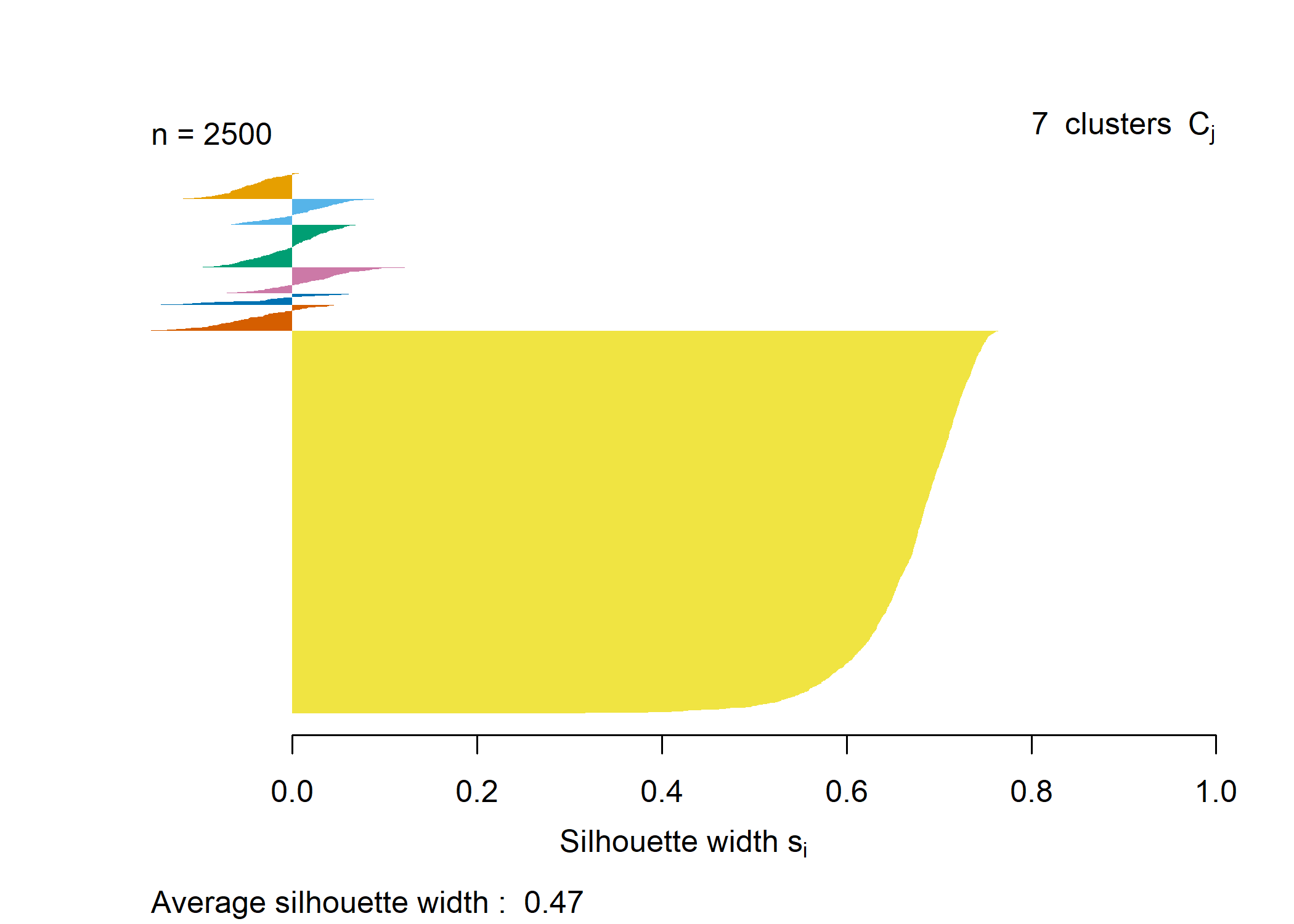}
         \caption{}
         \label{std_sil2}
     \end{subfigure}
     \caption[Comparison of silhouette score plots]{Comparison of silhouette scores for a representative simulation run with $n = 2500$, $p = 5$, and $\eta = 0.8$. a) Modified silhouette score incorporating spatial links;
(b) Standard silhouette score.}
     \label{fig_sil}
\end{figure}

 Across all settings, this procedure resulted in the selection of seven initial clusters and eight nearest neighbors to define spatial links. Figure~\ref{fig_sil} compares the modified silhouette score with the standard version for the representative run. Details of the selection procedure for this representative run are provided in \ref{app4}.   The results indicate that the modified silhouette score provides more reliable guidance for spatial clustering under CAHC.

\begin{figure}[t]
	\centering
	\includegraphics[width=0.8\textwidth]{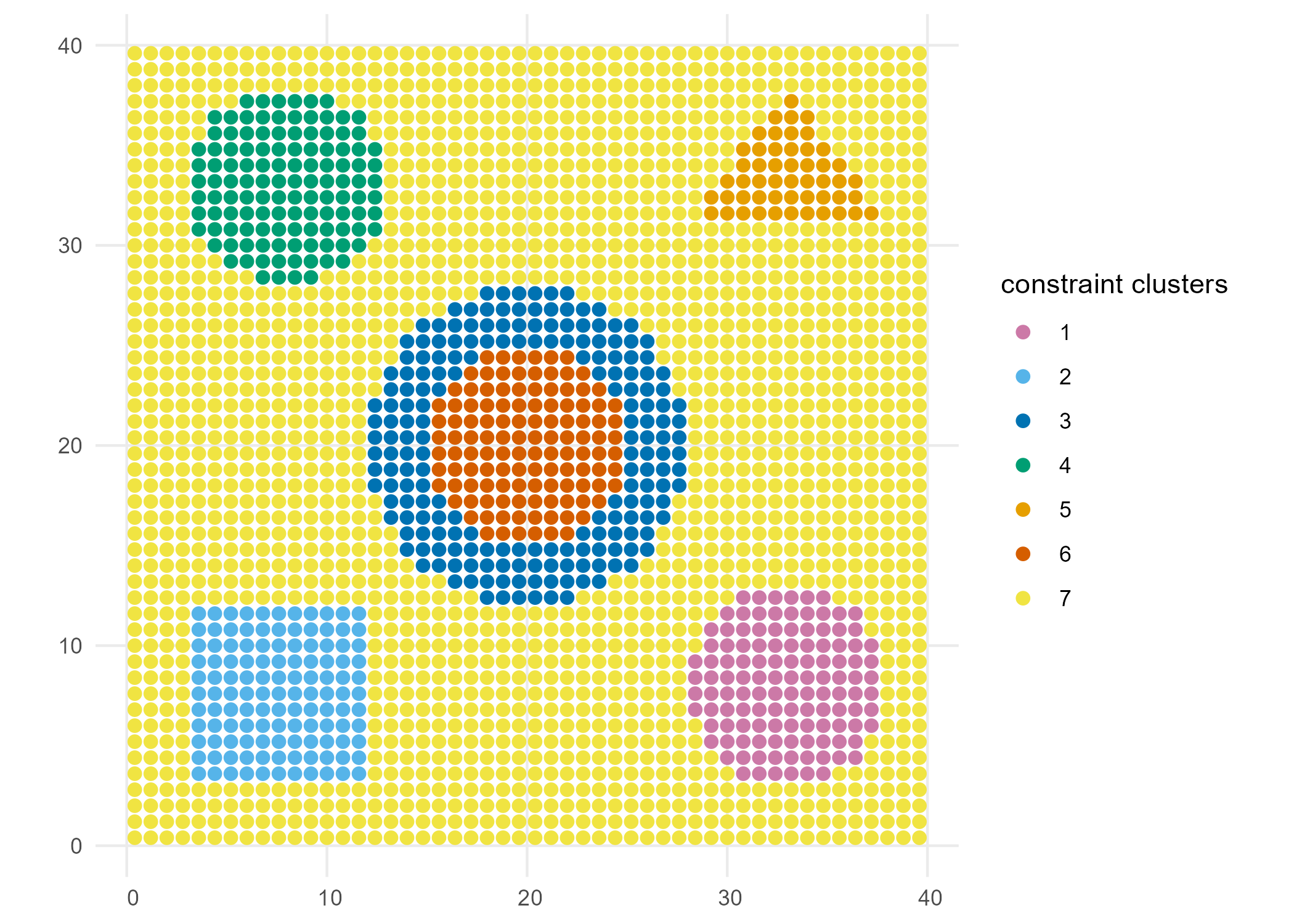}
	\caption[CAHC result for Setup 2]{Constrained clustering result from CAHC for the simulation setting $n = 2500$, $p = 5$, and $\eta = 0.8$.}
	\label{fig_sim_const_results}
\end{figure}

As illustrated in Figure~\ref{fig_sim_const_results}, CAHC detects the same seven clusters across all simulation settings. Thus, CAHC results in more clusters than the true number of clusters in the presence of RSP. To address this, we apply the \texttt{repSpat} framework to the CAHC output. Clusters are reassigned to a common label if they form a clique in the similarity graph.

 To evaluate the testing performance of \texttt{repSpat}, for each simulation run, we record two types of errors: (1) false positives, where clusters sharing the same underlying distribution are identified as different; and (2) false negatives, where clusters with different distributions are incorrectly identified as similar.

Table~\ref{tab_sim_results} summarizes the results across all simulation settings. Overall, \texttt{repSpat} demonstrates strong performance, with median false positive rates (FPR) and false negative rates (FNR) equal to zero across all settings. This indicates that, in typical runs, the method correctly identifies both similar and dissimilar cluster distributions.

To assess variability across runs, the final column reports the proportion of runs with F1 scores less than 1. While the median F1 score is equal to 1 in all settings, a small proportion of runs exhibit F1 scores below 1, indicating occasional deviations from perfect classification. These deviations are slightly more frequent under stronger spatial autocorrelation ($\eta = 0.8$) and high-dimensional settings ($p = 10$).

Additional results with one, two, and three noise features are provided in Tables~\ref{tab_sim_results2} -- \ref{tab_sim_results4}. Across these settings, the median FPR and FNR remain at zero for \texttt{repSpat}  and only a small proportion of runs show F1 scores below 1. This suggests that the method is robust to moderate levels of feature noise. 

\begin{table}[t]
	\caption{Performance of \texttt{repSpat} over 100 simulation runs for each setting with no noise. For each run, 21 pairwise tests are conducted, of which six pairs correspond to true distributional differences and 15 pairs correspond to no differences. A false positive (FP) occurs when a pair of clusters sharing the same underlying distribution is incorrectly identified as different, while a false negative (FN) occurs when a pair of clusters with different distributions is incorrectly identified as similar. We report the median and interquartile range (IQR) of the proportions of false positives (FPR) and false negatives (FNR) across runs. Precision is defined as the proportion of correctly identified differences among all detected differences, and recall (sensitivity, or true positive rate) is the proportion of true differences that are correctly identified. The F1 score is computed as the harmonic mean of precision and recall. The final column reports the proportion of runs with F1 score less than 1.}
	\centering
	\renewcommand{\arraystretch}{1.4}
	\begin{tabular}{@{} c c c c c c c @{}}
		\toprule
		\textbf{$n$} & \textbf{$p$} & \textbf{$\eta$} & \textbf{FNR} & \textbf{FPR} & \textbf{F1 Score} & \textbf{\% F1 $<$ 1} \\
		\midrule
		2500 & 5  & 0.3 & 0 (0) & 0 (0) & 1 (0) & 0 \\
		& 5  & 0.8 & 0 (0) & 0 (0) & 1 (0) & 0.02 \\
		& 10 & 0.3 & 0 (0) & 0 (0) & 1 (0) & 0 \\
		& 10 & 0.8 & 0 (0) & 0 (0) & 1 (0) & 0.04 \\
		\midrule
		4900 & 5  & 0.3 & 0 (0) & 0 (0) & 1 (0) & 0 \\
		& 5  & 0.8 & 0 (0) & 0 (0) & 1 (0) & 0 \\
		& 10 & 0.3 & 0 (0) & 0 (0) & 1 (0) & 0 \\
		& 10 & 0.8 & 0 (0) & 0 (0) & 1 (0) & 0 \\
		\bottomrule
	\end{tabular}
	\label{tab_sim_results}
\end{table}

To compute clustering accuracy, we use the adjusted Rand index (ARI), which quantifies the similarity between clustering results and the true class labels while correcting for chance agreement \citep{hubert1985comparing}.  The Rand index (RI) is defined as the ratio of the number of pairs of observations that are assigned either to the same cluster or to different clusters in both partitions to the total number of pairs. The ARI adjusts the RI by accounting for its expected value under random permutations of the cluster labels.  ARI values range from 0 (no agreement beyond chance) to 1 (perfect concordance). 
 Let $\lbrace \mathcal{C}^{(1)}, \ldots, \mathcal{C}^{(G)}\rbrace$ and $\lbrace \mathcal{U}^{(1)}, \ldots, \mathcal{U}^{(R)}\rbrace$ denote the clusters from the estimated partition and the true labels, respectively. Let $n_{gr} = |\mathcal{C}^{(g)} \cap \mathcal{U}^{(r)}|$, $a_g = \sum_r n_{gr}$, and $b_r = \sum_g n_{gr}$. The adjusted Rand index is defined as
\begin{equation}
	\text{ARI} = \dfrac{\sum_{g,r} \binom{n_{gr}}{2} - \dfrac{\sum_g \binom{a_g}{2} \sum_r \binom{b_r}{2}}{\binom{n}{2}}}{\dfrac{1}{2} \left[ \sum_g \binom{a_g}{2} + \sum_r \binom{b_r}{2} \right] - \dfrac{\sum_g \binom{a_g}{2} \sum_r \binom{b_r}{2}}{\binom{n}{2}}}.
\end{equation} 

We evaluate the performance of \texttt{repSpat} and the recently proposed \texttt{Banksy} algorithm \citep{singhal_banksy_2024}.  For \texttt{Banksy}, we use a fixed parameter configuration with $\lambda = 0.8$ and $k = 3$, reflecting strong spatial smoothing and the true number of underlying regions, respectively. Table~\ref{tab_sim_comp} reports the median ARI with interquartile range (IQR) across 100 simulation runs for each setting, along with the percentage of runs in which the ARI of \texttt{repSpat} is less than 1. Across all settings, \texttt{repSpat} achieves a median ARI of 1 with IQR equal to 0, indicating consistent recovery of the true clustering structure. The final column shows that a small proportion of runs yield ARI values below 1, reflecting minor variability across runs. In contrast, \texttt{Banksy} produces lower ARI values, ranging from 0.75 to 0.83, indicating less accurate recovery of the underlying clusters. 

Additional results with one, two, and three noise features are provided in Tables~\ref{tab_sim_comp1}--\ref{tab_sim_comp3}. Across these settings, \texttt{repSpat} continues to achieve a median ARI equal to 1 with IQR equal to 0, and only a small proportion of runs exhibit ARI values below 1. This indicates that clustering performance remains stable under moderate levels of feature noise.

\begin{figure}[t]
     \centering
     \begin{subfigure}[b]{0.46\textwidth}
         \centering
         \includegraphics[width=\textwidth]{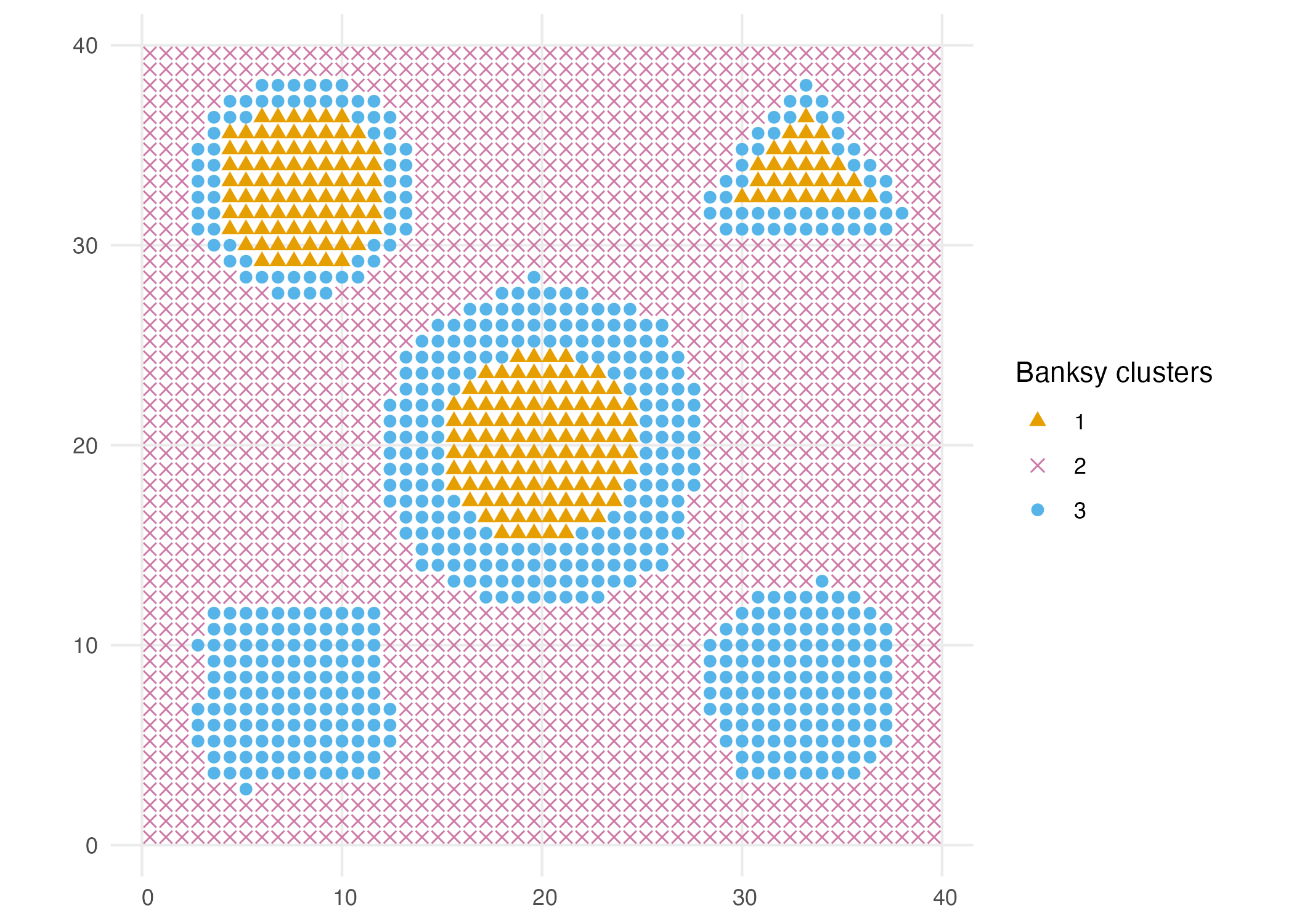}
         \caption{}
         \label{adj_sil}
     \end{subfigure}
     \hspace{0.05\textwidth}
     \begin{subfigure}[b]{0.46\textwidth}
         \centering
         \includegraphics[width=\textwidth]{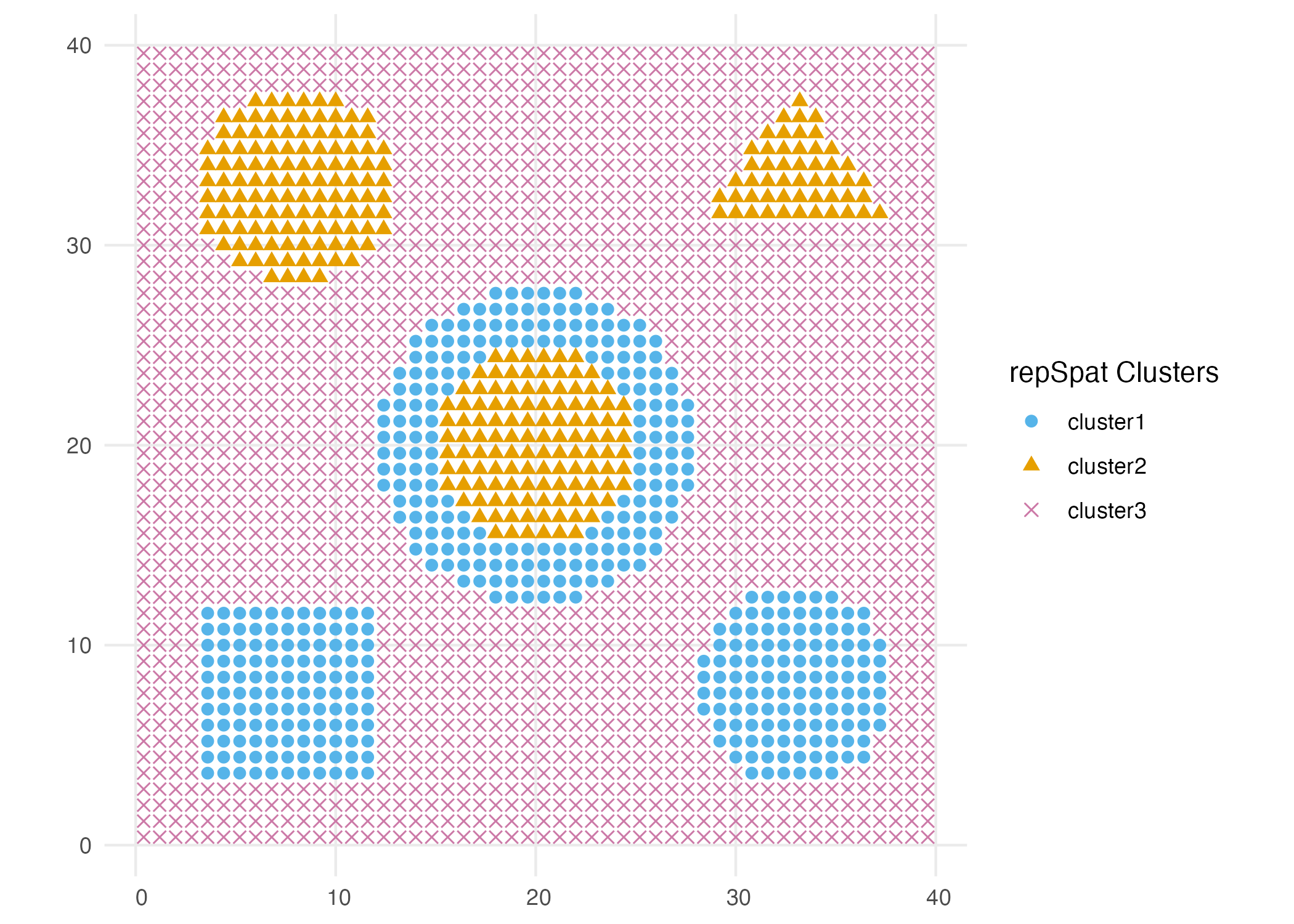}
         \caption{}
         \label{std_sil}
     \end{subfigure}
     \caption[Comparison of clustering results]{Example of (a) \texttt{Banksy} and (b) \texttt{repSpat} clustering results for the simulation setting with $n = 2500$, $p = 5$, and $\eta = 0.8$.}
     \label{bansky-repSpat}
\end{figure}

As shown in Figure~\ref{bansky-repSpat}(a), \texttt{Banksy} detects RSP but often misclassifies cluster boundaries, which reduces ARI performance. \texttt{Banksy} may  further improve performance if it is appropriately tuned or combined with alternative clustering algorithms.   However,  such  tuning may not be feasible in settings with many samples and irregularly shaped clusters. Figure~\ref{bansky-repSpat}(b) shows that \texttt{repSpat} correctly identifies three clusters and the cluster boundaries. Using \texttt{repSpat} with features learned by \texttt{Banksy} may improve the detection of RSP. Although our simulations focus on continuous multivariate attributes, the \texttt{repSpat} framework naturally extends to other data types, such as binary or multinomial features, through the use of  appropriate dissimilarity measures,  as demonstrated in the application.

\begin{table}[t]
	\caption{Clustering performance measured by the adjusted Rand index (ARI) for \texttt{repSpat} and \texttt{Banksy} over 100 simulation runs for each setting with no noise features. Entries report the median ARI with interquartile range (IQR) in parentheses. The final column reports the percentage of runs in which the ARI of \texttt{repSpat} is less than 1.}
	\centering
	\renewcommand{\arraystretch}{1.4}
	\begin{tabular}{@{} c c c c c c @{}} 
		\toprule
		\textbf{$n$} & \textbf{$p$} & \textbf{$\eta$} 
		& \textbf{repSpat (ARI)} 
		& \textbf{Banksy (ARI)} 
		& \textbf{\% ARI $<$ 1 (repSpat)} \\  
		\midrule
		2500 & 5  & 0.3 & 1 (0) & 0.81 (0.13) & 0.01 \\
		& 5  & 0.8 & 1 (0) & 0.81 (0.09) & 0.03 \\
		& 10 & 0.3 & 1 (0) & 0.82 (0.02) & 0 \\
		& 10 & 0.8 & 1 (0) & 0.82 (0.03) & 0.06 \\
		\midrule
		4900 & 5  & 0.3 & 1 (0) & 0.75 (0.09) & 0.01 \\
		& 5  & 0.8 & 1 (0) & 0.75 (0.09) & 0 \\
		& 10 & 0.3 & 1 (0) & 0.83 (0.10) & 0 \\
		& 10 & 0.8 & 1 (0) & 0.76 (0.09) & 0 \\
		\bottomrule
	\end{tabular}
	\label{tab_sim_comp}
\end{table}

These simulation results demonstrate that the \texttt{repSpat} framework is robust in identifying RSP across varying numbers of spatial points, multivariate attributes, and spatial configurations.  This robustness persists in the presence of additional noise features, where performance remains stable with only a small proportion of runs exhibiting imperfect classification.

\section{Real Data Application}
\label{appli}

We demonstrate \texttt{repSpat} to spatial proteomics data obtained through multiplexed ion beam imaging by time-of-flight (MIBI-TOF) technology from patients diagnosed with triple negative breast cancer (TNBC) \citep{keren_structured_2018}. These TNBC tissue samples may contain multiple regions of similar spatial patterns that are mixtures of biomarkers. In particular, we focus on the repeated tumor microenvironments (TMEs).  

The data set contains tissue samples from  39 patients, and each sample consists of spatial coordinates of cells within the tissue and expression levels for 36 protein markers. We use protein marker intensities and multivariate binary data derived from the marker threshold of these intensities \citep{bressan_dawn_2023} to uncover TMEs in each tissue sample. Binary markers can be considered to remove noise and are expected to enhance clustering results.  Descriptive summaries, including the mean expression of each protein marker within each tissue sample and the percentage of cells in which each marker is present, are provided in \ref{append_descriptive}.

\begin{figure}
    \centering
    \begin{tabular}{cc} 
        \begin{subfigure}{0.45\textwidth}
            \centering
            \includegraphics[width=\linewidth]{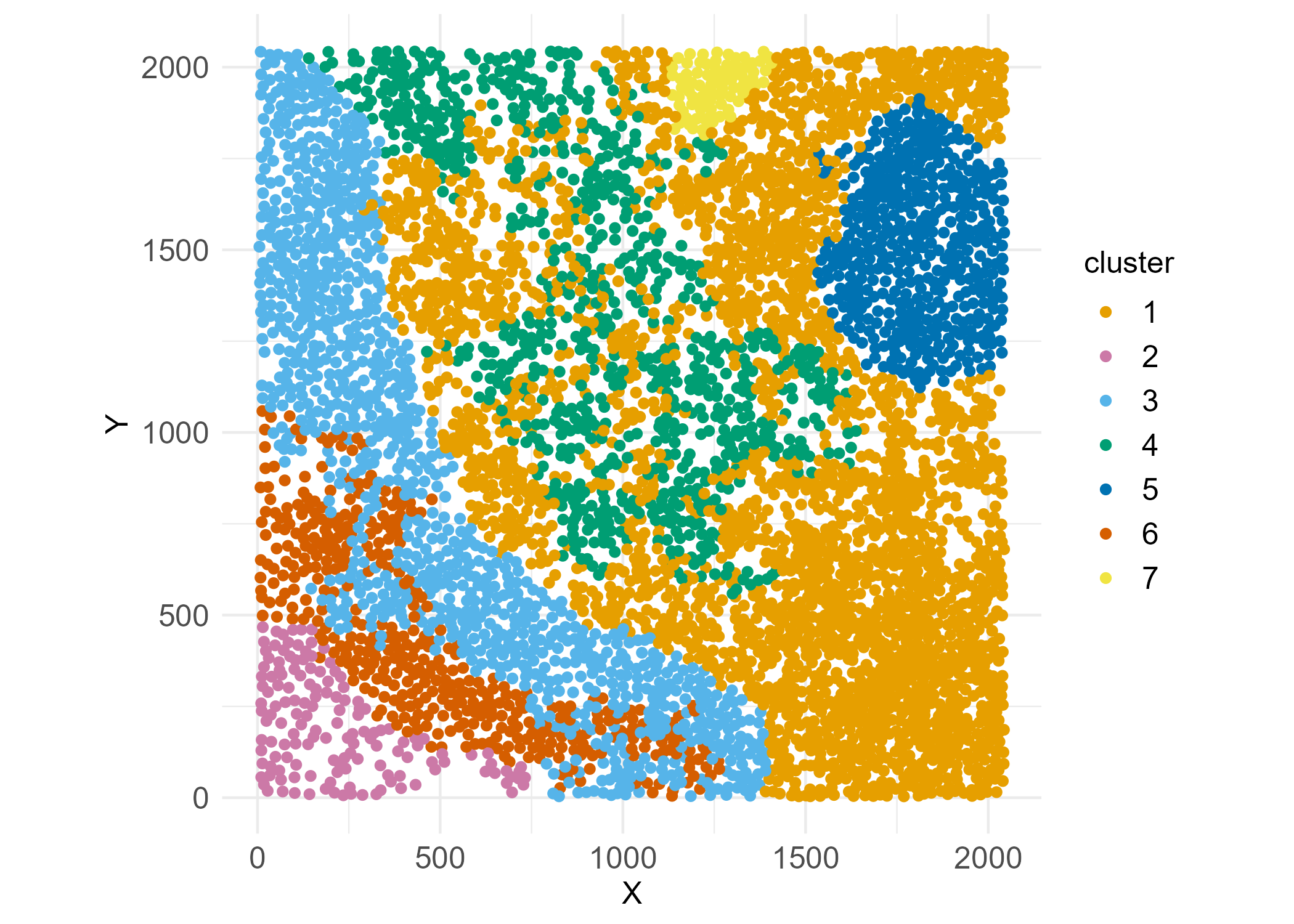}
            \caption{}
        \end{subfigure} &
        \begin{subfigure}{0.45\textwidth}
            \centering
            \includegraphics[width=\linewidth]{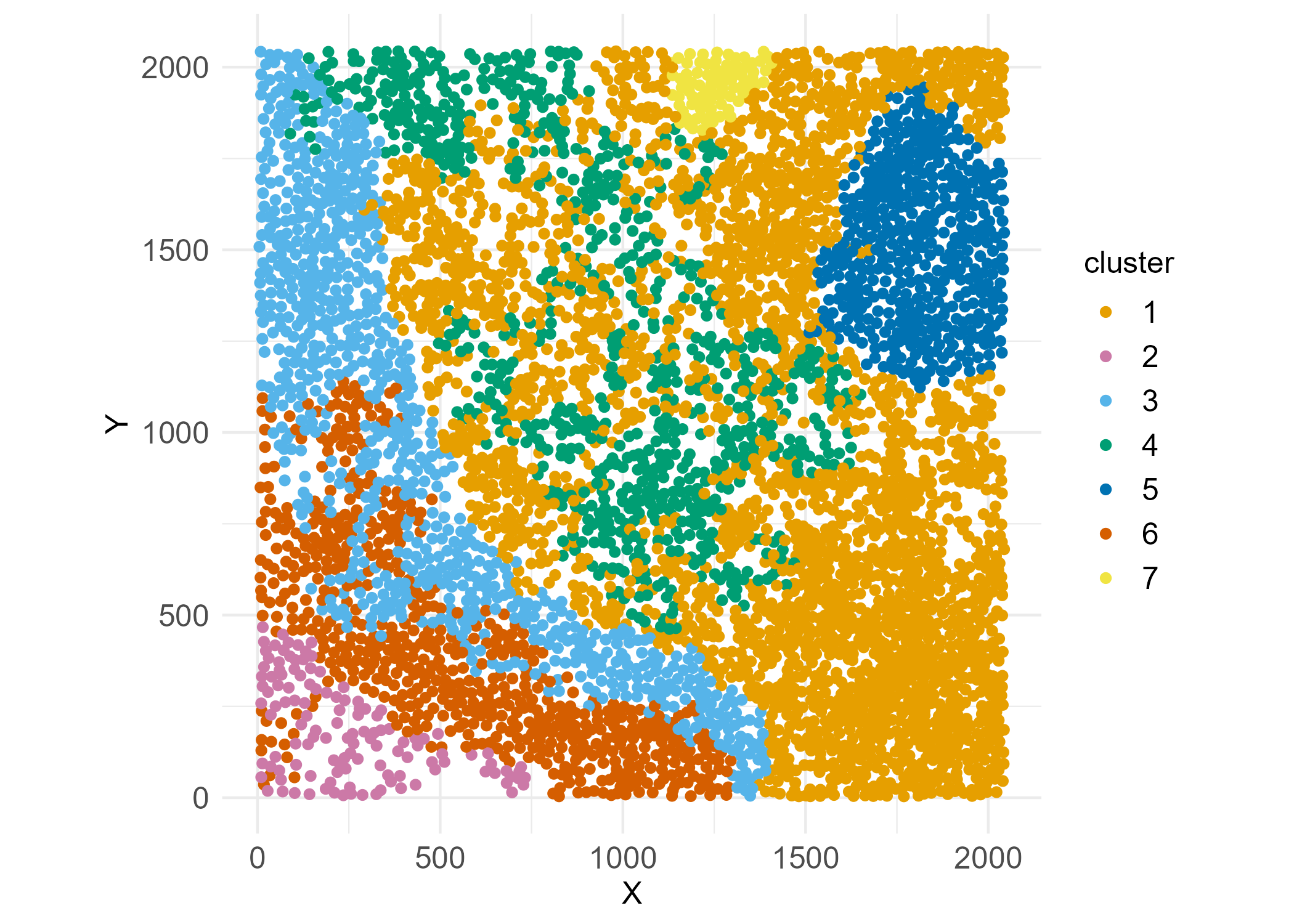}
            \caption{}
        \end{subfigure} \\
        \begin{subfigure}{0.45\textwidth}
            \centering
            \includegraphics[width=\linewidth]{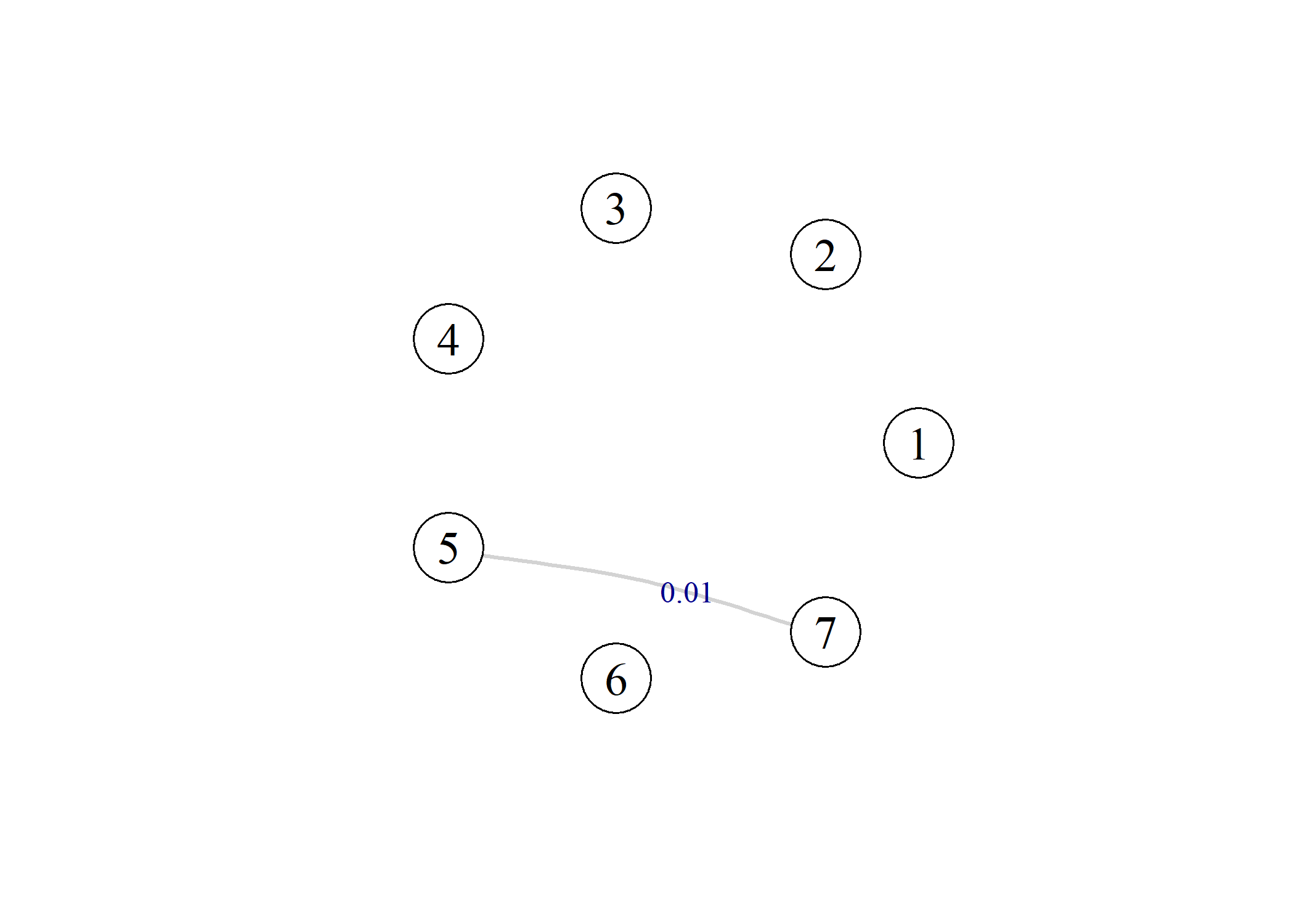}
            \caption{}
        \end{subfigure} &
        \begin{subfigure}{0.45\textwidth}
            \centering
            \includegraphics[width=\linewidth]{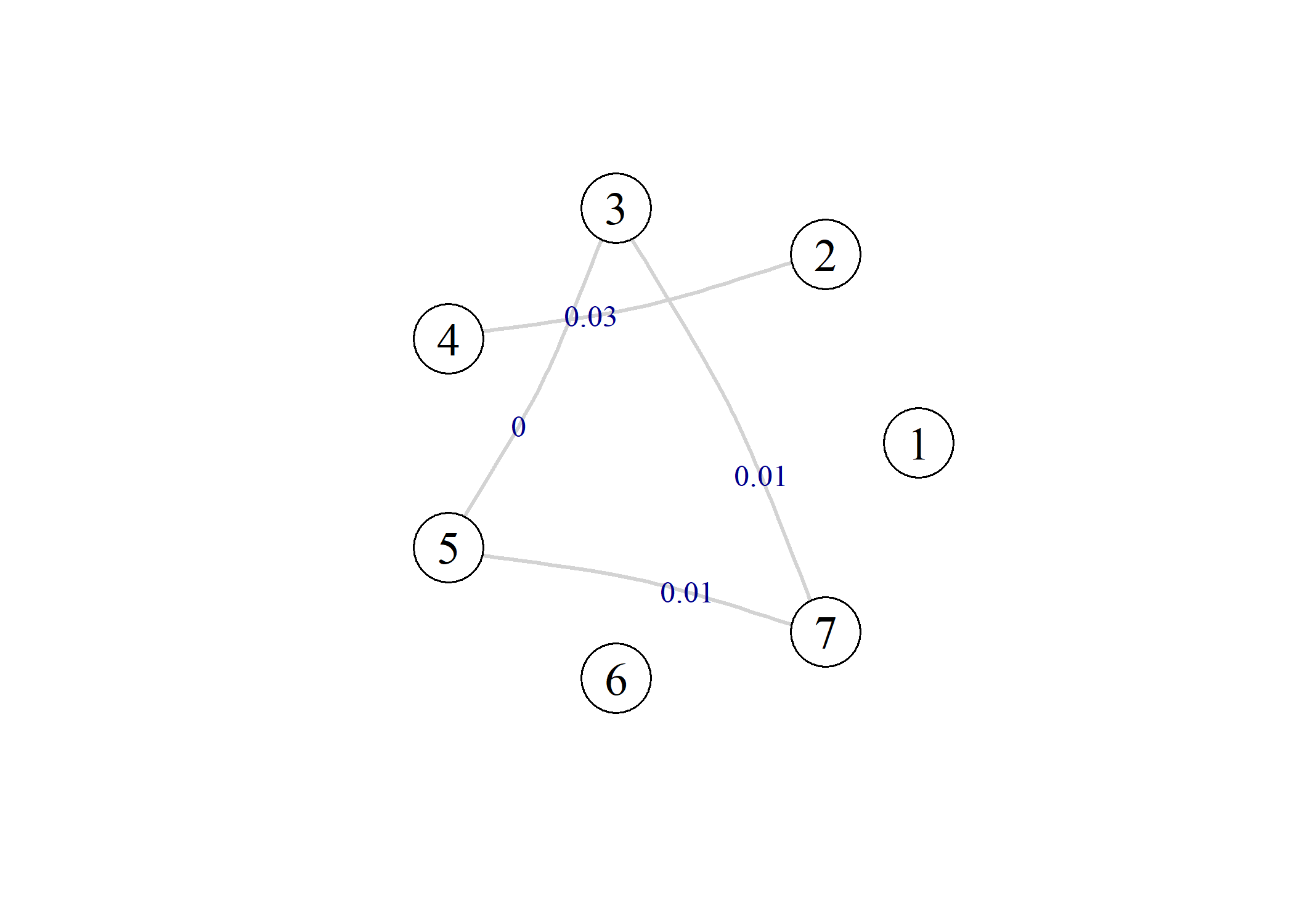}
            \caption{}
        \end{subfigure} \\
        \begin{subfigure}{0.45\textwidth}
            \centering
            \includegraphics[width=\linewidth]{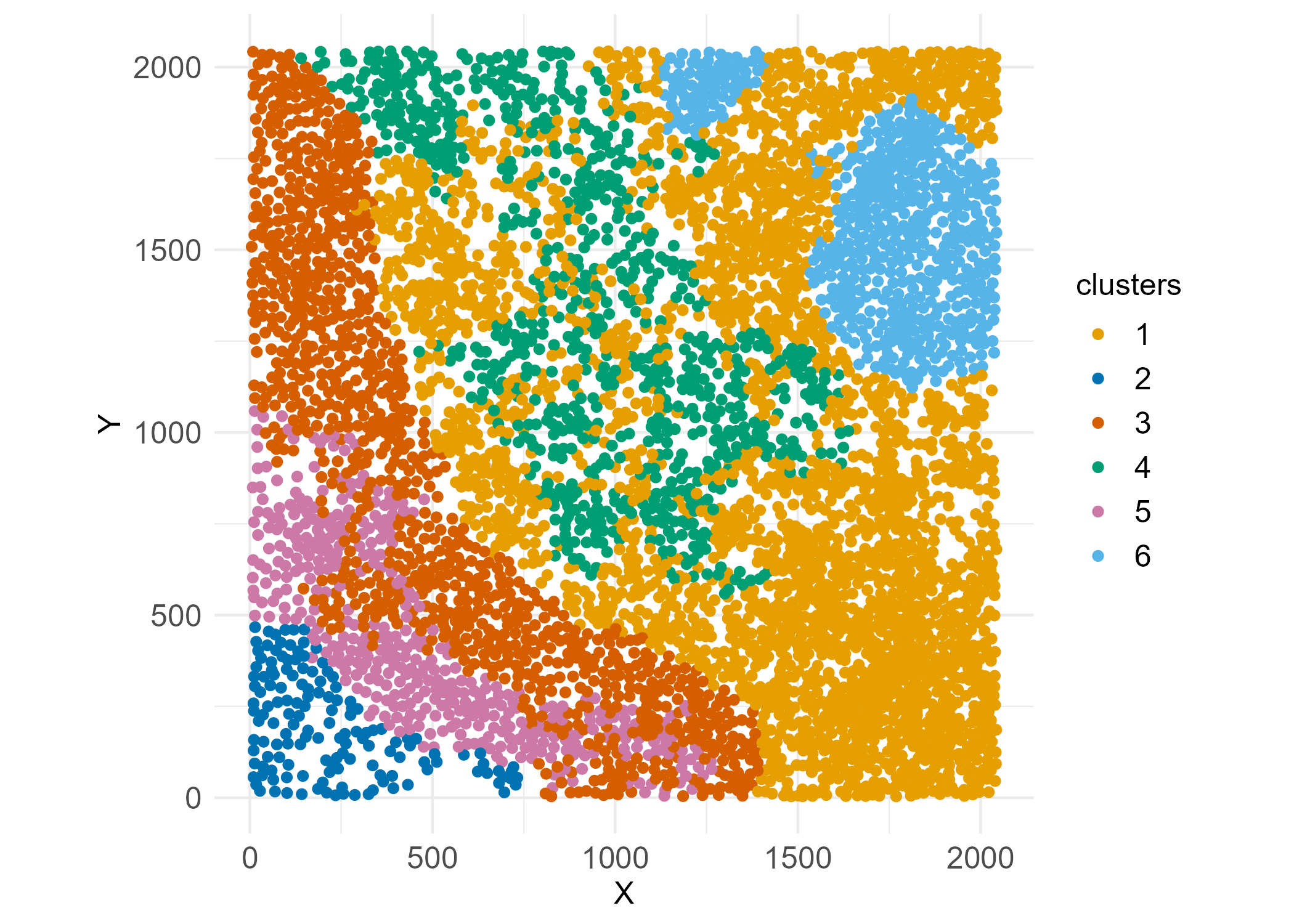}
            \caption{}
        \end{subfigure} &
        \begin{subfigure}{0.45\textwidth}
            \centering
            \includegraphics[width=\linewidth]{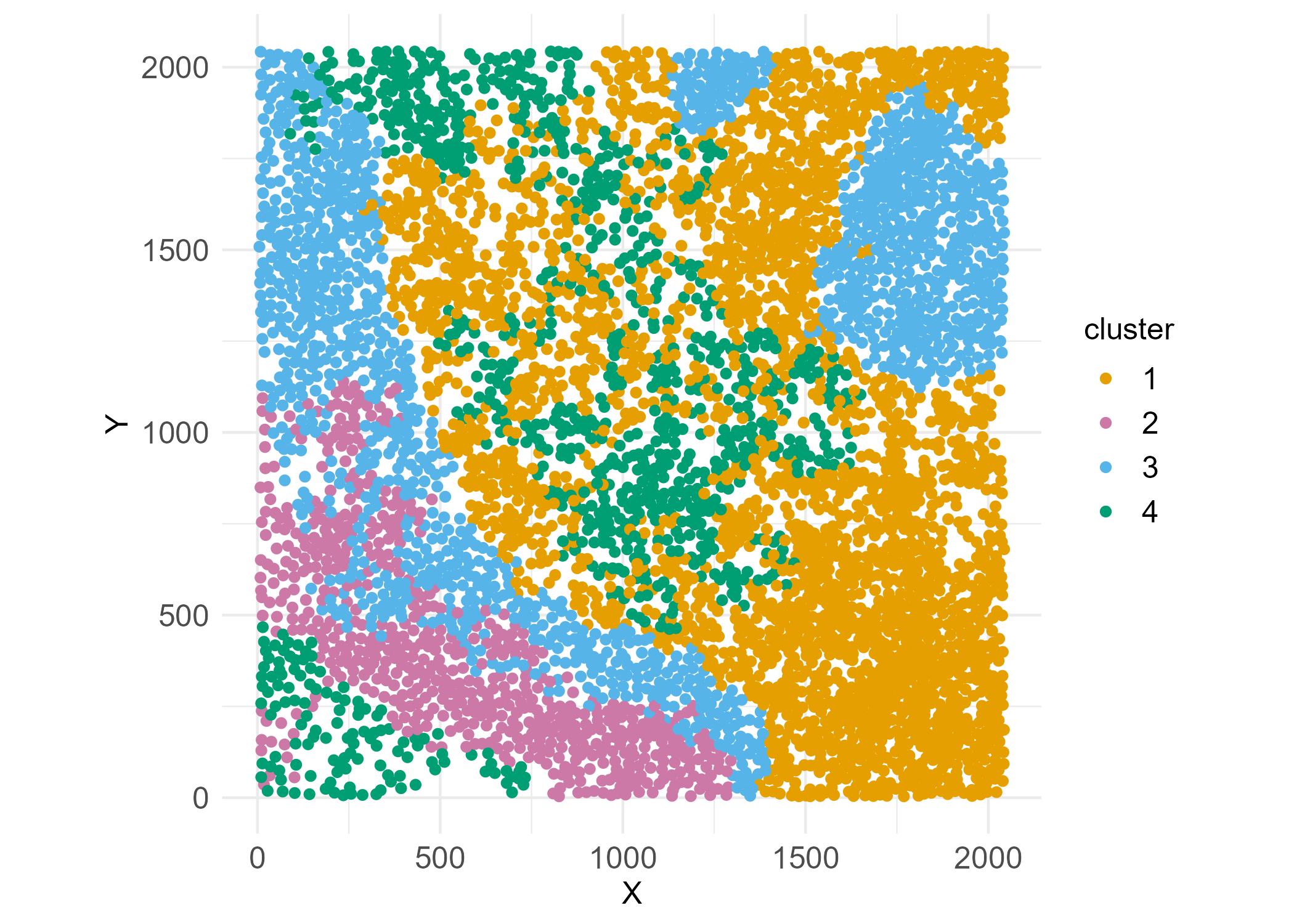}
            \caption{}
        \end{subfigure} \\
    \end{tabular}
    \caption{CAHC and \texttt{repSpat} results for Sample 04. CAHC clusters with (a) marker intensities and (b) binary markers. \texttt{repSpat} pairwise test results with (c) marker intensities and (d) binary markers. Reassigned cluster labels with (e) marker intensities and (f) binary markers.}
    \label{results-04}
\end{figure}

\begin{figure}
    \centering
    \begin{tabular}{cc} 
        \begin{subfigure}{0.45\textwidth}
            \centering
            \includegraphics[width=\linewidth]{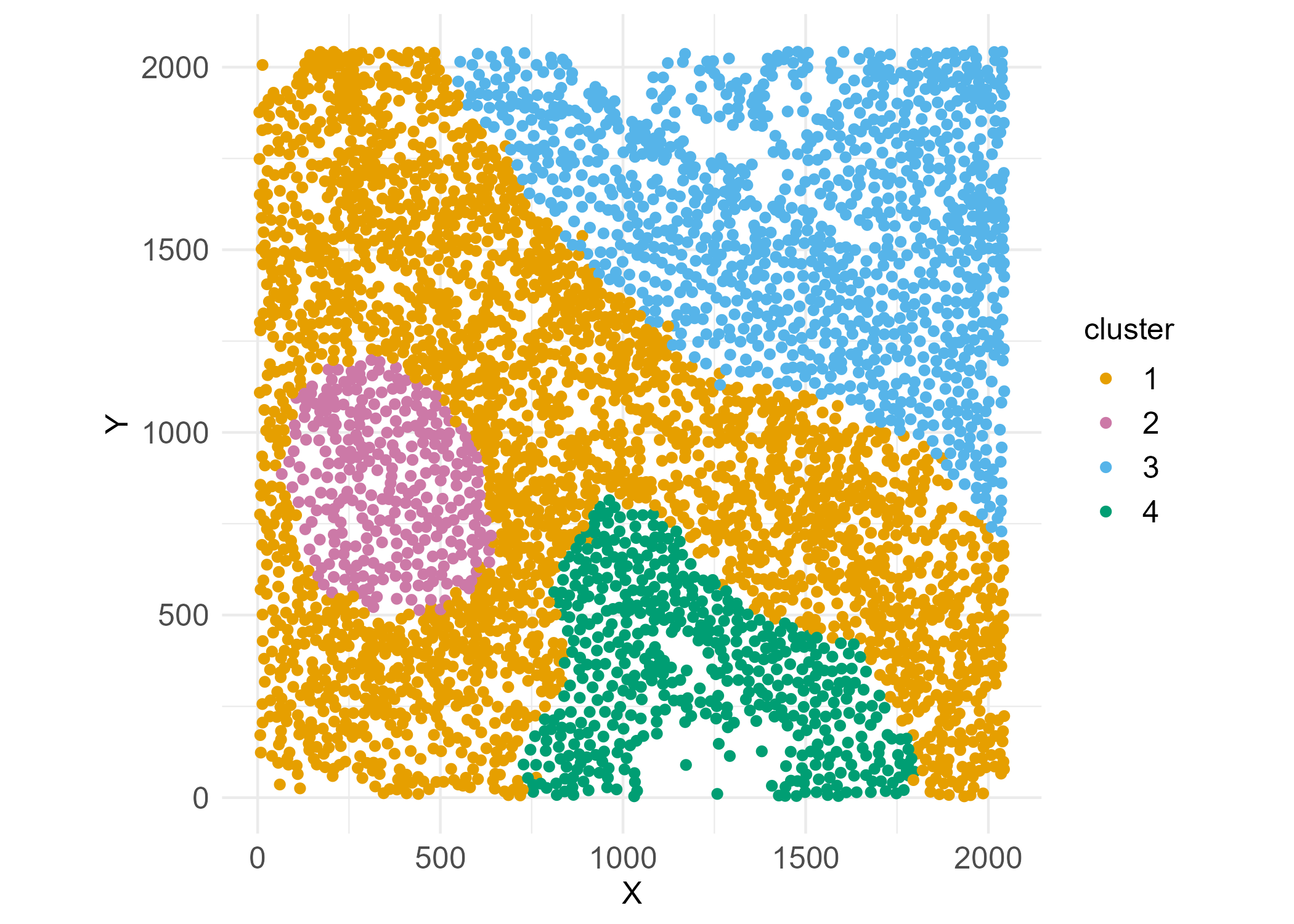}
            \caption{}
        \end{subfigure} &
        \begin{subfigure}{0.45\textwidth}
            \centering
            \includegraphics[width=\linewidth]{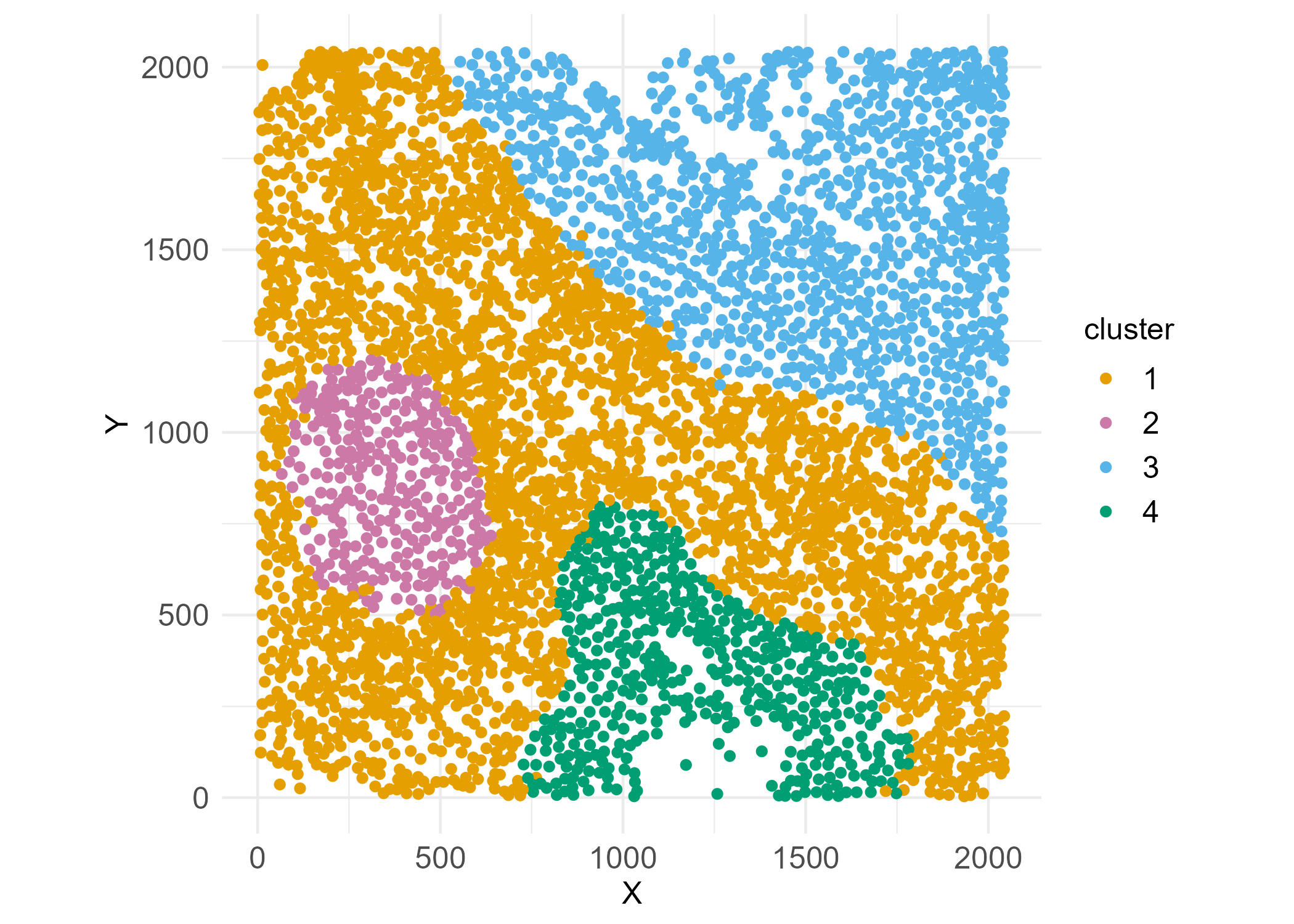}
            \caption{}
        \end{subfigure} \\
        \begin{subfigure}{0.45\textwidth}
            \centering
            \includegraphics[width=\linewidth]{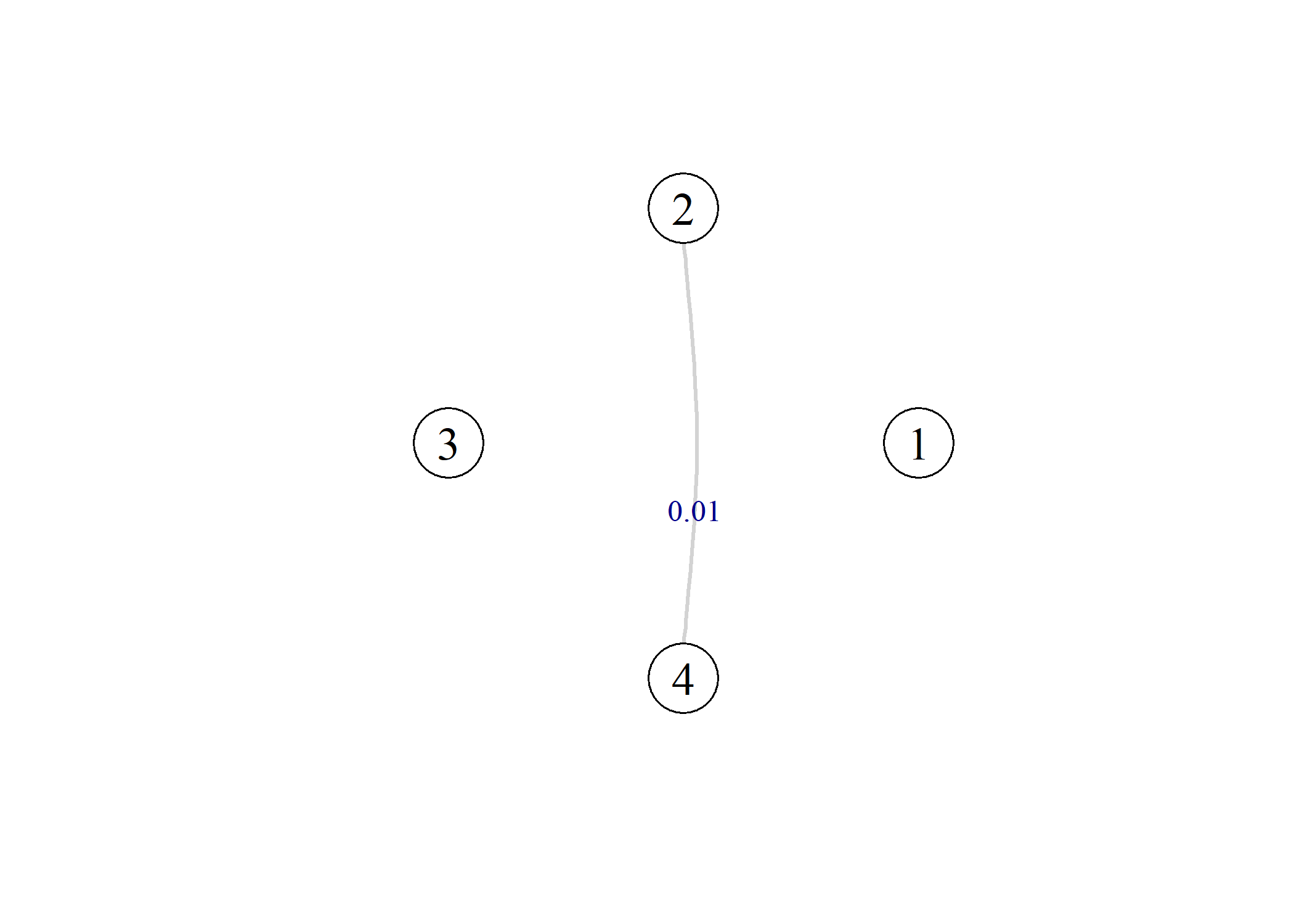}
            \caption{}
        \end{subfigure} &
        \begin{subfigure}{0.45\textwidth}
            \centering
            \includegraphics[width=\linewidth]{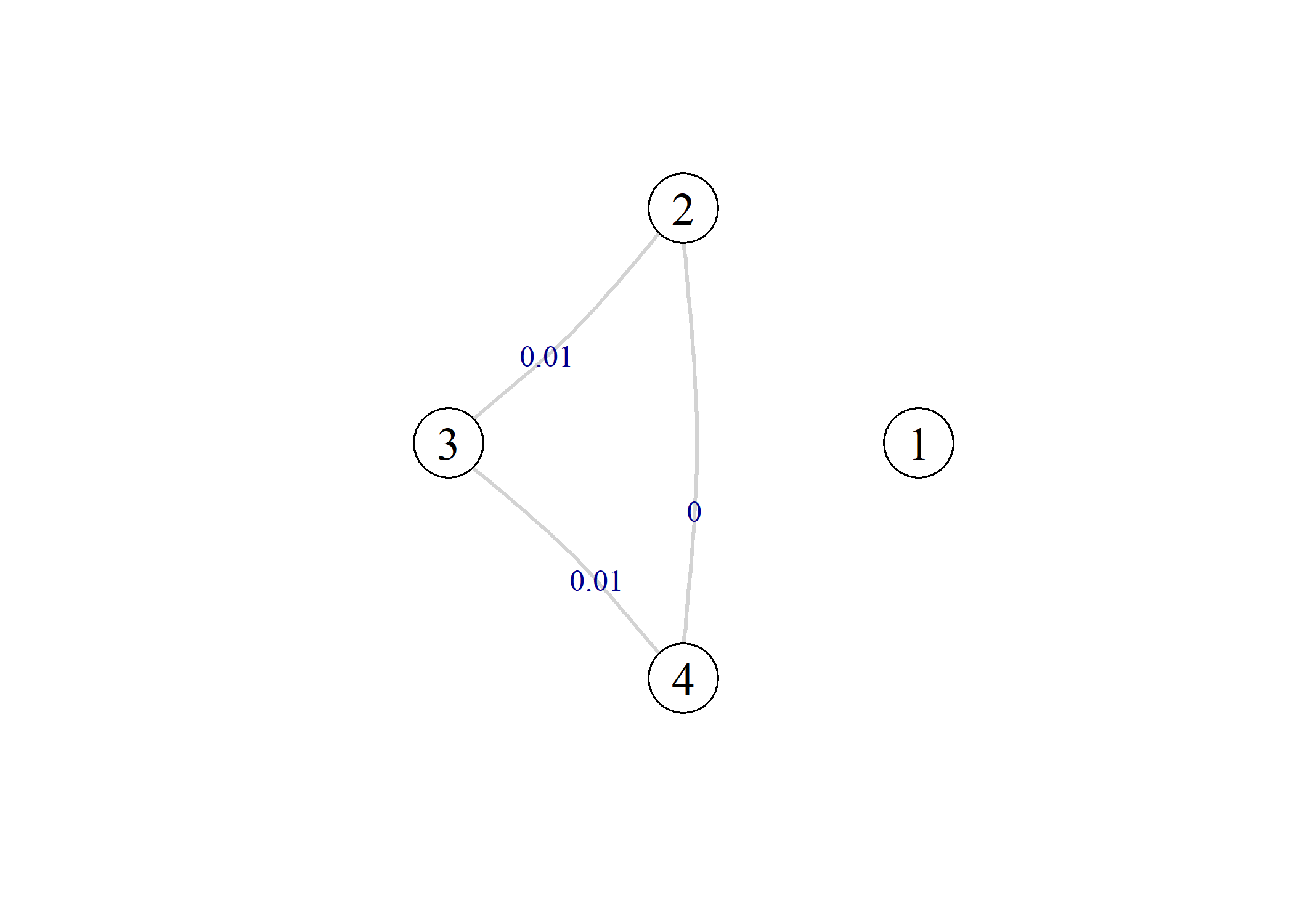}
            \caption{}
        \end{subfigure} \\
        \begin{subfigure}{0.45\textwidth}
            \centering
            \includegraphics[width=\linewidth]{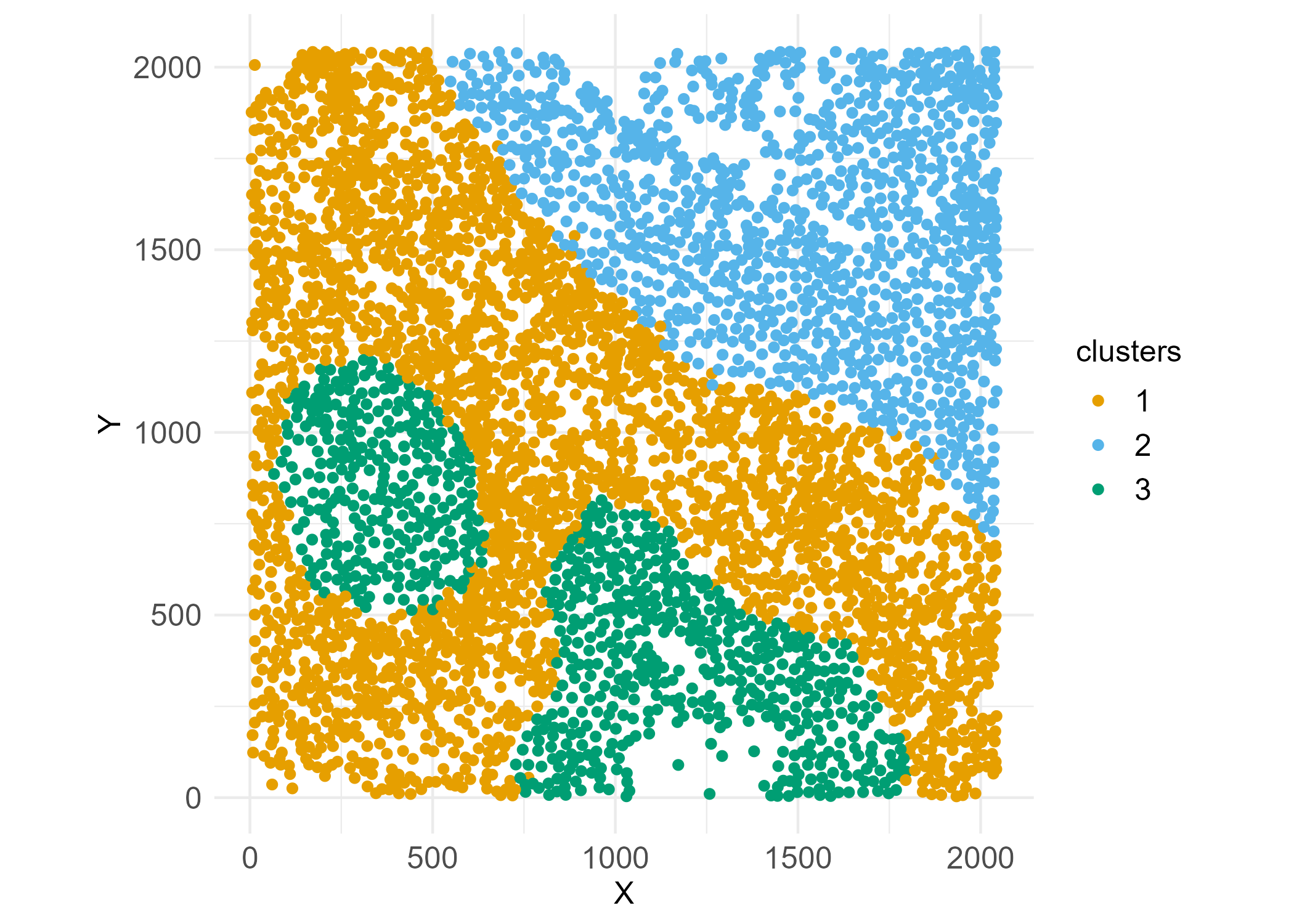}
            \caption{}
        \end{subfigure} &
        \begin{subfigure}{0.45\textwidth}
            \centering
            \includegraphics[width=\linewidth]{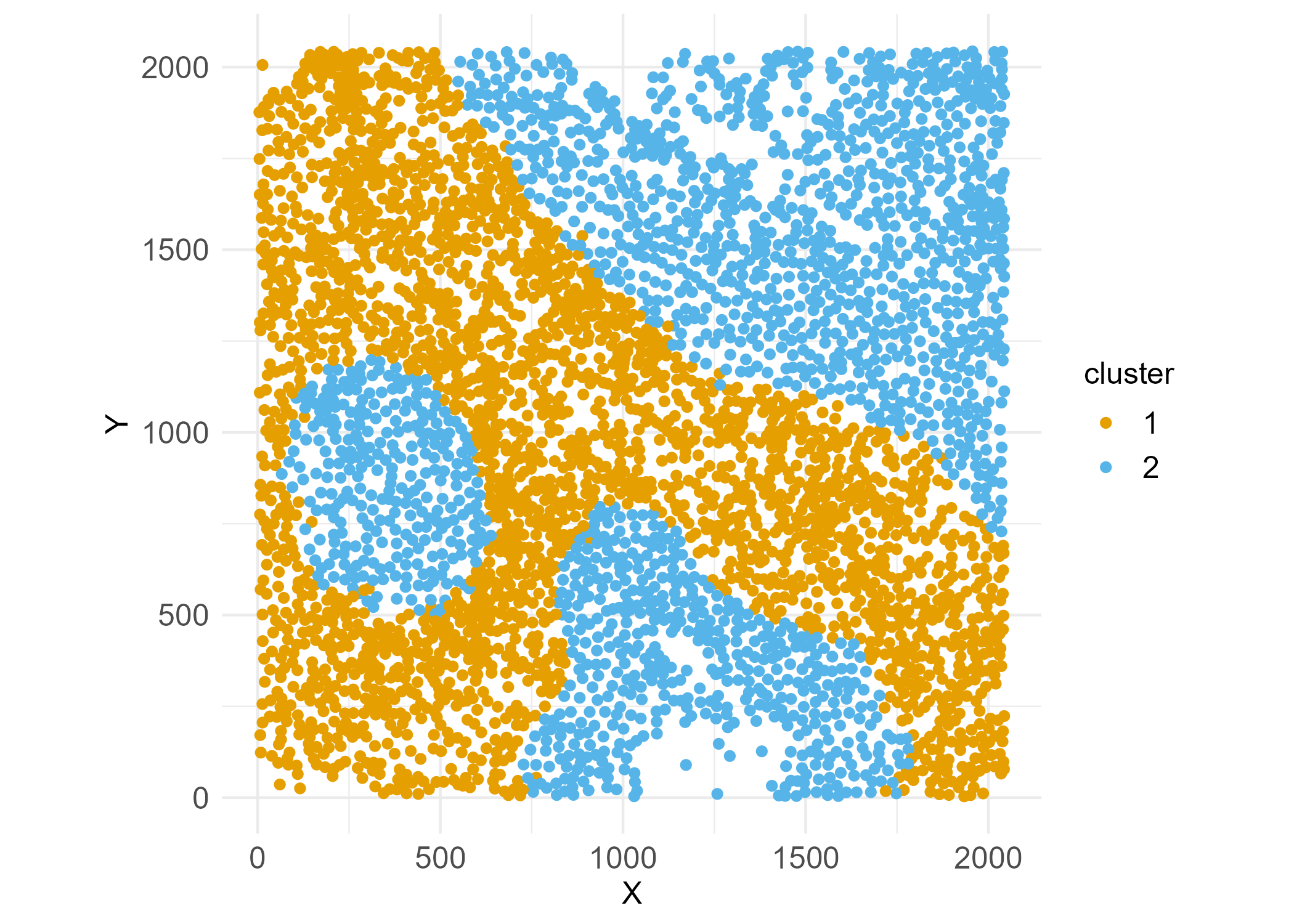}
            \caption{}
        \end{subfigure} \\
    \end{tabular}
    \caption{CAHC and \texttt{repSpat} results for Sample 05. CAHC clusters with (a) marker intensities and (b) binary markers. \texttt{repSpat} pairwise test results with (c) marker intensities and (d) binary markers. Reassigned cluster labels with (e) marker intensities and (f) binary markers.}
    \label{results-05}
\end{figure}

\begin{figure}
    \centering
    \begin{tabular}{cc} 
        \begin{subfigure}{0.45\textwidth}
            \centering
            \includegraphics[width=\linewidth]{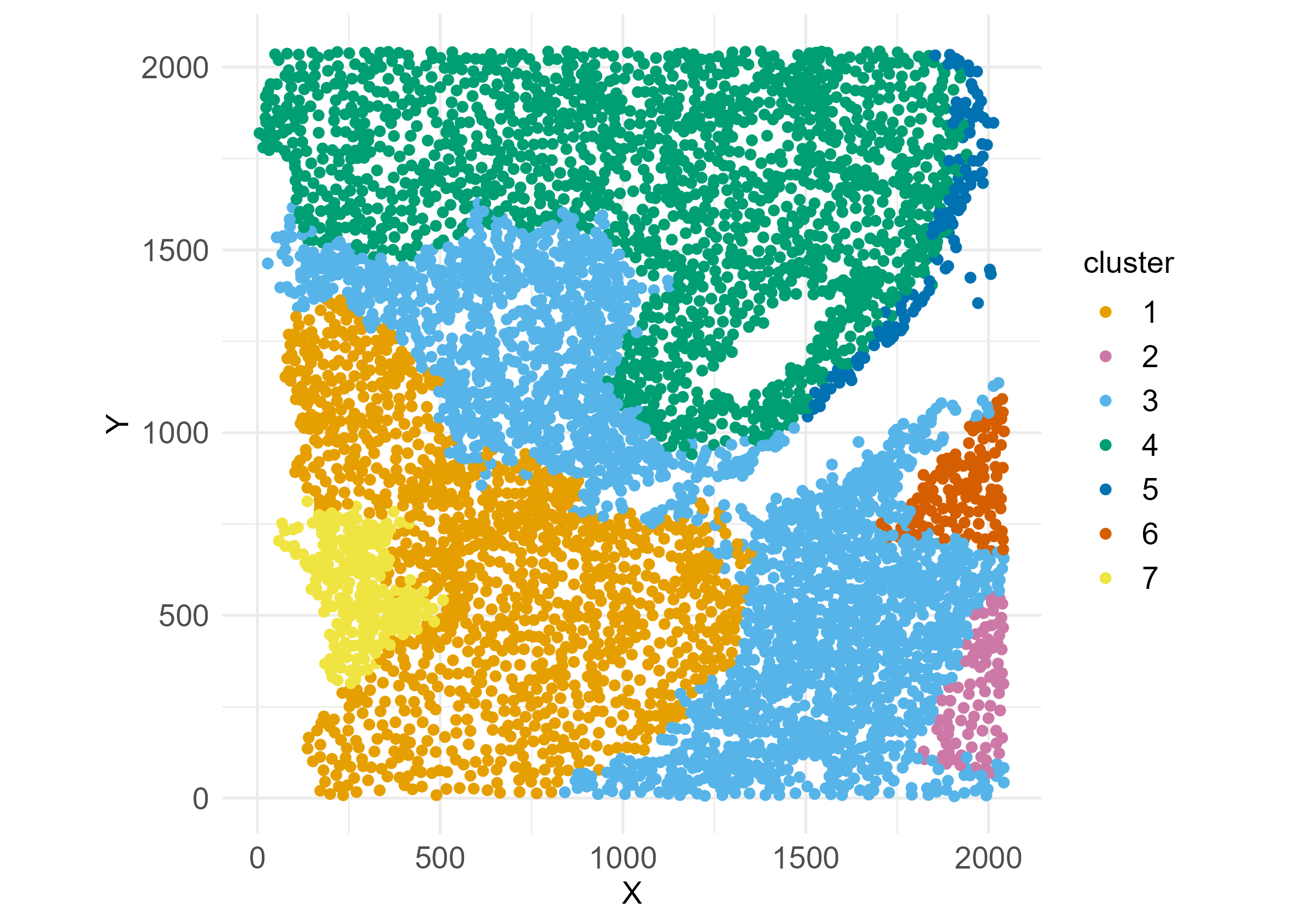}
            \caption{}
        \end{subfigure} &
        \begin{subfigure}{0.45\textwidth}
            \centering
            \includegraphics[width=\linewidth]{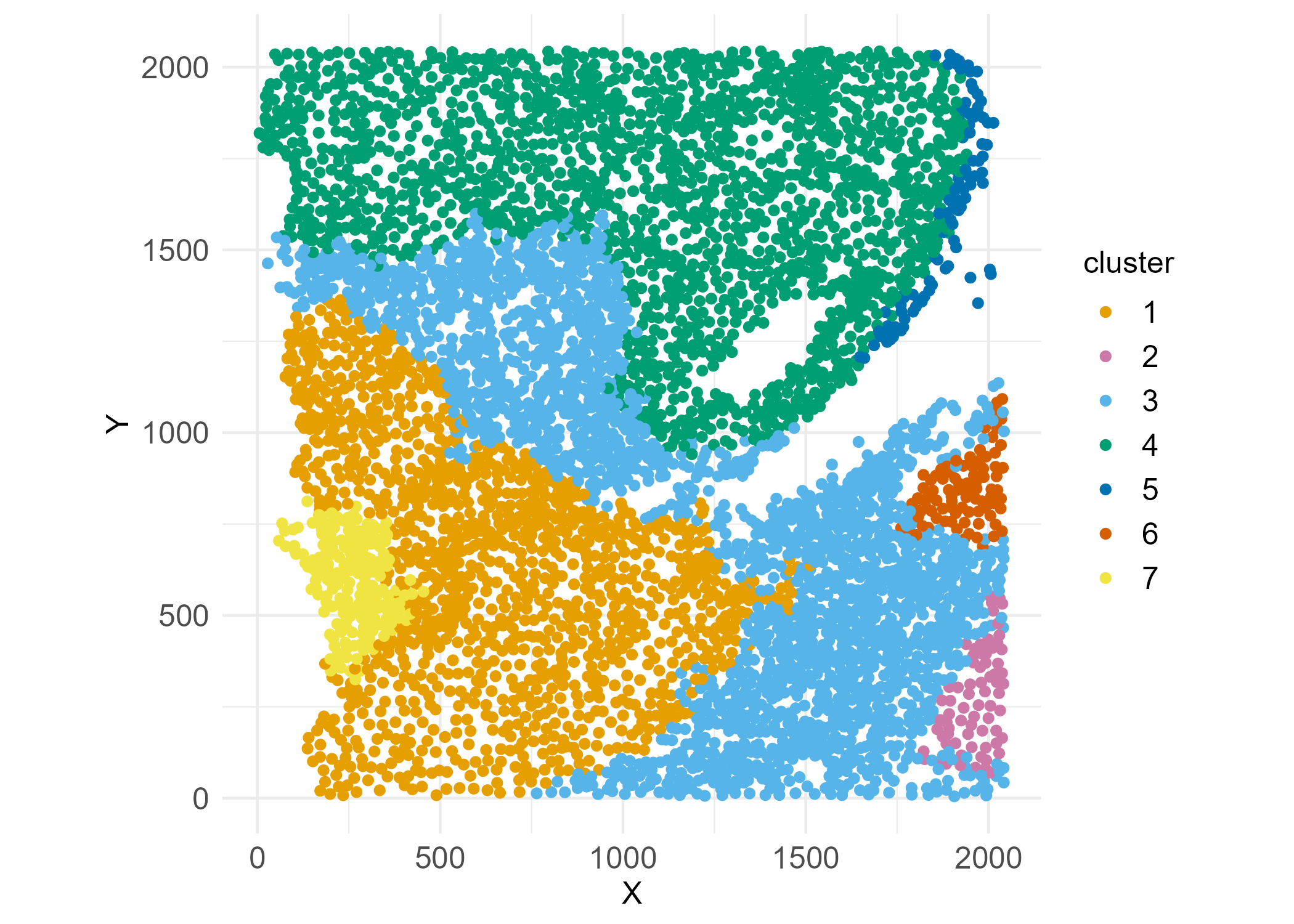}
            \caption{}
        \end{subfigure} \\
        \begin{subfigure}{0.45\textwidth}
            \centering
            \includegraphics[width=\linewidth]{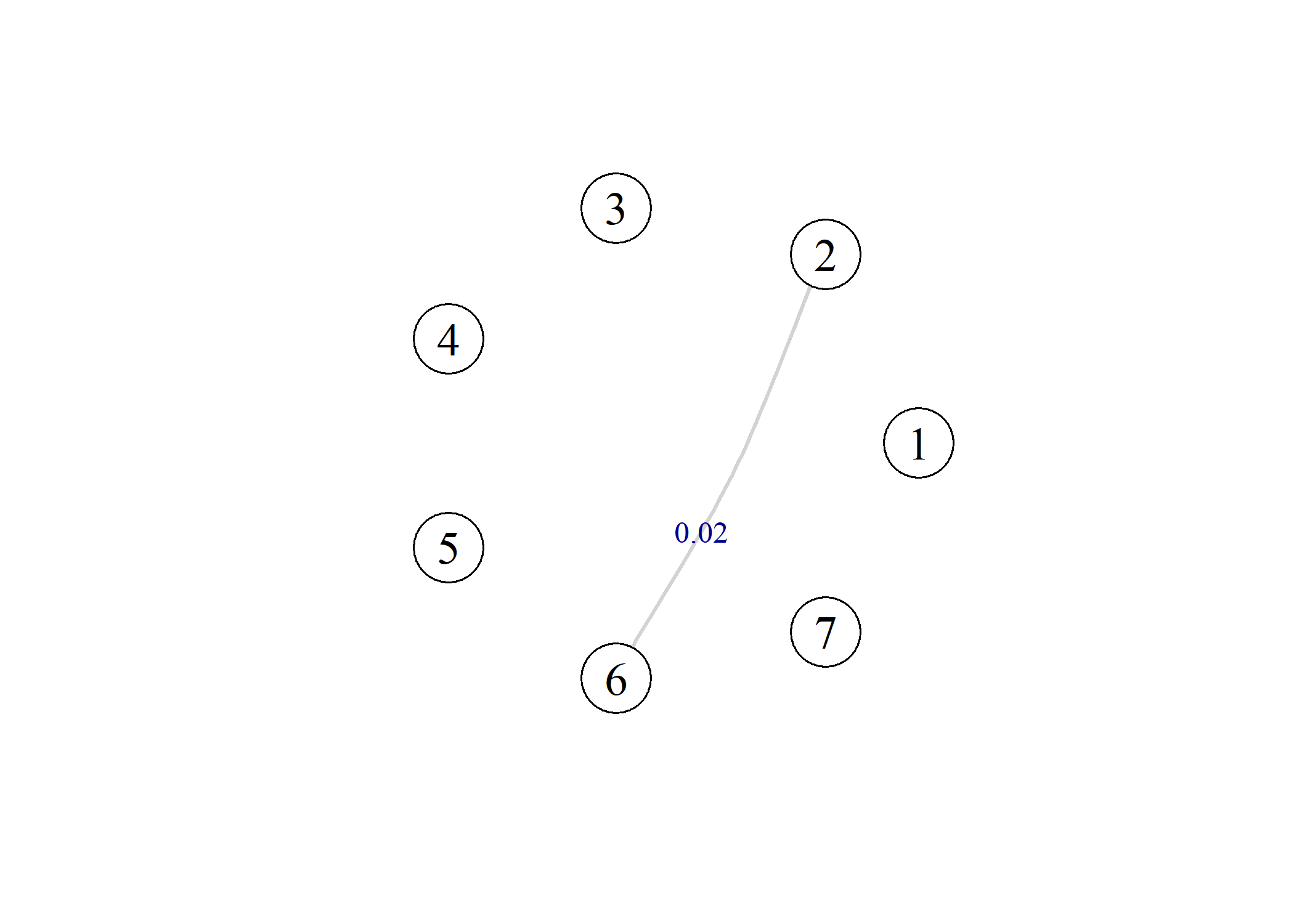}
            \caption{}
        \end{subfigure} &
        \begin{subfigure}{0.45\textwidth}
            \centering
            \includegraphics[width=\linewidth]{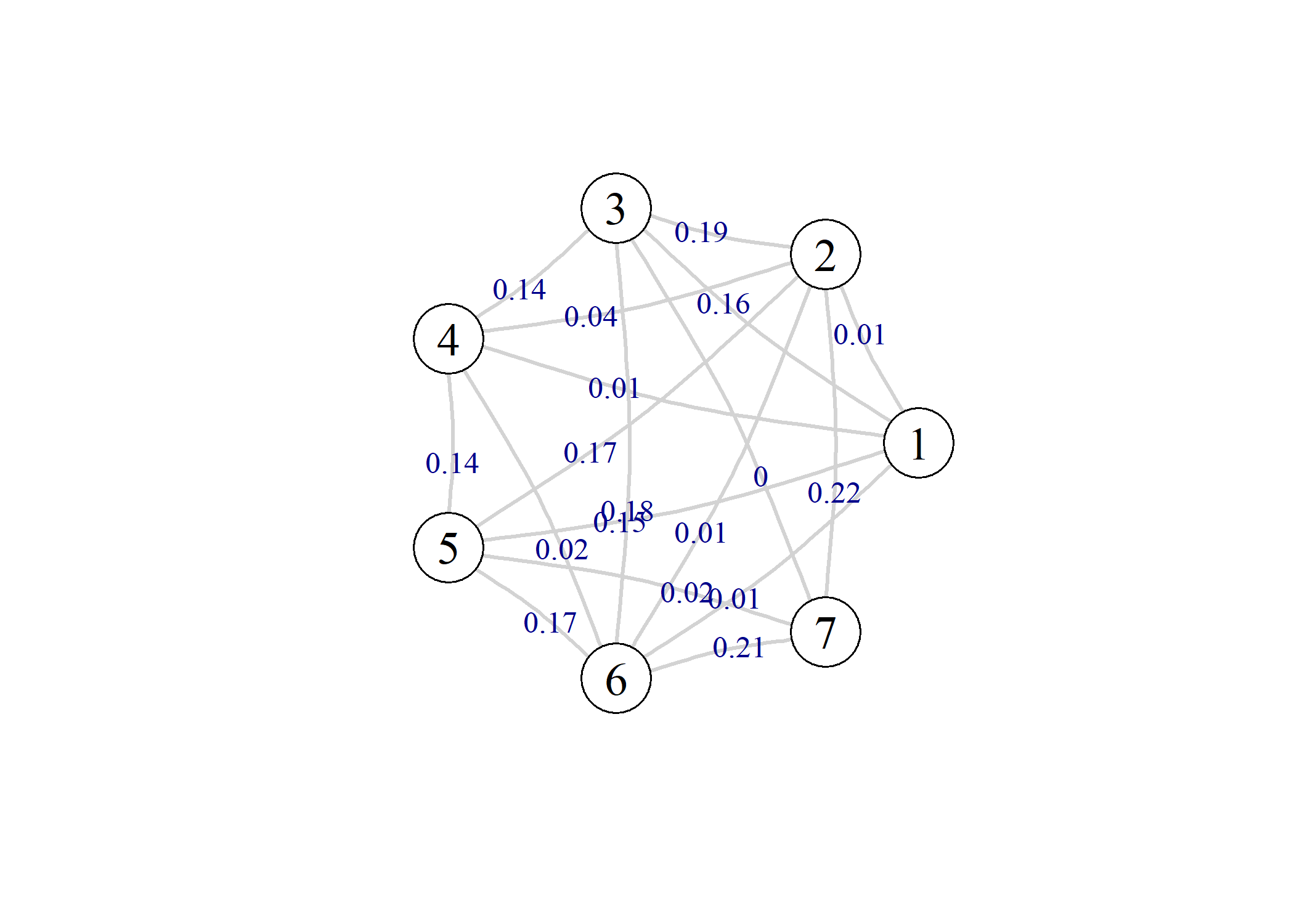}
            \caption{}
        \end{subfigure} \\
        \begin{subfigure}{0.45\textwidth}
            \centering
            \includegraphics[width=\linewidth]{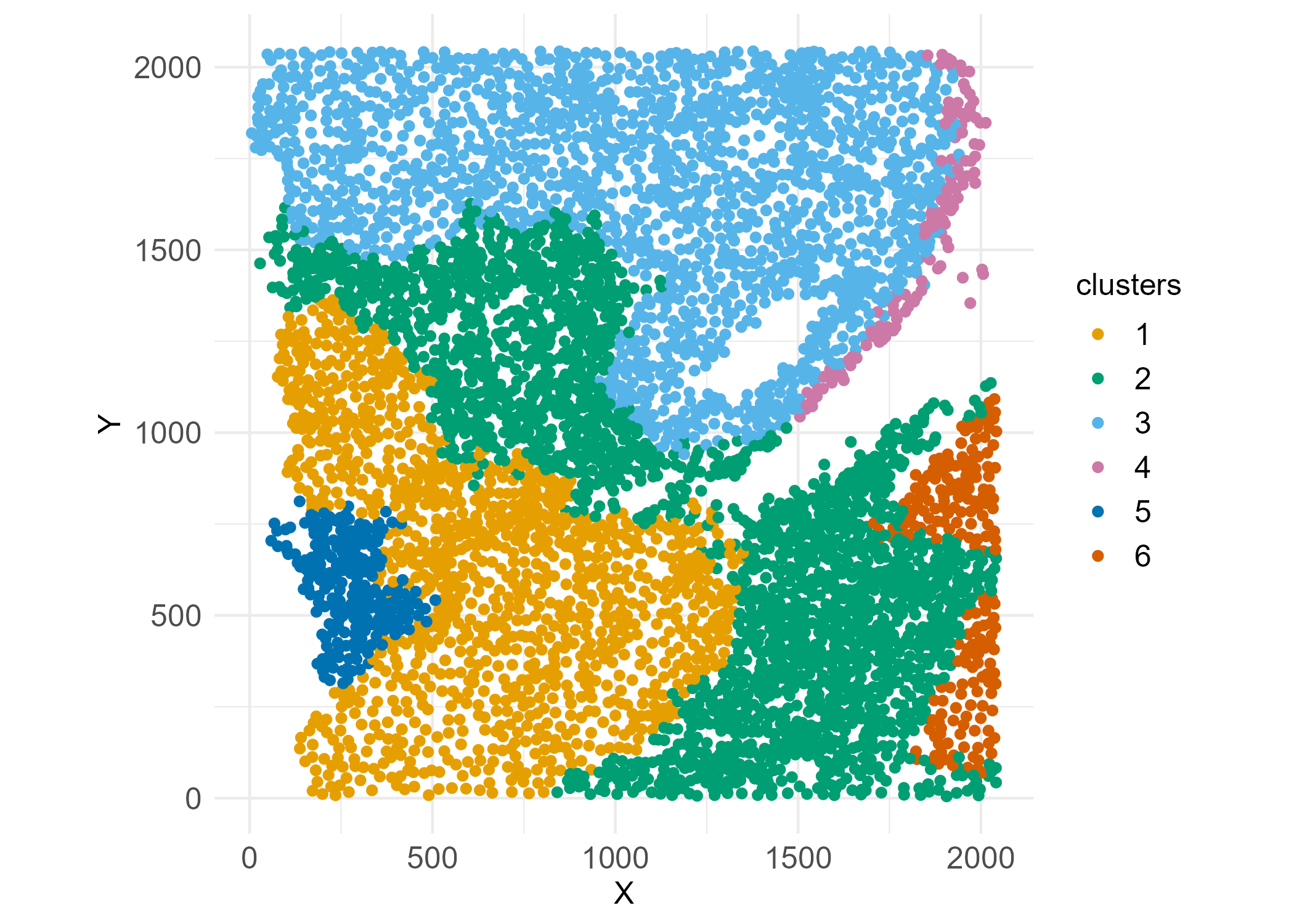}
            \caption{}
        \end{subfigure} &
        \begin{subfigure}{0.45\textwidth}
            \centering
            \includegraphics[width=\linewidth]{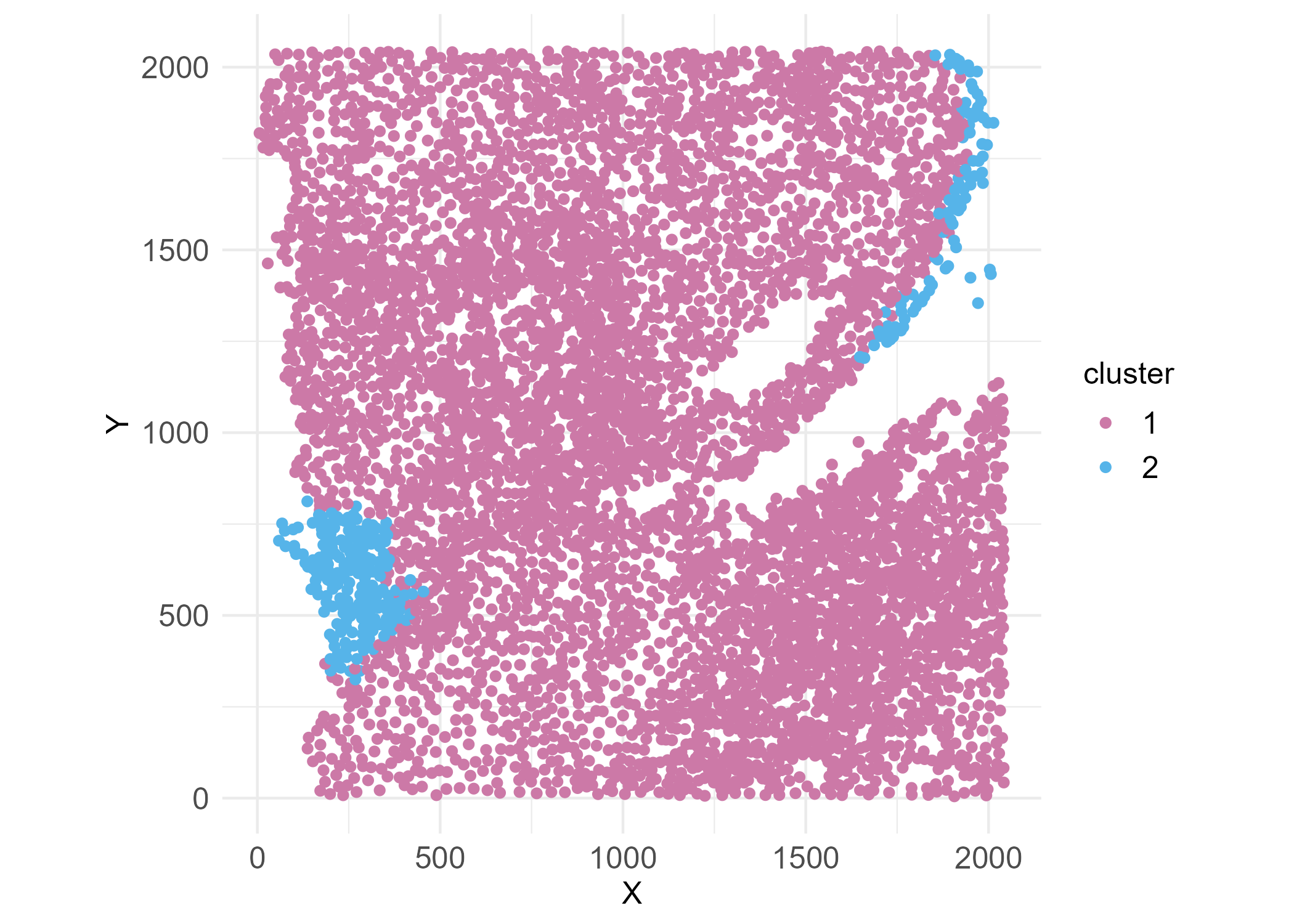}
            \caption{}
        \end{subfigure} \\
    \end{tabular}
    \caption{CAHC and \texttt{repSpat} results for Sample 26. CAHC clusters with (a) marker intensities and (b) binary markers. \texttt{repSpat} pairwise test results with (c) marker intensities and (d) binary markers. Reassigned cluster labels with (e) marker intensities and (f) binary markers.}
    \label{results-26}
\end{figure}

\begin{figure}
    \centering
    \begin{tabular}{cc} 
        \begin{subfigure}{0.45\textwidth}
            \centering
            \includegraphics[width=\linewidth]{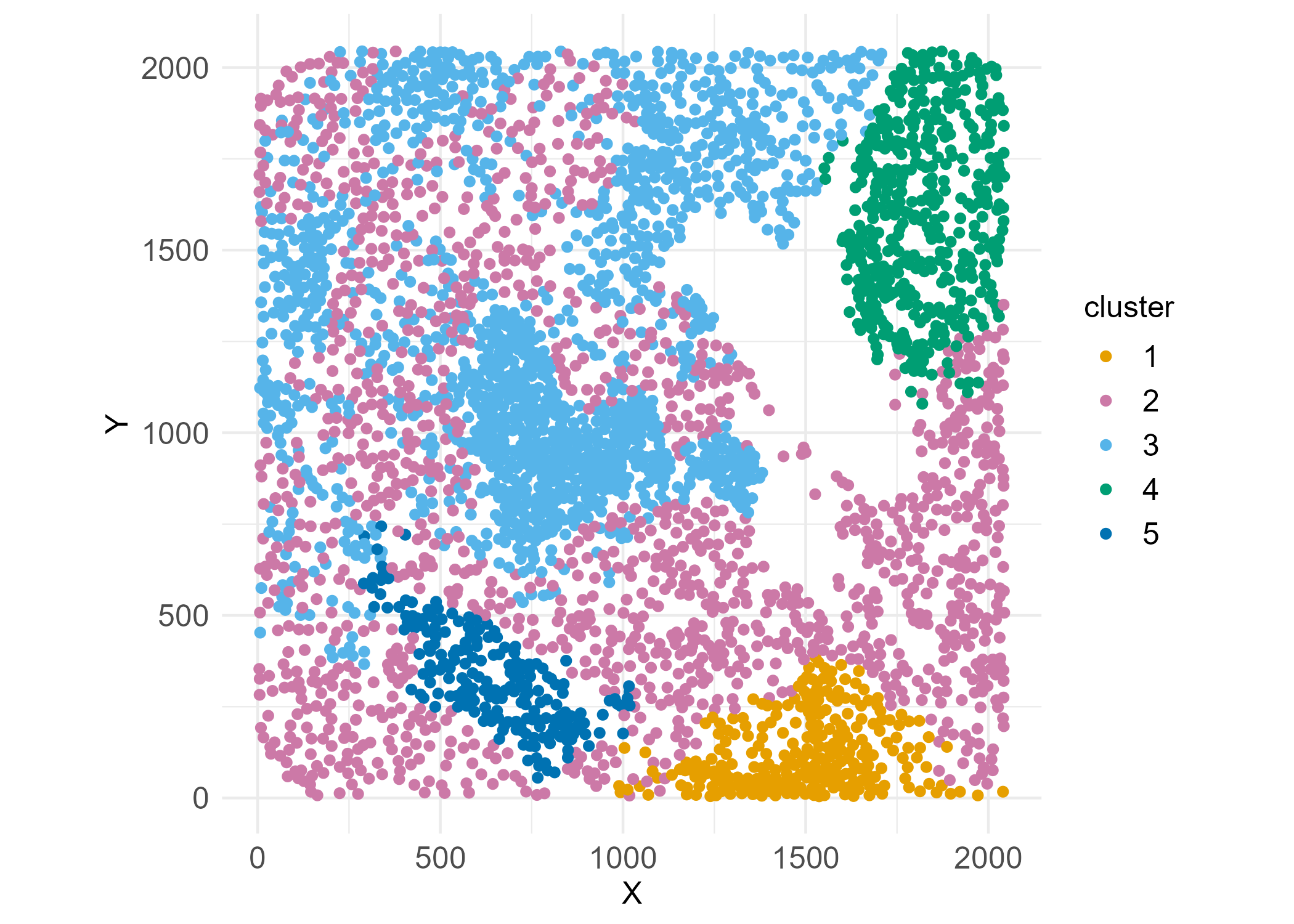}
            \caption{}
        \end{subfigure} &
        \begin{subfigure}{0.45\textwidth}
            \centering
            \includegraphics[width=\linewidth]{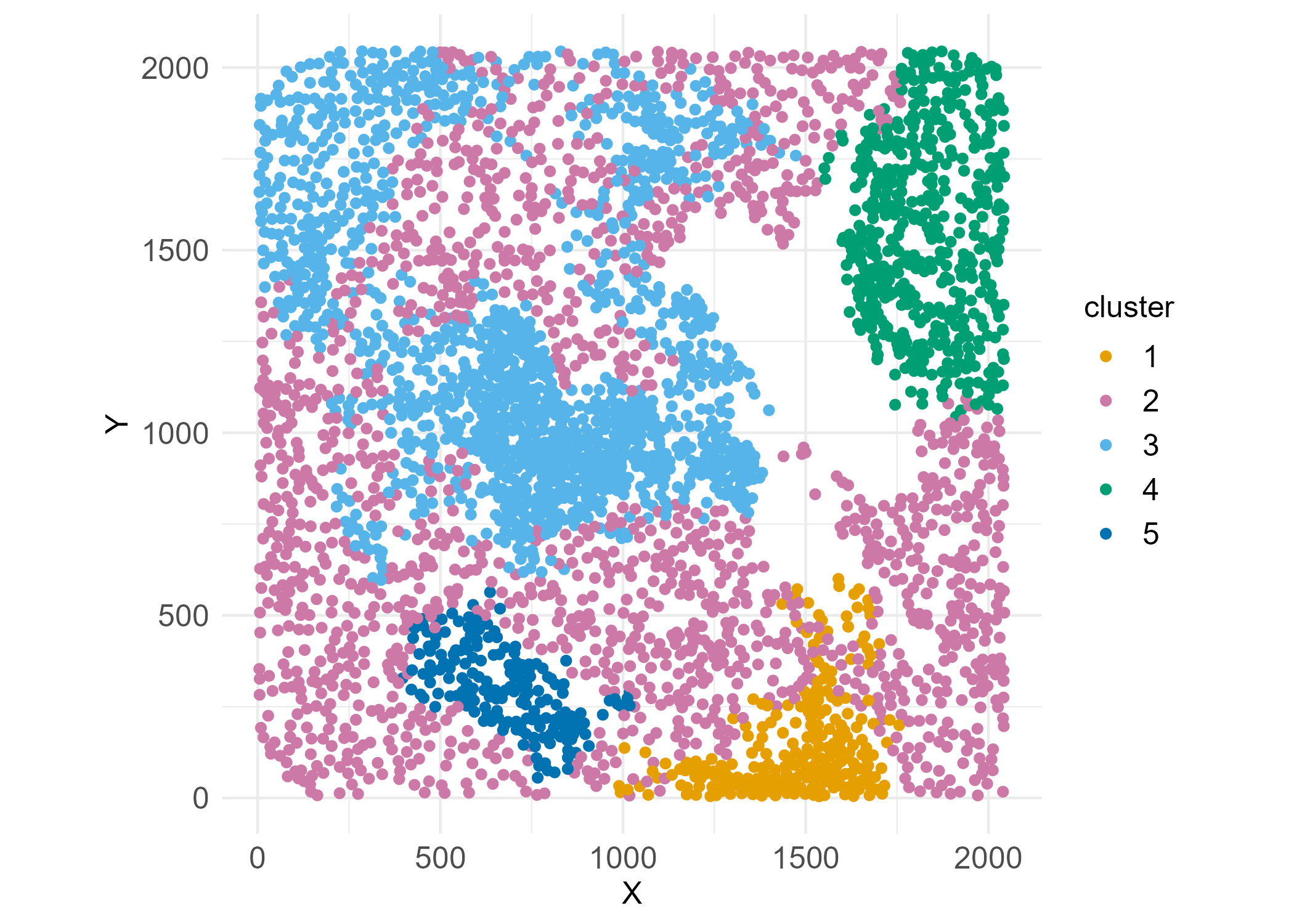}
            \caption{}
        \end{subfigure} \\
        \begin{subfigure}{0.45\textwidth}
            \centering
            \includegraphics[width=\linewidth]{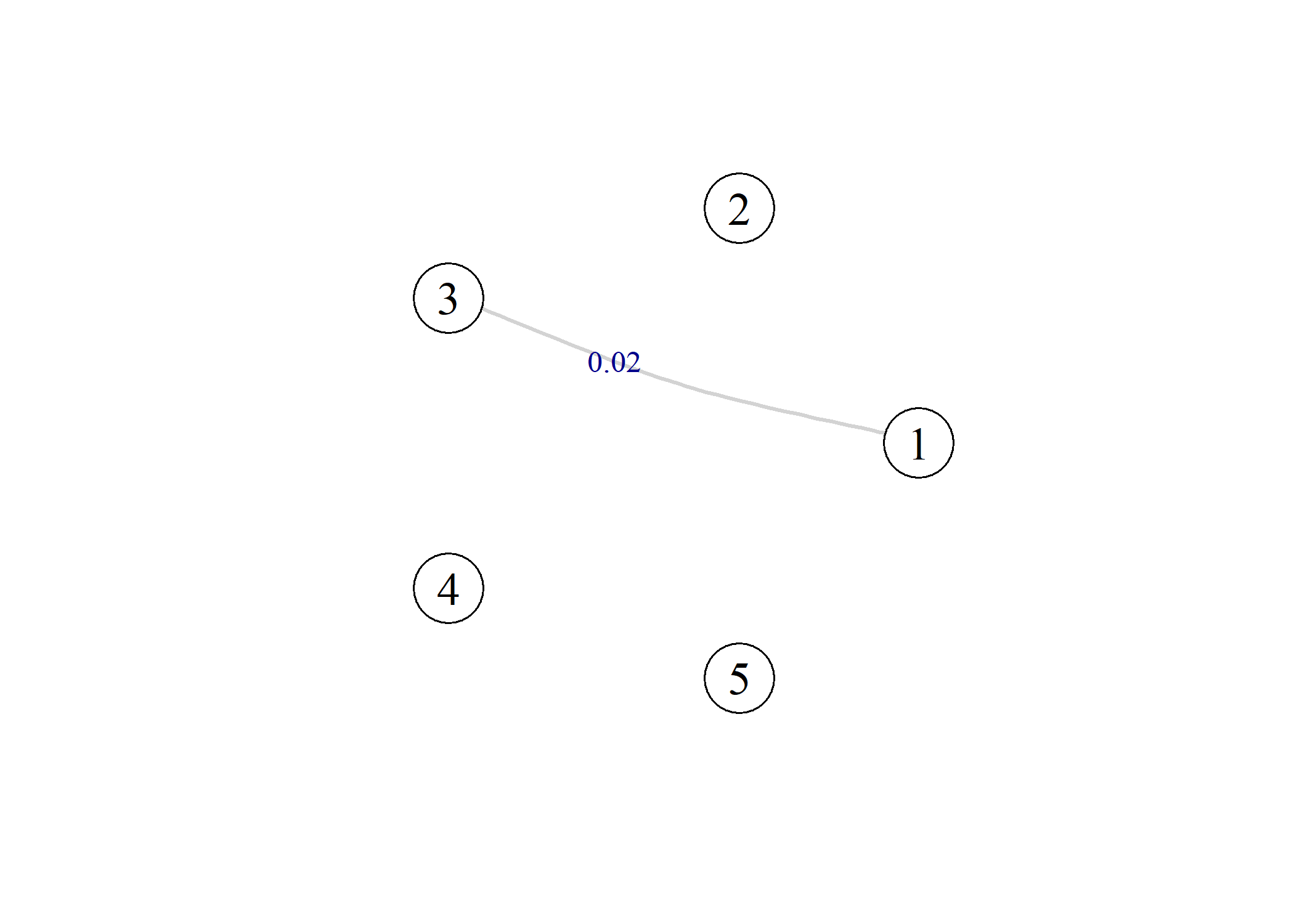}
            \caption{}
        \end{subfigure} &
        \begin{subfigure}{0.45\textwidth}
            \centering
            \includegraphics[width=\linewidth]{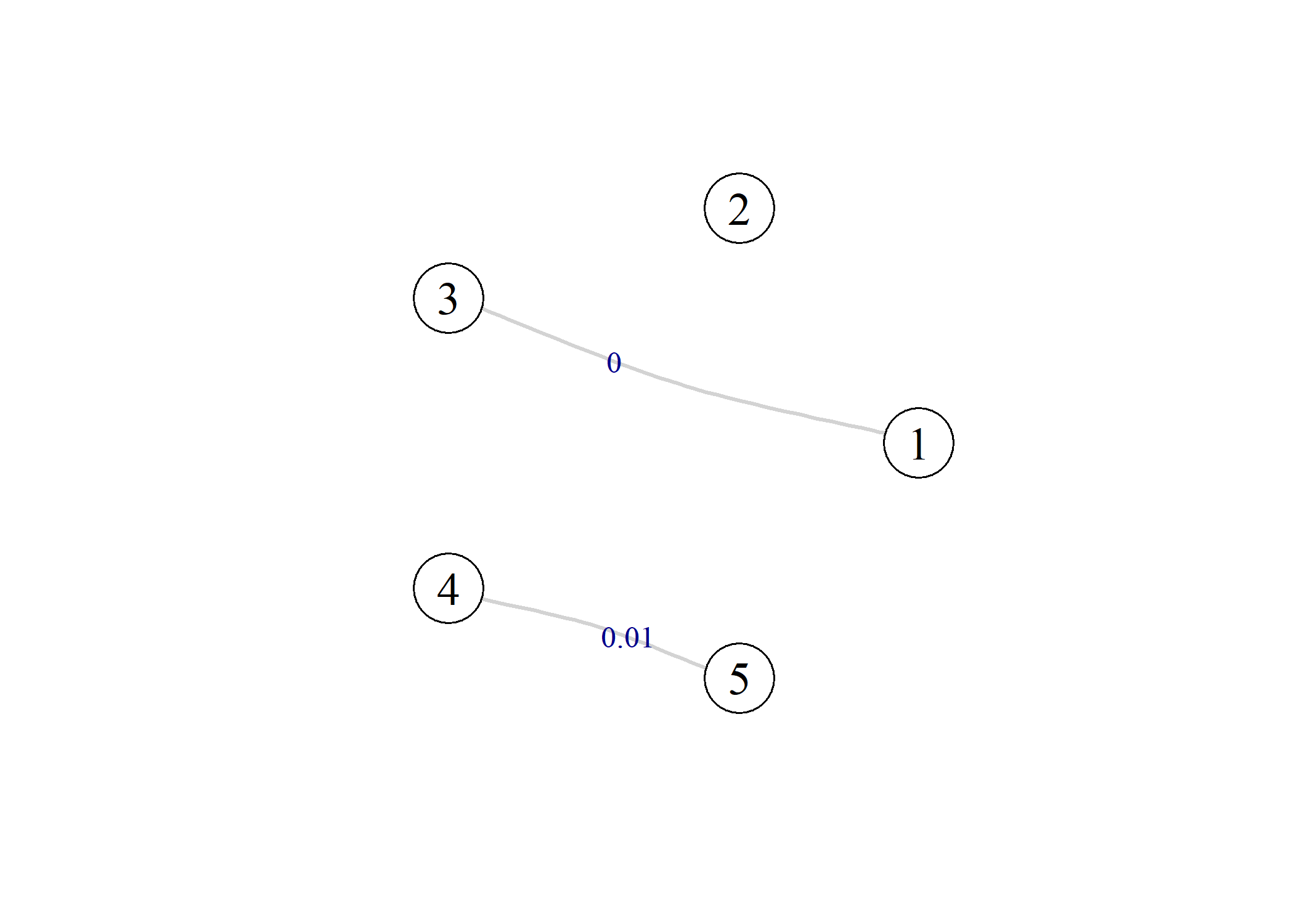}
            \caption{}
        \end{subfigure} \\
        \begin{subfigure}{0.45\textwidth}
            \centering
            \includegraphics[width=\linewidth]{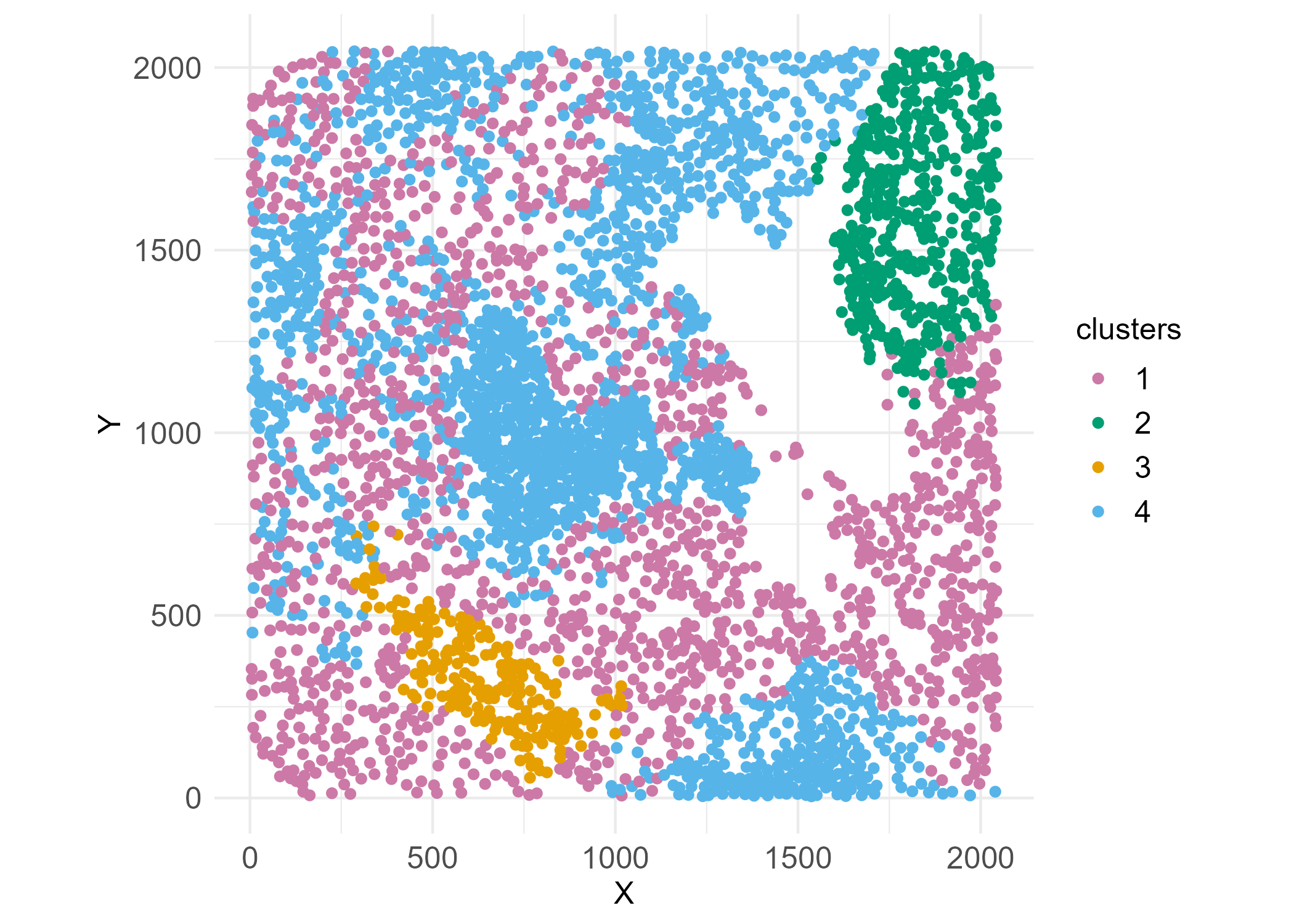}
            \caption{}
        \end{subfigure} &
        \begin{subfigure}{0.45\textwidth}
            \centering
            \includegraphics[width=\linewidth]{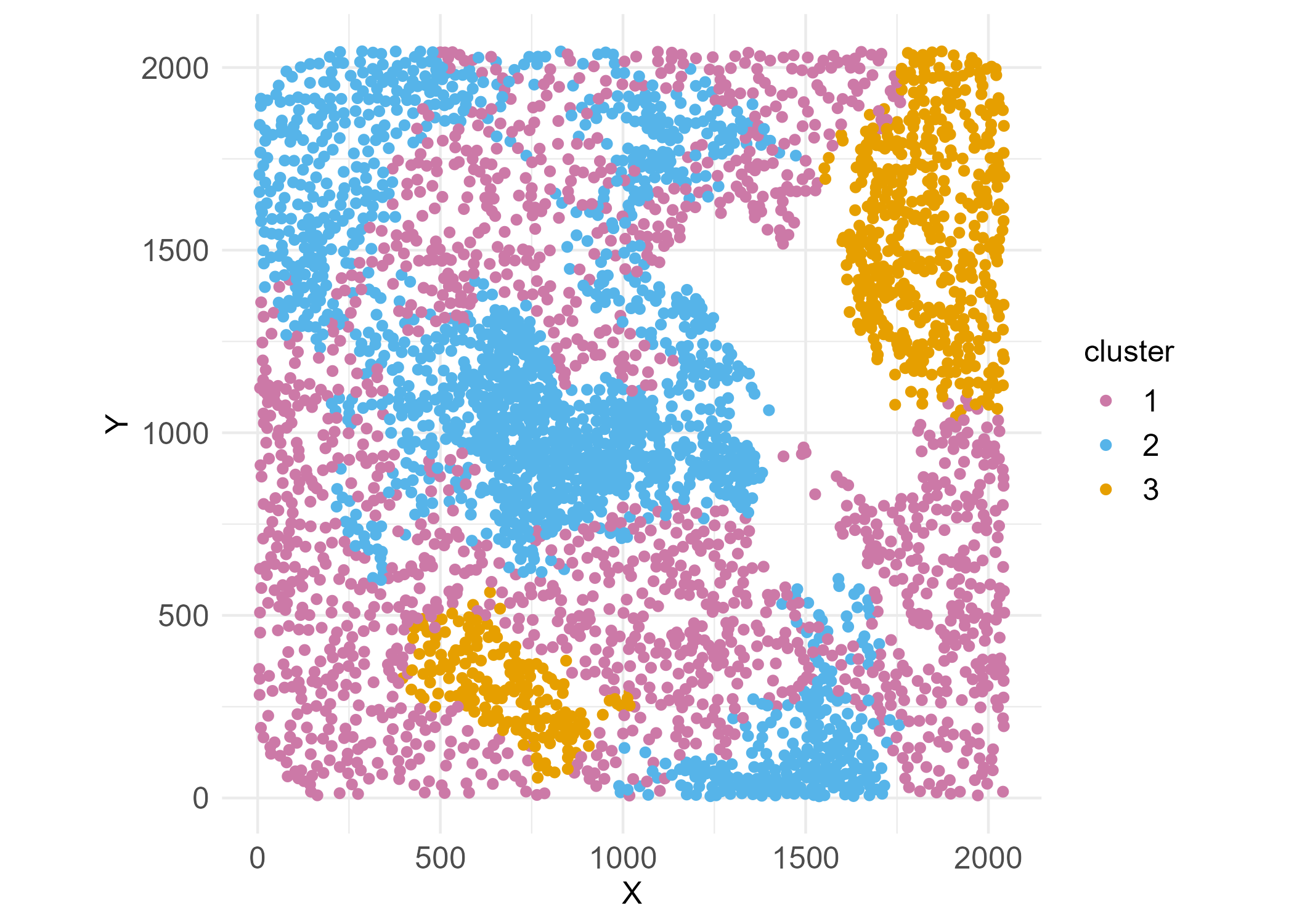}
            \caption{}
        \end{subfigure} \\
    \end{tabular}
    \caption{CAHC and \texttt{repSpat} results for Sample 39. CAHC clusters with (a) marker intensities and (b) binary markers. \texttt{repSpat} pairwise test results with (c) marker intensities and (d) binary markers. Reassigned cluster labels with (e) marker intensities and (f) binary markers.}
    \label{results-39}
\end{figure}

We apply \texttt{repSpat} with CAHC to all tissue samples, separately, and present the results of four patients. The remaining results are available in \ref{app4}.  The four samples presented in this section were selected to illustrate a range of clustering behaviors observed across the dataset, including differences in the number of clusters, neighbourhood size, spatial compactness, and number of cells, thereby representing heterogeneous configurations in both spatial structure and cluster granularity.  Figures \ref{results-04}(a) and \ref{results-26}(a) exhibit more spatially contiguous clustering patterns based on marker intensities compared to Figures \ref{results-05}(a) and \ref{results-39}(a). Among them, only Figure \ref{results-04}(a) displays a dense distribution of cell locations without large spatial gaps.

As the first step, we computed the Euclidean and Jaccard distance matrices for the marker intensities and binary markers, respectively. We applied CAHC with Ward linkage.  Ward linkage merges clusters by minimizing the increase in total within-cluster variance at each step. In CAHC, we use the Lance--Williams update formula with parameters defined in Table~\ref{tab_param} for Ward linkage.  Although Ward linkage typically favors compact, equally sized clusters, incorporating spatial constraints allows the detection of clusters with irregular shapes and varying sizes. We derived the spatial constraints using the k-nearest neighbor algorithm on the spatial coordinates of cells. In the second step, we varied the neighborhood size and the number of clusters based on the modified silhouette score to obtain the optimal CAHC clusters. Table \ref{tab_con_clust_tnbc} shows these two values for four selected samples based on marker intensities and binary markers. We observe that the number of clusters is the same using both data types. This might be because the spatial constraints are similar for both attributes, and the optimal clustering is based on the mean distance within and between clusters. 

The results of the CAHC cluster with these optimal values are in Figures \ref{results-04} - \ref{results-39} (a) and (b) with marker intensities and binary markers, respectively. Figures \ref{results-04} (a) and (b) show many small, well-separated clusters interspersed with larger regions, such as the dark blue and orange clusters, indicating a heterogeneous spatial pattern. In contrast, Figures \ref{results-05} (a) and (b) contain fewer clusters, with the spatial domain dominated by three large, contiguous regions. Figures \ref{results-26} (a) and (b) display a mixture of regional continuity and transitional zones, with the green cluster forming a curved structure. Figures \ref{results-39} (a) and (b) present the most diffuse clustering, characterized by low spatial compactness, particularly evident in the widespread pink cluster.

\begin{table}[t]
\centering
\captionsetup{font=normalsize, labelfont=normal, labelsep=colon}
\caption[CAHC - Neighborhood size and number of clusters]{Optimal neighbourhood size and the number of clusters based on the modified silhouette score for the CAHC in four patient samples.}
\label{tab_con_clust_tnbc}
\renewcommand{\arraystretch}{1.4}
\begin{tabular}{@{} c  cc  cc @{}}
\toprule
\multirow{2}{*}{\textbf{Samples}} & \multicolumn{2}{c}{\textbf{Marker intensities}} & \multicolumn{2}{c}{\textbf{Binary markers}} \\
& Neighbourhood Size & Clusters & Neighbourhood Size & Clusters \\
\midrule
04 & 8 & 7 & 8 & 7 \\
05 & 8 & 4 & 8 & 4 \\
26 & 6 & 7 & 6 & 7 \\
39 & 8 & 5 & 6 & 5 \\
\bottomrule
\end{tabular}
\end{table}

Next, we applied \texttt{repSpat} to the CAHC clusters, followed by the cluster label reassignment step. As shown in Figures~\ref{results-04}–\ref{results-39} (c) and (d), clustering based on binary markers reveals more RSP. Thresholding the marker intensities reduces noise, enabling \texttt{repSpat} to identify repeated patterns more effectively across all tissue samples. In this analysis, we used the inverse multiquadratic (IMQ) kernel with $c = 1$, which provides stronger separation between distributions when paired with the Jaccard distance bounded within $[0,1]$. This choice enhances the sensitivity of kernel to differences in distributions in the binary data setting.

In Figure \ref{results-04} (e), the red cluster shares a similar spatial distribution with the blue clusters identified using binary markers in Figure \ref{results-04} (f). These overlapping regions correspond to tumour cell populations based on cell type annotations \citep{keren_structured_2018}. Similarly, in Figures \ref{results-05}-\ref{results-39} (f), the identified repeated clusters are predominantly enriched in mixed tumour cell populations, with relatively lower representation of immune cells.

\section{Conclusion}
\label{sec:conclusion}

 We proposed a robust nonparametric framework for detecting repeated spatial patterns (RSP) in geostatistical spatial processes.  By integrating constrained agglomerative hierarchical clustering (CAHC) with a kernel-based maximum mean discrepancy squared (MMD$^2$) test and a block permutation procedure, \texttt{repSpat} addresses the common issue of over-segmentation in constrained clustering and enables the re-identification of spatially disjoint regions that share a similar distribution.

Our framework is nonparametric and supports a variety of attribute types through distance-based kernels. Simulation studies demonstrate that \texttt{repSpat} achieves high clustering accuracy in the presence of RSP while effectively controlling the false discovery rate under weak to moderate spatial dependence. When applied to spatial proteomics data, \texttt{repSpat} successfully uncovers repeated patterns, highlighting its utility for real-world applications, particularly in the spatial omics domain.

While \texttt{repSpat} is effective in detecting repeated patterns, it exhibits limited sensitivity to slight changes in spatial dependence alone, such as variations in autocorrelation strength. This points to an opportunity for future research to explore dependence-sensitive kernels or strategies for selecting kernel parameters.

In future work, we aim to extend this framework to: (1) incorporate other types of constrained clustering algorithms followed by \texttt{repSpat}; (2) perform variable selection to enhance clustering interpretability and power in detecting dissimilarity between cluster distribution; and (3) develop Bayesian models that provide uncertainty quantification for cluster assignments.

Although this study focused on continuous and binary attributes, the proposed methodology is readily extensible to other data types such as compositional or count-valued observations through suitable dissimilarity measures and kernel functions.  In addition, while the framework is presented for continuous spatial domains, it can be extended to lattice or areal data by representing each spatial unit using a reference location (e.g., centroid), so that spatial contiguity can be defined using the same distance-based construction with minimal modification to the procedure. This is particularly relevant for spatial omics applications where measurements are collected on a fixed array of spatial locations (e.g., spot-based transcriptomics) \citep{crowell2025orchestrating}, and clustering is used to identify spatial domains or tissue regions, which may appear in multiple disjoint locations.  Given its robustness across different sample sizes, attribute dimensions, and spatial dependence levels, \texttt{repSpat} offers a flexible and general-purpose framework for discovering repeated spatial patterns in complex spatial datasets.

\section*{Funding}
This work was supported by the Natural Sciences and Engineering Research Council of Canada (NSERC) through the Discovery Grant [RGPIN-2022-05272] and Discovery Launch Supplement [DGECR-2022-00465]. 

\section*{Acknowledgement}
Partial support for computational resources was provided by SHARCNET (www.sharcnet.ca) and the Digital Research Alliance of Canada (www.alliancecan.ca). Jamie D. McNicol (Department of Medicine, McMaster Immunology Research Center, Hamilton, Ontario, Canada) provided us with guidance on accessing the dataset and the threshold for protein markers used in the application.

\newpage
\appendix

\section{Methodolgoy Details}
\subsection{Constrained Clustering Details}\label{app-cahc}

\begin{table}[h]
	\centering
	\caption{Parameter values for the Lance--Williams update formula in Eq.~\eqref{eq:lw-update} under common linkage methods. Here, $n_i$, $n_j$, $n_k$, and $n_h$ denote the number of elements in clusters $i$, $j$, $k$, and $h$, respectively, where cluster $h$ is formed by merging clusters $i$ and $j$, and the formula updates the distance between cluster $h$ and another cluster $k$.}
	\renewcommand{\arraystretch}{1.4}
	\begin{tabular}{@{} l c c c c @{}}
		\toprule
		Linkage Method & $\alpha_i$ & $\alpha_j$ & $\beta$ & $\gamma$ \\
		\midrule
		Single linkage & $1/2$ & $1/2$ & $0$ & $-1/2$ \\
		Complete linkage & $1/2$ & $1/2$ & $0$ & $1/2$ \\
		Average linkage & $\dfrac{n_i}{n_i+n_j}$ & $\dfrac{n_j}{n_i+n_j}$ & $0$ & $0$ \\
		Ward's method & $\dfrac{n_i+n_k}{n_i+n_j+n_k}$ & $\dfrac{n_j+n_k}{n_i+n_j+n_k}$ & $-\dfrac{n_k}{n_i+n_j+n_k}$ & $0$ \\
		\bottomrule
	\end{tabular}
	\label{tab_param}
\end{table}

\subsection{Block Permutation Procedure}
\label{blk-perm}
\begin{enumerate}
  \item Create attribute-based blocks within $\mathcal{C}^{(g}$ and $\mathcal{C}^{(h)}$ using k-Means clustering. The number of blocks $b^{(g)}$ and $b^{(h)}$ for each cluster is determined based on the optimal neighborhood size $m$ and the number of observations $n^{(g)}$ and $n^{(h)}$ in each cluster from the CAHC:
  \begin{equation}\label{num_blocks}
  \text{Number of blocks} = \frac{\text{Number of observations in the cluster}}{\text{Neighborhood size}}.
  \end{equation}
  If the neighborhood size exceeds the number of observations in a cluster, we assign a single block.

  \item Randomly select blocks from both clusters without replacement until the number of observations in the selected blocks exceeds the size of the smaller cluster (assume $\mathcal{C}^{(g)}$).

  \item Assign the selected blocks to $\mathcal{C}^{(g)}$, and assign the remaining blocks to $\mathcal{C}^{(h)}$.
\end{enumerate}

\subsection{Asymptotic Properties of $\widehat{\operatorname{MMD}}^{2}$}\label{proof-mmd-con}

We use the biased version of $\widehat{\operatorname{MMD}}^{2}$ in Eq.~\eqref{eq:mmd} to test differences in the distribution of $\vX(s)$ between pairs of clusters obtained from constrained agglomerative hierarchical clustering (CAHC).  We show consistency for the unbiased version of 
$\widehat{\operatorname{MMD}}_{N}^{2}$ in Eq.~\eqref{eq:bmmd}. Consistency of the biased version then follows because the two estimators differ by
\begin{equation}
\widehat{\operatorname{MMD}}^{2}_{\text{biased}} - \widehat{\operatorname{MMD}}^{2}_{N}
	= \frac{1}{n_g(n_g-1)}\sum_{i=1}^{n_g} k\left(\vx^{(g)}_i, \vx^{(g)}_i\right)
	+ \frac{1}{n_h(n_h-1)}\sum_{i=1}^{n_h} k\left(\vx^{(h)}_i, \vx^{(h)}_i\right),
\end{equation}
which is $O(1/n)$ and converges to zero as $n_g, n_h \to \infty$ since $k$ is bounded. Consistency of the biased version therefore follows by Slutsky's theorem.

Recall that $\mathcal{C}^{(g)} = \{\vs_1^{(g)}, \ldots, \vs_{n_g}^{(g)}\}$ and $\mathcal{C}^{(h)} = \{\vs_1^{(h)}, \ldots, \vs_{n_h}^{(h)}\}$ denote two such clusters, each corresponding to a spatial subregion. The associated multivariate attributes are given by
$\vX^{(g)} = \{ X(\vs_i^{(g)}) \in \mathbb{R}^p : i = 1, \ldots, n_g \}$ and
$\vX^{(h)} = \{ X(\vs_j^{(h)}) \in \mathbb{R}^p : j = 1, \ldots, n_h \}$. 

We assume that all clusters $\mathcal{C}^{(g)}$  are disjoint. Recall that $k: \mathcal{X} \times \mathcal{X} \to \mathbb{R}$ is a positive definite kernel, where $\mathcal{X} \subseteq \mathbb{R}^p$. 
Moreover, the RKHS inner product satisfies $\langle f, k(\cdot, \vx) \rangle_{\mathcal{H}_k} = f(\vx)$. This also implies that for any $\vx, \vy \in \mathcal{X}$, we have $k(\vx, \vy) = \langle k(\cdot, \vx), k(\cdot, \vy) \rangle_{\mathcal{H}_k}$.

We now define the unbiased version of $\widehat{\operatorname{MMD}}^{2}_{N}$ following \cite{gretton2012kernel}, where $i$ and $j$ are spatial indices:
\begin{align}
\widehat{\operatorname{MMD}}^{2}(X^{(g)}, X^{(h)}; \mathcal{H}_k) =\; &
\frac{1}{n_g (n_g-1)} \sum_{\substack{i,j = 1 \\ i \ne j}}^{n_g}  k\left(\vx_i^{(g)}, \vx_j^{(g)}\right) 
+ \frac{1}{n_h(n_h-1)} \sum_{\substack{i,j = 1 \\ i \ne j}}^{n_h}  k\left(\vx_i^{(h)}, \vx_j^{(h)}\right) \notag \\
&- \frac{2}{n_g n_h} \sum_{i=1}^{n_g} \sum_{j=1}^{n_h} k\left(\vx_i^{(g)}, \vx_j^{(h)}\right). \label{eq:bmmd}
\end{align}

Following the asymptotics of U-statistic \purple{s} in Chapter 12 of \cite{van2000asymptotic}, the formulation of $\widehat{\operatorname{MMD}}^{2}_{N}$ as a U-statistic in \cite{gretton2012kernel},  the definition of  dependence coefficient based on the kernel mean
embedding (or covariance expressed in the RKHS $\mathcal{H}_k$)   from \cite{cherief-abdellatif_finite_2022}, and the spatial mixing conditions to estimate the nonparametric variance of a statistic from \cite{sherman1996variance}, we show that the unbiased $\widehat{\operatorname{MMD}}^{2}_{N}$ in Eq.~\eqref{eq:bmmd} is asymptotically consistent.

 For the proof, we take  $n_g = n_h = N$ (see the Remarks at the end for the general case). Then, Eq.~\eqref{eq:bmmd} becomes a one-sample U-statistic of order two \citep{gretton2012kernel}. Let $\vz_i = \left(\vx_{i}^{(g)}, \vx_{i}^{(h)}\right)^\top$  denote a sample of bivariate vector for $i=1, \ldots, N$ and define 
$$h\left(\vz_i, \vz_j\right) = k\left(\vx_i^{(g)}, \vx_j^{(g)}\right) + k\left(\vx_i^{(h)}, \vx_j^{(h)}\right) - k\left(\vx_i^{(g)}, \vx_j^{(h)}\right) - k\left(\vx_j^{(g)}, \vx_i^{(h)}\right) ,$$
 so that $\widehat{\operatorname{MMD}}^2_{N} = \frac{1}{N(N-1)} \sum_{i \neq j} h\left(\vz_i, \vz_j\right).$

 Since $\widehat{\operatorname{MMD}}^2_N$ is unbiased, we now show that  $\E[(\widehat{\operatorname{MMD}}^2_N)^2] = O\!\left(\frac{1}{N}\right) $ ,  which implies consistency. Following the proof of Theorem 12.3 in \cite{van2000asymptotic}, we have
\begin{equation}
    \E [(\widehat{\operatorname{MMD}}^2_N)^2] \leq \frac{1}{N^{2}(N-1)^{2}}   \sum_{i \neq j} \sum_{u \neq v} \text{Cov} \left(h(\vz_i, \vz_j), h(\vz_u, \vz_v)\right). \label{eq:emmdsq}
\end{equation}

The covariance $\text{Cov}(h(\vz_i, \vz_j), h(\vz_u, \vz_v))$ depends on the spatial lags $|s_i - s_u|$ and $|s_j - s_v|$ for $s_i, s_u \in \mathcal{C}^{(g)}$ and $s_j, s_v \in \mathcal{C}^{(h)}$. The covariance is defined over the RKHS $\mathcal{H}_k$. Following Definition 2.1 in \cite{cherief-abdellatif_finite_2022}, we define the spatial dependence coefficient via the kernel mean embedding  as 
\begin{equation}
    Q_\ell = \left|\E \left[  \langle k(\vx_{s}, \cdot) - \mu_\text{P}, k(\vx_{s+\ell} , \cdot) - \mu_\text{P} \rangle_{\mathcal{H}_k}\right] \right|,
    \label{eq:ql}
\end{equation}
where $\mu_p := \E_{\vx \sim P}[k(\cdot, \vx)]$. Since $\mathcal{C}^{(g)}$ and $\mathcal{C}^{(h)}$ are disjoint, the spatial lags in their respective domains are distinct.  For the collection of spatial lags $\mathcal{L}$ over either process
$\vX^{(g)}$ or $\vX^{(h)}$, we assume the following:

\begin{assumption}\label{ass1}
There exists a $\Sigma < \infty$ such that $\sum_{\ell \in \mathcal{L}} Q_\ell \le \Sigma$.
\end{assumption}
We now show that this assumption is satisfied under model-free mixing conditions. Following \cite{sherman1996variance}, for $\kappa \ge 1$ define
\begin{align}
	\alpha_{\kappa}(\ell) := \sup \big\{ &\mathbb{P}(AB) - \mathbb{P}(A)\mathbb{P}(B): A \in \mathcal{F}(\Delta_1), B \in \mathcal{F}(\Delta_2), \notag\\
	& |\Delta_1| \leq \kappa, |\Delta_2| \leq \kappa, d(\Delta_1, \Delta_2) \geq \ell \big\}, \label{eq:alphal}
\end{align}
where $\Delta_j \subset \mathcal{C}^{(g)}$, $\mathcal{F}(\Delta_j)$ is the $\sigma$-field generated by $\{ \vX^{(g)}(s_i): i \in \Delta_j \}$, and $d(A,B) = \inf \{ |s_i - s_j| : s_i \in A, s_j \in B \}$.

Let $m$ denote the neighborhood size used in CAHC. We assume that this induces
a finite effective dependence range $\ell_0<\infty$, in the sense that for lags
$\ell>\ell_0$, the dependence between the corresponding subsets becomes negligible. Under this finite-range dependence assumption, we have $\alpha_{\kappa}(\ell)=0,  \ell>\ell_0,$ for every finite $\kappa$. Hence Sherman's  \citep{sherman1996variance} model-free mixing condition holds.

Since $Q_\ell$ is the RKHS covariance term at lag $\ell$, $Q_\ell$ inherits the same finite-range property. Thus,
\begin{equation}
	Q_\ell = 0, \qquad \ell>\ell_0.
\end{equation}
It follows that
\begin{equation}
\sum_{\ell\in\mathcal L} Q_\ell
=
\sum_{\ell\in\mathcal L:\,\ell\le \ell_0} Q_\ell
<\infty,
\end{equation}
since only finitely many terms are nonzero. Thus, Assumption~\ref{ass1} $\sum_{\ell\in\mathcal L} Q_\ell \le \Sigma$ is satisfied.

Eq.~\eqref{eq:emmdsq} together with
$\sum_{\ell \in \mathcal{L}} Q_\ell < \infty$ implies
\begin{equation} 
	\E [(\widehat{\operatorname{MMD}}^2_N)^2] = O\!\left(\frac{1}{N}\right) \to 0. 
\end{equation} 
It follows that
\begin{equation} 
	\widehat{\operatorname{MMD}}^2_N \xrightarrow{p} \operatorname{MMD}^{2} \quad \text{as} \quad N \rightarrow \infty. \label{eq:consistency} 
\end{equation} 
Thus, under model-free mixing conditions, the unbiased $\operatorname{MMD}^2_N$ is asymptotically consistent. Under $\text{H}_0: \text{P}^{(g)} = \text{P}^{(h)}$ and model-free mixing conditions, $\operatorname{MMD}^2_N \xrightarrow{p} 0$, which establishes consistency of the test under $\mathrm{H}_0$.

Consequently, under the same conditions, the biased version of Eq.~\eqref{eq:bmmd} is also asymptotically consistent. This follows because
the difference between the biased and unbiased estimators is $O(1/N)$ and the
kernel $k$ is bounded. Hence, by Slutsky's theorem, the biased estimator is
also asymptotically consistent.

 \begin{remark}[Case $n_g \neq n_h$]  When $n_g \neq n_h$, $\operatorname{MMD}^2$ is a two-sample von Mises functional. the same dependence and summability conditions apply separately to each of the
 three double sums in Eq.~\eqref{eq:bmmd}, giving $\widehat{\operatorname{MMD}}^2 \xrightarrow{p} \operatorname{MMD}^{2} \quad \text{as} \quad N \rightarrow \infty$ under the same regularity conditions.
 \end{remark}

\section{Illustrative Example }
\subsection{Testing Spatial Invariance ($\widehat{\operatorname{MMD}}^{2}$ Results)}
The results reveal two groups of clusters with small $\widehat{\operatorname{MMD}}^{2}$ values. 
Specifically, clusters $\{\mathcal{C}_1, \mathcal{C}_2, \mathcal{C}_3\}$ and 
$\{\mathcal{C}_4, \mathcal{C}_5, \mathcal{C}_6\}$ exhibit pairwise similarities, as indicated by the bold entries in Table~\ref{tab_mmd_example}.

\begin{table}[h]
	\centering
	\caption{Observed $\widehat{\operatorname{MMD}}^{2}$ values for all pairs of clusters in the illustrative example. Bold values indicate pairs for which the null hypothesis is not rejected.}
	\label{tab_mmd_example}
	\renewcommand{\arraystretch}{1.2}
	\begin{tabular}{cccc}
		\toprule
		\textbf{Cluster Pair} & $\widehat{\operatorname{MMD}}^{2}$ & \textbf{Cluster Pair} & $\widehat{\operatorname{MMD}}^{2}$ \\
		\midrule
		$(\mathcal{C}_1, \mathcal{C}_2)$ & \textbf{0.02} & $(\mathcal{C}_3, \mathcal{C}_4)$ & 0.51 \\
		$(\mathcal{C}_1, \mathcal{C}_3)$ & \textbf{0.01} & $(\mathcal{C}_3, \mathcal{C}_5)$ & 0.50 \\
		$(\mathcal{C}_1, \mathcal{C}_4)$ & 0.53 & $(\mathcal{C}_3, \mathcal{C}_6)$ & 0.51 \\
		$(\mathcal{C}_1, \mathcal{C}_5)$ & 0.52 & $(\mathcal{C}_3, \mathcal{C}_7)$ & 0.44 \\
		$(\mathcal{C}_1, \mathcal{C}_6)$ & 0.52 & $(\mathcal{C}_4, \mathcal{C}_5)$ & \textbf{0.03} \\
		$(\mathcal{C}_1, \mathcal{C}_7)$ & 0.45 & $(\mathcal{C}_4, \mathcal{C}_6)$ & \textbf{0.01} \\
		$(\mathcal{C}_2, \mathcal{C}_3)$ & \textbf{0.02} & $(\mathcal{C}_4, \mathcal{C}_7)$ & 0.59 \\
		$(\mathcal{C}_2, \mathcal{C}_4)$ & 0.52 & $(\mathcal{C}_5, \mathcal{C}_6)$ & \textbf{0.02} \\
		$(\mathcal{C}_2, \mathcal{C}_5)$ & 0.51 & $(\mathcal{C}_5, \mathcal{C}_7)$ & 0.58 \\
		$(\mathcal{C}_2, \mathcal{C}_6)$ & 0.51 & $(\mathcal{C}_6, \mathcal{C}_7)$ & 0.58 \\
		$(\mathcal{C}_2, \mathcal{C}_7)$ & 0.46 &  &  \\
		\bottomrule
	\end{tabular}
\end{table}

\subsection{Block Permutation Results}

\begin{figure}[h]
	\centering
	\includegraphics[width=\textwidth]{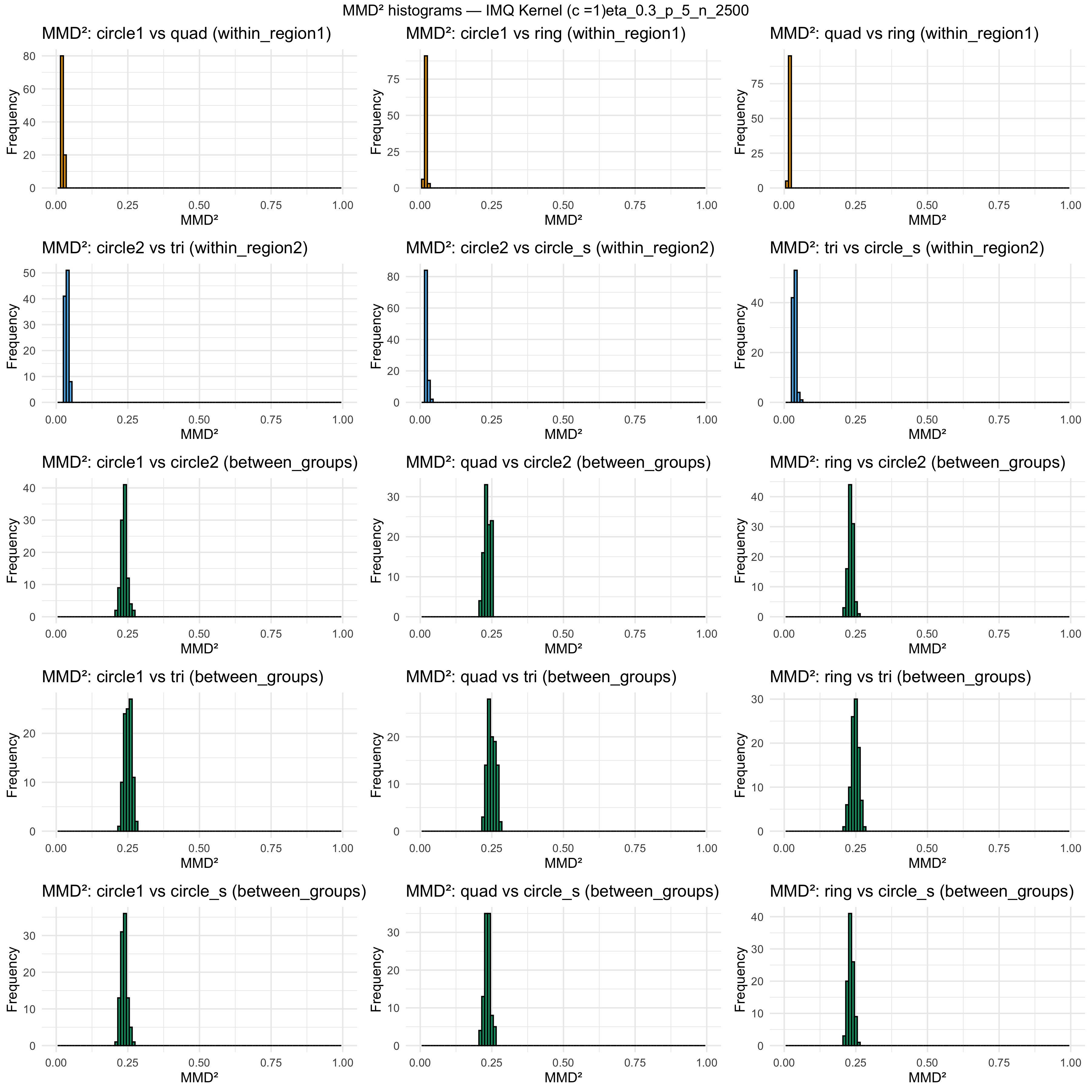}
	\caption{Sampling distribution of $\widehat{\operatorname{MMD}}^{2}$ using the IMQ kernel with $c = 1$, spatial autocorrelation $\eta = 0.8$, number of attributes $p = 5$, and $n = 2500$. Rows 1--2: comparisons between clusters from regions simulated under the same distribution. Rows 3--5: comparisons between clusters from regions simulated under different distributions. We omitted other six pairs of clusters with noise regions. }
	\label{blk_perm_illus_example}
\end{figure}

Figure~\ref{blk_perm_illus_example} shows the sampling distributions of the $\widehat{\operatorname{MMD}}^{2}$ statistic obtained from the block permutation procedure for 15 pairs of clusters. The remaining six pairs involving noise regions are omitted for clarity. Table~\ref{tab:pvalues} summarizes the corresponding $p$-values and adjusted $p$-values. 

\begin{table}[h]
	\centering
	\caption{Hypothesis test results for all cluster pairs. Bold adjusted $p$-values indicate pairs for which the null hypothesis is not rejected.}
	\label{tab:pvalues}
	\renewcommand{\arraystretch}{1.2}
	\begin{tabular}{ccc}
		\toprule
		\textbf{Cluster Pair} & \textbf{$p$-value} & \textbf{Adjusted $p$-value} \\
		\midrule
		$(\mathcal{C}_1, \mathcal{C}_2)$ & 0.92 & \textbf{1.00} \\
		$(\mathcal{C}_1, \mathcal{C}_3)$ & 1.00 & \textbf{1.00} \\
		$(\mathcal{C}_1, \mathcal{C}_4)$ & 0.00 & 0.00 \\
		$(\mathcal{C}_1, \mathcal{C}_5)$ & 0.00 & 0.00 \\
		$(\mathcal{C}_1, \mathcal{C}_6)$ & 0.00 & 0.00 \\
		$(\mathcal{C}_1, \mathcal{C}_7)$ & 0.00 & 0.00 \\
		$(\mathcal{C}_2, \mathcal{C}_3)$ & 0.80 & \textbf{1.00} \\
		$(\mathcal{C}_2, \mathcal{C}_4)$ & 0.00 & 0.00 \\
		$(\mathcal{C}_2, \mathcal{C}_5)$ & 0.00 & 0.00 \\
		$(\mathcal{C}_2, \mathcal{C}_6)$ & 0.00 & 0.00 \\
		$(\mathcal{C}_2, \mathcal{C}_7)$ & 0.00 & 0.00 \\
		$(\mathcal{C}_3, \mathcal{C}_4)$ & 0.00 & 0.00 \\
		$(\mathcal{C}_3, \mathcal{C}_5)$ & 0.00 & 0.00 \\
		$(\mathcal{C}_3, \mathcal{C}_6)$ & 0.00 & 0.00 \\
		$(\mathcal{C}_3, \mathcal{C}_7)$ & 0.00 & 0.00 \\
		$(\mathcal{C}_4, \mathcal{C}_5)$ & 0.98 & \textbf{1.00} \\
		$(\mathcal{C}_4, \mathcal{C}_6)$ & 1.00 & \textbf{1.00} \\
		$(\mathcal{C}_4, \mathcal{C}_7)$ & 0.00 & 0.00 \\
		$(\mathcal{C}_5, \mathcal{C}_6)$ & 1.00 & \textbf{1.00} \\
		$(\mathcal{C}_5, \mathcal{C}_7)$ & 0.00 & 0.00 \\
		$(\mathcal{C}_6, \mathcal{C}_7)$ & 0.00 & 0.00 \\
		\bottomrule
	\end{tabular}
\end{table}

\subsection{Graph Construction and Reassignment}\label{graph_illustration}

\begin{figure}[h]
	\centering
	\includegraphics[width=\textwidth]{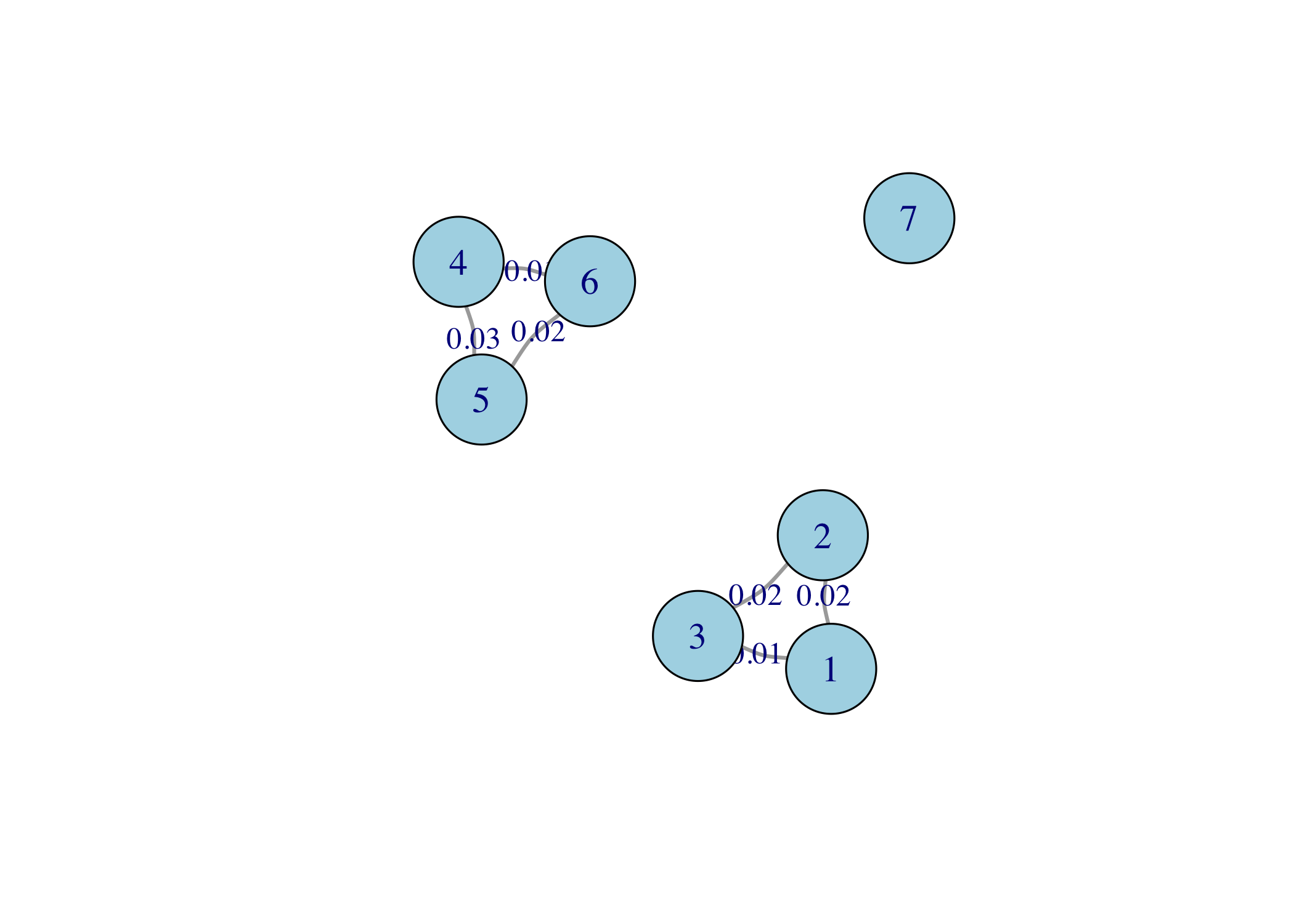}
	\caption[Graph Based on Pairwise Tests]{Graph constructed from pairwise $\widehat{\operatorname{MMD}}^{2}$ tests across seven clusters. Edges represent not enough evidence to find differences, with edge weights corresponding to observed $\widehat{\operatorname{MMD}}^{2}$ values.}
	\label{sample_pwr_graph}
\end{figure}

Figure~\ref{sample_pwr_graph} illustrates a graph constructed from pairwise test results among seven clusters. Two distinct fully connected subgraphs emerge: one comprising nodes $\{1,2,3\}$ and another comprising nodes $\{4,5,6\}$. These cliques correspond to groups of clusters with similar underlying distributions. Following the merging rule in the proposed framework, clusters $\{1,2,3\}$ are combined into a single cluster, and clusters $\{4,5,6\}$ are combined into another. Cluster 7 remains separate, as it does not form a clique with any other cluster. Consequently, the final number of clusters is three. 

\section{Sampling distribution of MMD$^2$ Using Monte Carlo Method}
\label{app1}

We conducted simulations by varying the number of spatial points $n$, attribute dimension $p$, spatial autocorrelation $\eta$, kernel type, and kernel-specific parameters $c$ to investigate the sampling distribution of the MMD$^2$ statistic under scenarios where the underlying distributions of two clusters are either identical or different. Notably, we found that the inverse multiquadratic (IMQ) kernel with shape parameter $c \geq 1$ shifts the distribution of MMD$^2$ further away from zero when the cluster distributions differ, thereby enhancing sensitivity. The IMQ kernel also increases the range of MMD$^2$ values under the alternative hypothesis, improving the ability to detect differences in distributions.

\subsection{Sampling Distribution of MMD$^2$ Under Differences in Mean Function, $\mu$}
\label{app2}

As described in in Section \ref{sim-sett}, we simulate random fields for patches in two regions with a varying means $\mu_{1} = 4$, $\mu_{2} = 10$ and variance $\tau^{2} = 1$, and the same covariance function with spatial autocorrelation $\eta = 0.3$ or $\eta = 0.8$. We approximate the sampling distribution of MMD$^2$ for each setting using the Monte Carlo method. 

We demonstrate that the IMQ kernel with parameter $c\geq1$ exhibits the greatest statistical power in detecting differences between the underlying distributions of clusters that are spatially separated.

\begin{figure}[h]
	\centering
	\includegraphics[width=\linewidth]{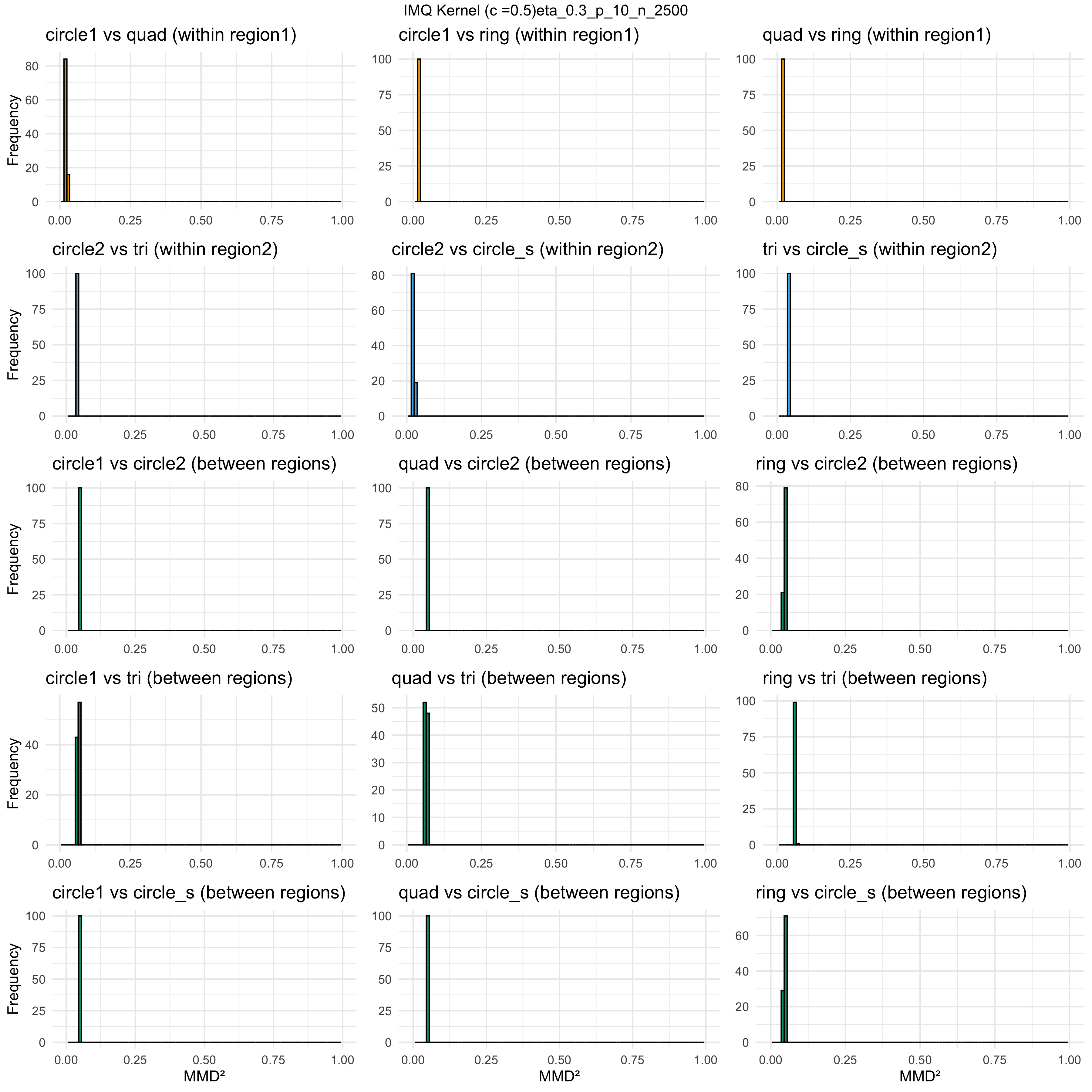}
	\caption{Sampling distribution of MMD$^2$ using the IMQ kernel with $c = 0.5$, spatial autocorrelation $\eta = 0.3$, number of attributes $p = 10$, and $n = 2500$. Rows 1--2: comparisons between clusters from regions simulated under the same distribution. Rows 3--5: comparisons between clusters from regions simulated under different distributions.}
	\label{fig:mmd-prop1}
\end{figure}

\begin{figure}[h]
	\centering
	\includegraphics[width=\linewidth]{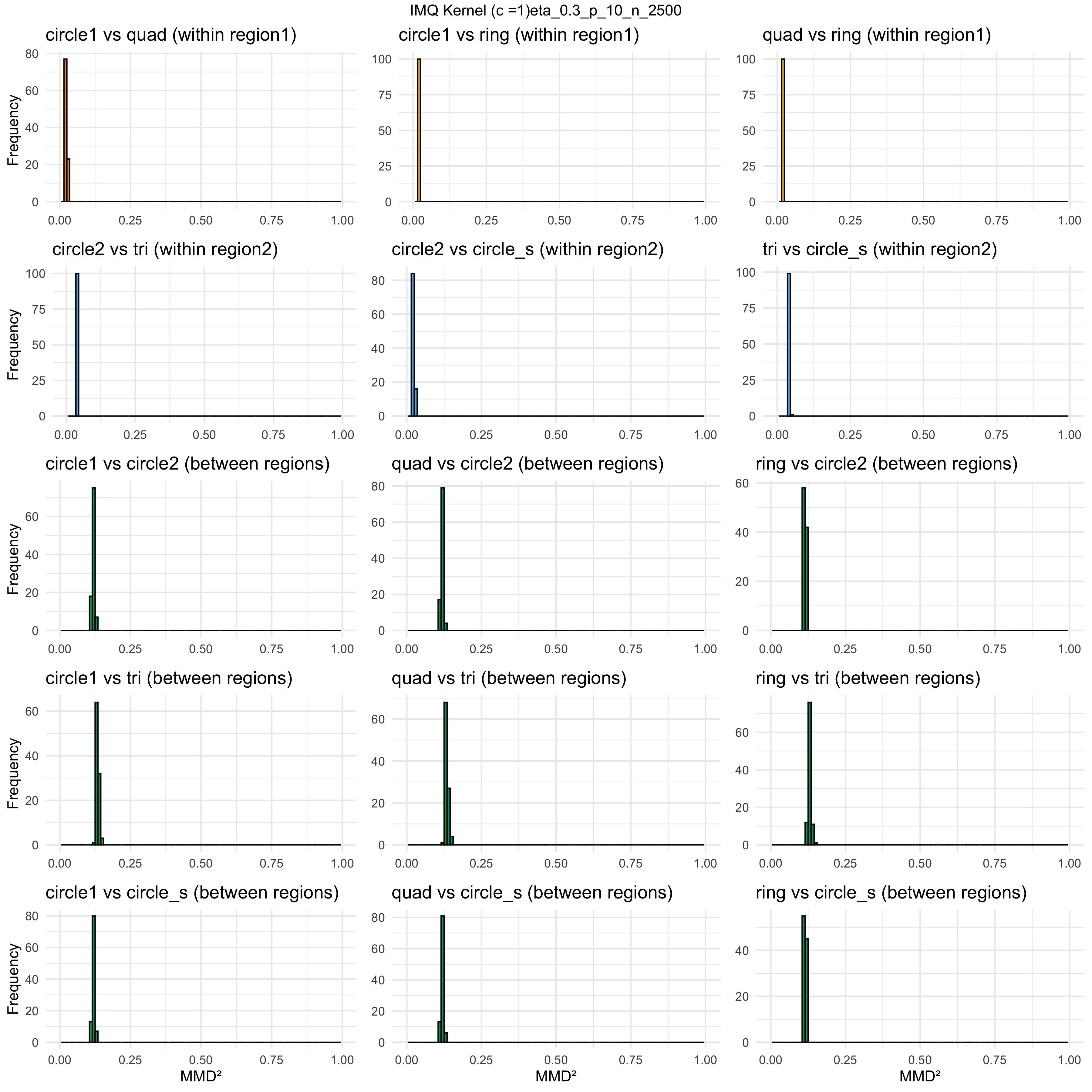}
	\caption{Sampling distribution of MMD$^2$ using the IMQ kernel with $c = 1$, spatial autocorrelation $\eta = 0.3$, number of attributes $p = 10$, and $n = 2500$. Rows 1--2: comparisons between clusters from regions simulated under the same distribution. Rows 3--5: comparisons between clusters from regions simulated under different distributions.}
	\label{fig:mmd-prop2}
\end{figure}

\begin{figure}[h]
	\centering
	\includegraphics[width=\linewidth]{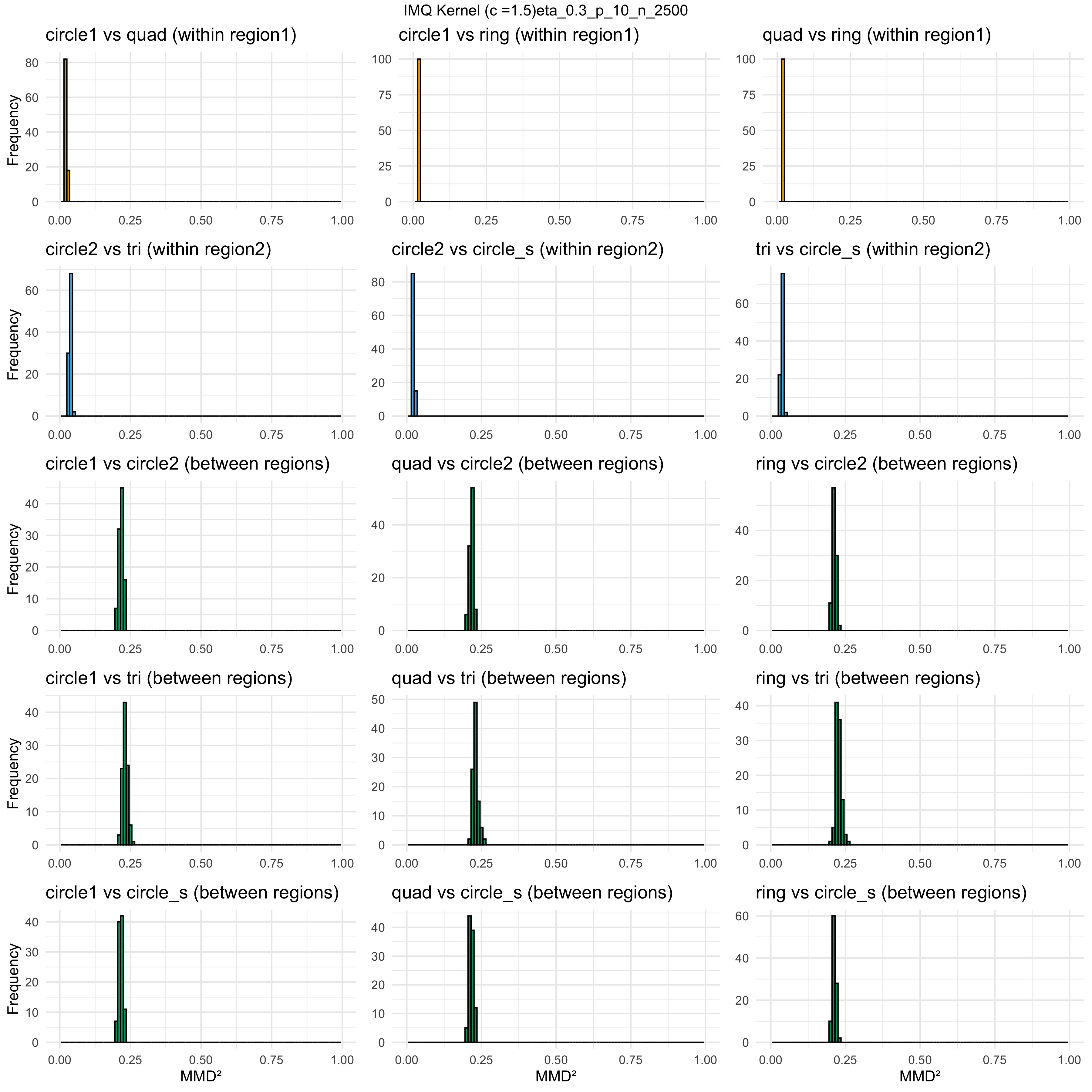}
	\caption{Sampling distribution of MMD$^2$ using the IMQ kernel with $c = 1.5$, spatial autocorrelation $\eta = 0.3$, number of attributes $p = 10$, and $n = 2500$. Rows 1--2: comparisons between clusters from regions simulated under the same distribution. Rows 3--5: comparisons between clusters from regions simulated under different distributions.}
	\label{fig:mmd-prop3}
\end{figure}

\begin{figure}[h]
	\centering
	\includegraphics[width=\linewidth]{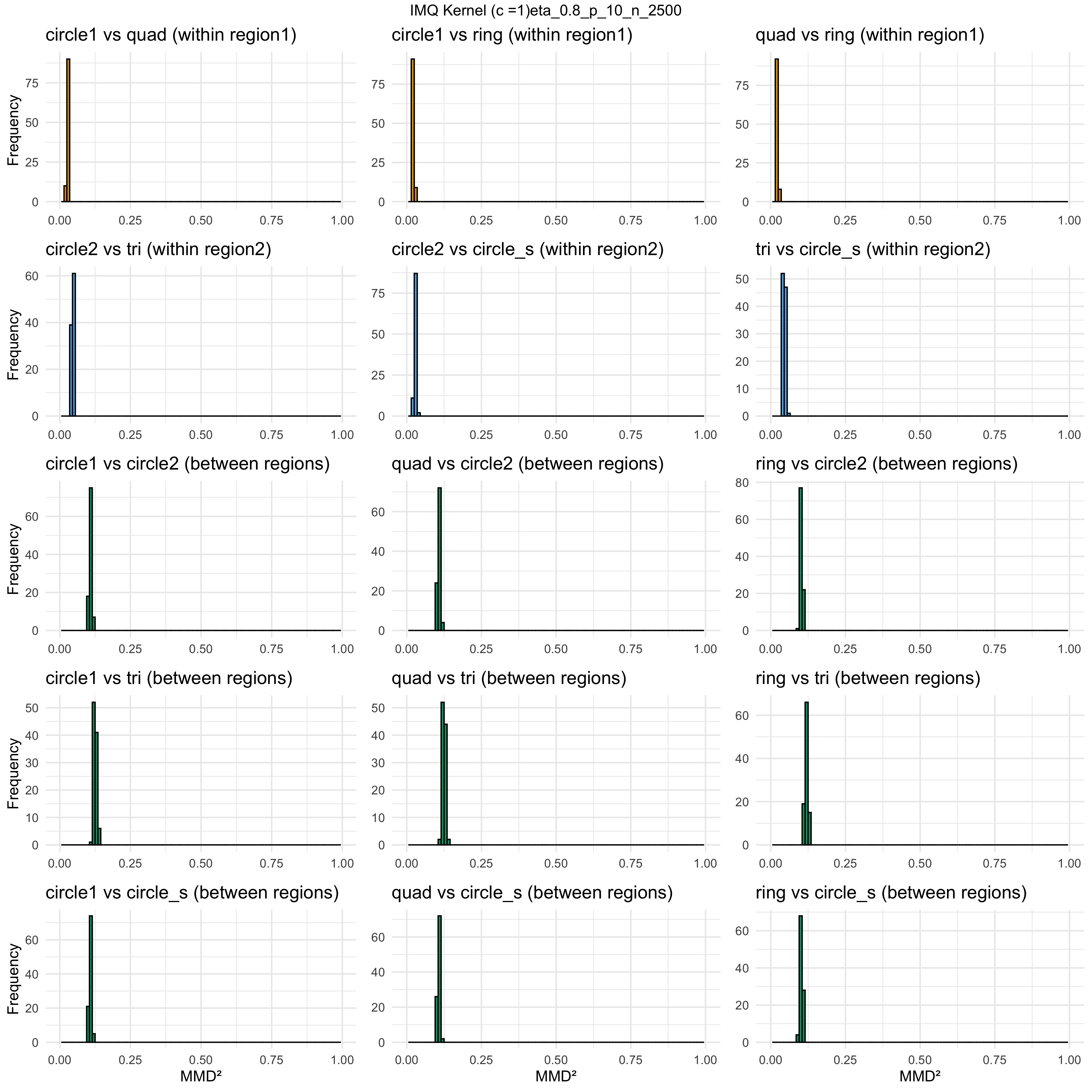}
	\caption{Sampling distribution of MMD$^2$ using the IMQ kernel with $c = 1$, spatial autocorrelation $\eta = 0.8$, number of attributes $p = 10$, and $n = 2500$. Rows 1--2: comparisons between clusters from regions simulated under the same distribution. Rows 3--5: comparisons between clusters from regions simulated under different distributions.}
	\label{fig:mmd-prop8}
\end{figure}

\begin{figure}[h]
	\centering
	\includegraphics[width=\linewidth]{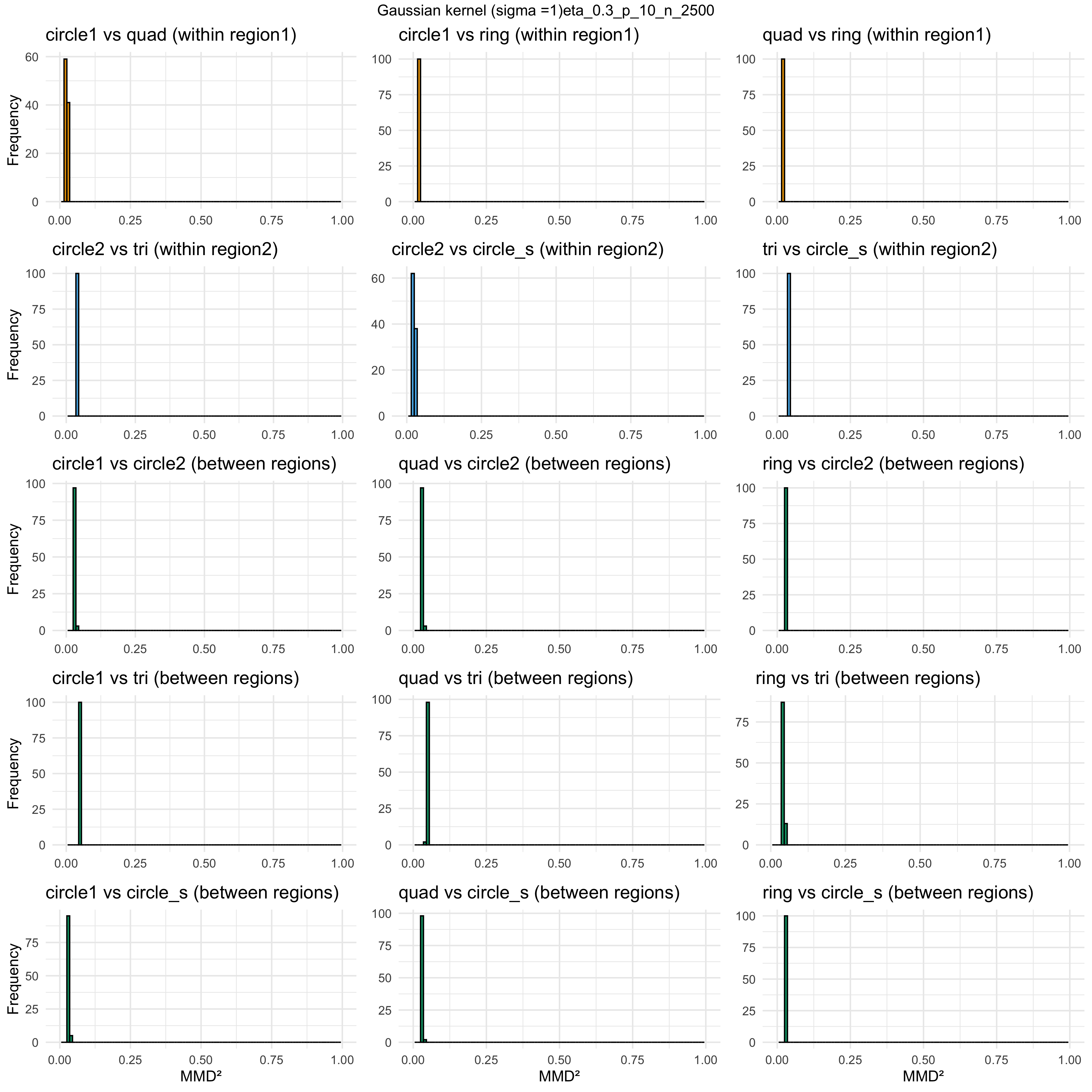}
	\caption{Sampling distribution of MMD$^2$ using the Gaussian kernel with $\sigma = 1$, spatial autocorrelation $\eta = 0.3$, number of attributes $p = 10$, and $n = 2500$. Rows 1--2: comparisons between clusters from regions simulated under the same distribution. Rows 3--5: comparisons between clusters from regions simulated under different distributions.}
	\label{fig:mmd-prop4}
\end{figure}

\begin{figure}[h]
	\centering
	\includegraphics[width=\linewidth]{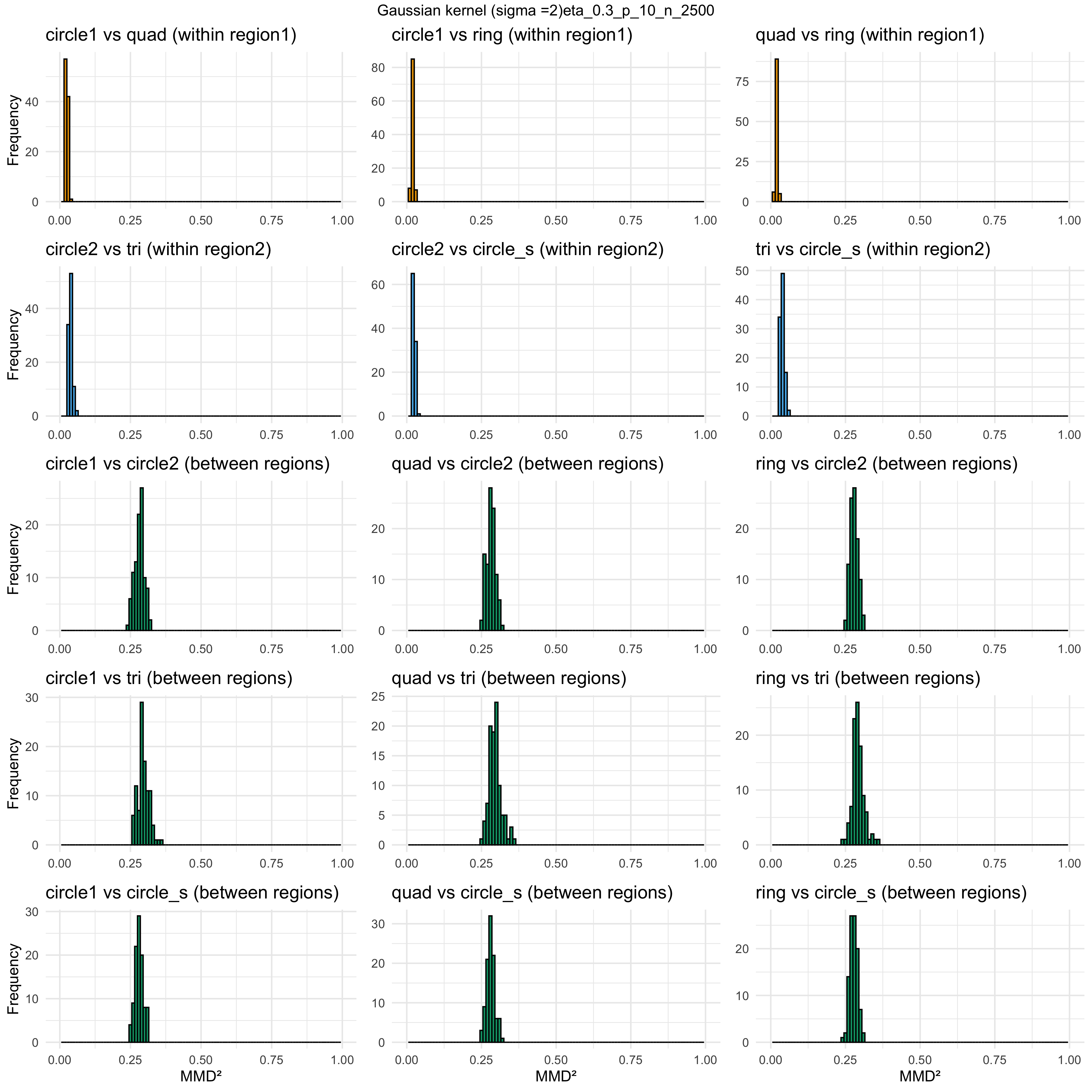}
	\caption{Sampling distribution of MMD$^2$ using the Gaussian kernel with $\sigma = 2$, spatial autocorrelation $\eta = 0.3$, number of attributes $p = 10$, and $n = 2500$. Rows 1--2: comparisons between clusters from regions simulated under the same distribution. Rows 3--5: comparisons between clusters from regions simulated under different distributions.}
	\label{fig:mmd-prop5}
\end{figure}

\begin{figure}[h]
	\centering
	\includegraphics[width=\linewidth]{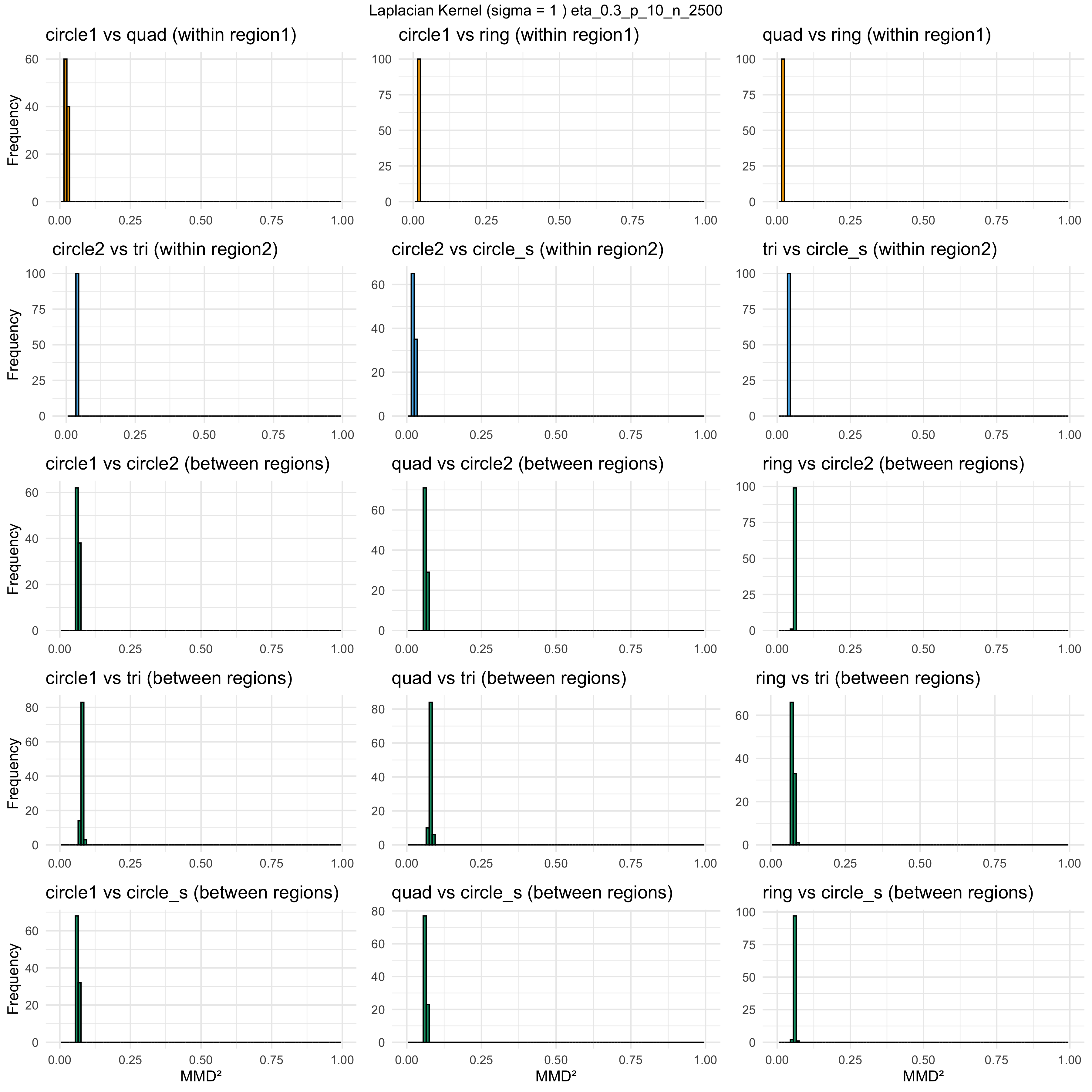}
	\caption{Sampling distribution of MMD$^2$ using the Laplacian kernel with $\sigma = 1$, spatial autocorrelation $\eta = 0.3$, number of attributes $p = 10$, and $n = 2500$. Rows 1--2: comparisons between clusters from regions simulated under the same distribution. Rows 3--5: comparisons between clusters from regions simulated under different distributions.}
	\label{fig:mmd-prop6}
\end{figure}

\begin{figure}[h]
	\centering
	\includegraphics[width=\linewidth]{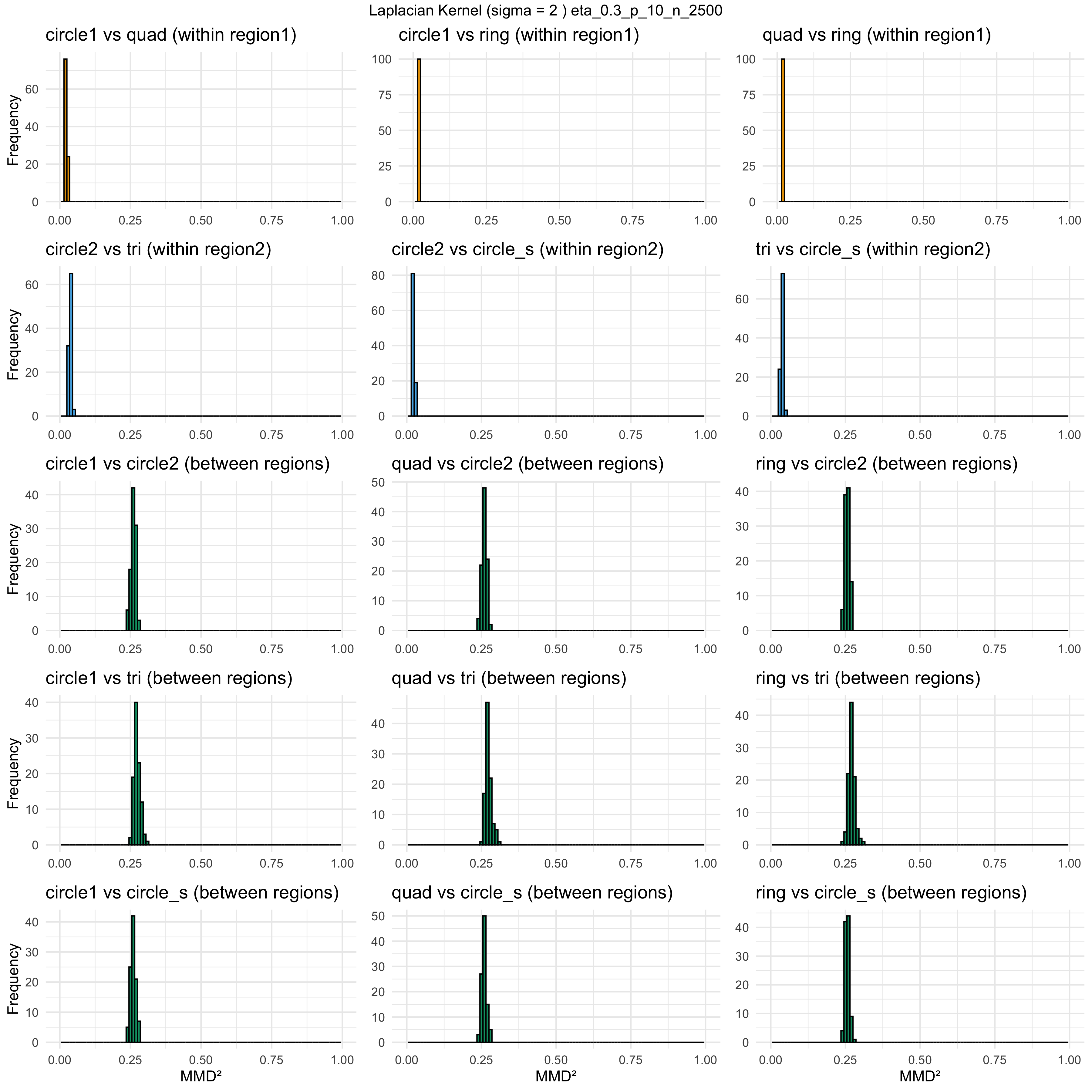}
	\caption{Sampling distribution of MMD$^2$ using the Laplacian kernel with $\sigma = 2$, spatial autocorrelation $\eta = 0.3$, number of attributes $p = 10$, and $n = 2500$. Rows 1--2: comparisons between clusters from regions simulated under the same distribution. Rows 3--5: comparisons between clusters from regions simulated under different distributions.}
	\label{fig:mmd-prop7}
\end{figure}

\subsection{Sampling distribution of MMD$^2$ under differences in spatial covariance (controlled by $\eta$)}
\label{app3}

We simulate spatial random fields over patches in two regions, each with a constant mean $\mu = 10$ and marginal variance $\tau^2 = 1$. To introduce differences in distributions, we vary the spatial autocorrelation parameters: $\eta_1 = 0.3$ for one region and $\eta_2 = 0.8$ for the other.

Although differences in spatial autocorrelation affect the covariance structure of the underlying spatial distributions, the MMD$^2$ statistic exhibits limited sensitivity to such differences in this setting. This is because the marginal distributions remain nearly identical and the shift in spatial dependence is relatively subtle in a conditional autoregressive model, making it difficult for MMD$^2$ to detect distributional differences that are driven solely by changes in spatial autocorrelation.

\section{Descriptive Summary of the Real Dataset}\label{append_descriptive}

The dataset consists of 179,194 cells across 39 tissue samples, with the number of cells per sample ranging from 1,716 to 9,886. We focus on four samples used for the \texttt{repSpat} application and presented in the main text (Samples 04, 05, 26, and 39), which contain 5,381, 4,252, 5,409, and 3,663 cells, respectively.

\begin{figure}[h]
	\centering
	\includegraphics[width=\linewidth]{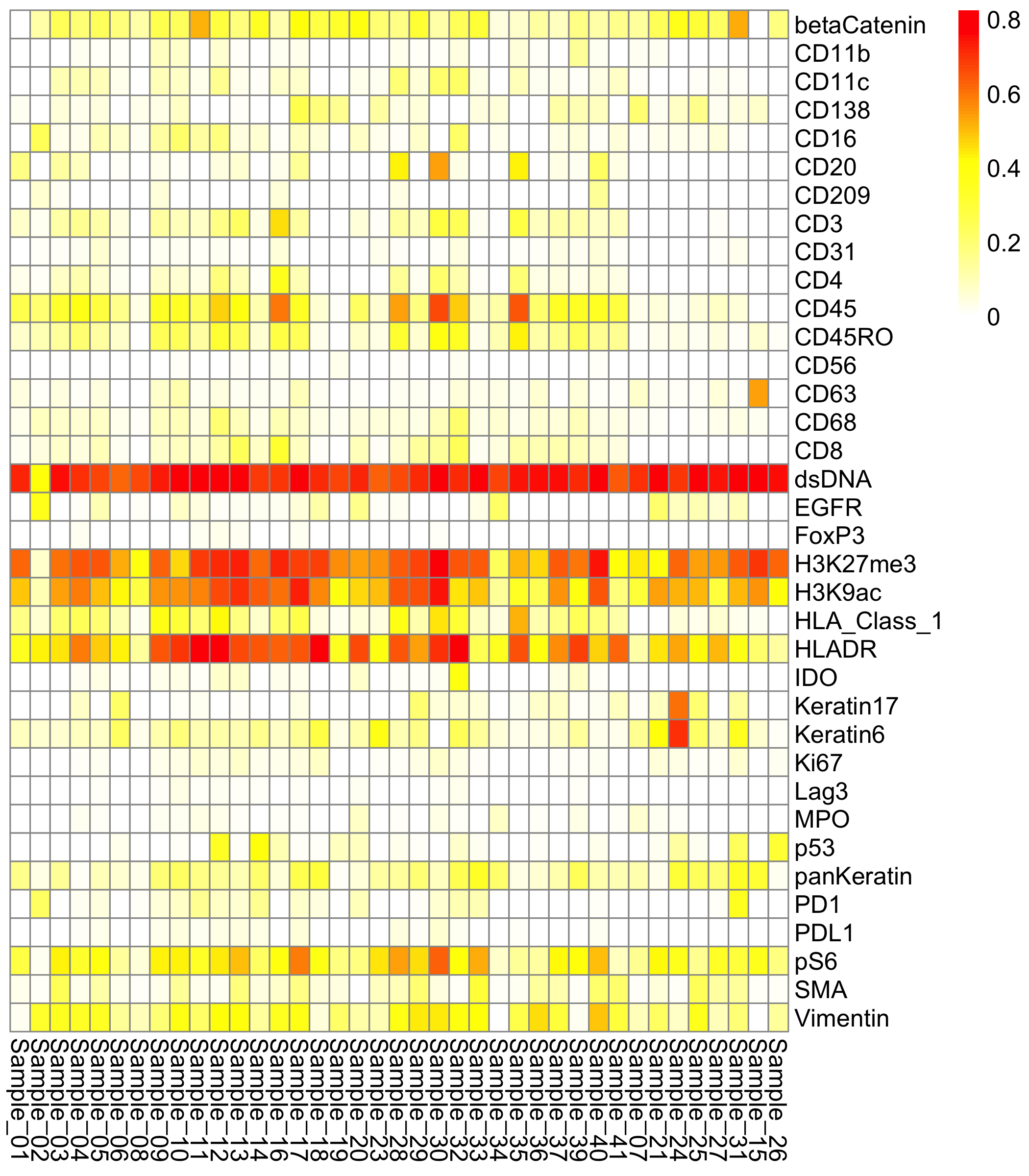}
	\caption{Heatmap of mean protein marker expression across tissue samples. Each row corresponds to a marker and each column represents a tissue sample. The values denote the mean expression of each marker within a sample. Colors range from white (low expression) to yellow and red (high expression). The heatmap illustrates variability in mean marker expression across tissues.}
	\label{fig:protein_mean}
\end{figure}

Figure~\ref{fig:protein_mean} indicates variability in marker expression across tissue samples, with several markers exhibiting consistently high or low mean expression levels. In contrast, other markers display substantial heterogeneity across samples, indicating differences in their distribution across tissues.

\begin{figure}[h]
	\centering
	\includegraphics[width=\linewidth]{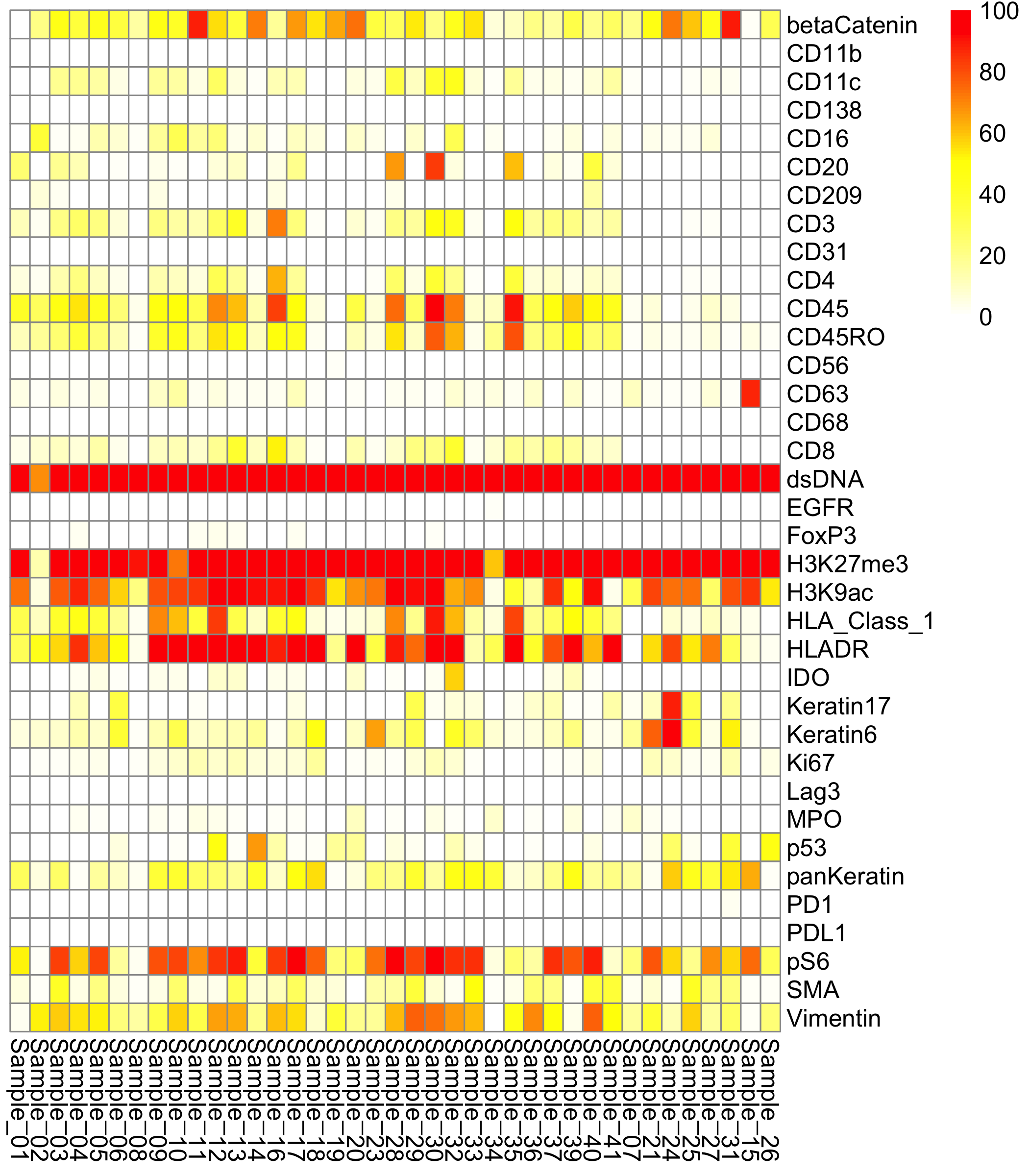}
	\caption{Heatmap of the percentage of cells in which each marker is present across tissue samples. Each row corresponds to a marker and each column represents a tissue sample. Values denote the proportion of cells (in percentage) with marker expression above the predefined threshold within each sample. Percentages are used to account for varying numbers of cells across samples. Colors range from white (0\% percentage) to yellow and red (high percentage).}
	\label{fig:protein_binary_percent}
\end{figure}

Figure~\ref{fig:protein_binary_percent} shows that several markers have zero presence across all tissue samples, while others exhibit nonzero and varying proportions of cells exceeding the threshold. This reflects feature selection through thresholding and the variability retained among the remaining markers. The thresholds are presented in the computational workflow for descriptive analysis in \ref{app4}.

\section{Additonal Simulation Results}

This section presents additional simulation results evaluating the performance of \texttt{repSpat} under increasing numbers of noise features.

\begin{table}[h]
	\caption{Performance of \texttt{repSpat} over 100 simulation runs for each setting with one additional noise feature. For each run, 21 pairwise tests are conducted, of which six pairs correspond to true distributional differences and 15 pairs correspond to no differences. A false positive (FP) occurs when a pair of clusters sharing the same underlying distribution is incorrectly identified as different, while a false negative (FN) occurs when a pair of clusters with different distributions is incorrectly identified as similar. We report the median and interquartile range (IQR) of the proportions of false positives (FPR) and false negatives (FNR) across runs. Precision is defined as the proportion of correctly identified differences among all detected differences, and recall (sensitivity, or true positive rate) is the proportion of true differences that are correctly identified. The F1 score is computed as the harmonic mean of precision and recall. The final column reports the proportion of runs with F1 score less than 1.}
	\centering
	\renewcommand{\arraystretch}{1.4}
	\begin{tabular}{@{} c c c c c c c @{}}
		\toprule
		\textbf{$n$} & \textbf{$p$} & \textbf{$\eta$} & \textbf{FNR} & \textbf{FPR} & \textbf{F1 Score} & \textbf{\% F1 $<$ 1} \\
		\midrule
		2500 & 5  & 0.3 & 0 (0) & 0 (0) & 1 (0) & 0.02 \\
		& 5  & 0.8 & 0 (0) & 0 (0) & 1 (0) & 0.02 \\
		& 10 & 0.3 & 0 (0) & 0 (0) & 1 (0) & 0.07 \\
		& 10 & 0.8 & 0 (0) & 0 (0) & 1 (0) & 0.04 \\
		\midrule
		4900 & 5  & 0.3 & 0 (0) & 0 (0) & 1 (0) & 0 \\
		& 5  & 0.8 & 0 (0) & 0 (0) & 1 (0) & 0.02 \\
		& 10 & 0.3 & 0 (0) & 0 (0) & 1 (0) & 0 \\
		& 10 & 0.8 & 0 (0) & 0 (0) & 1 (0) & 0.02 \\
		\bottomrule
	\end{tabular}
	\label{tab_sim_results2}
\end{table}

\begin{table}[h]
	\caption{Performance of \texttt{repSpat} over 100 simulation runs for each setting with two additional noise features. For each run, 21 pairwise tests are conducted, of which six pairs correspond to true distributional differences and 15 pairs correspond to no differences. A false positive (FP) occurs when a pair of clusters sharing the same underlying distribution is incorrectly identified as different, while a false negative (FN) occurs when a pair of clusters with different distributions is incorrectly identified as similar. We report the median and interquartile range (IQR) of the proportions of false positives (FPR) and false negatives (FNR) across runs. Precision is defined as the proportion of correctly identified differences among all detected differences, and recall (sensitivity, or true positive rate) is the proportion of true differences that are correctly identified. The F1 score is computed as the harmonic mean of precision and recall. The final column reports the proportion of runs with F1 score less than 1.}
	\centering
	\renewcommand{\arraystretch}{1.4}
	\begin{tabular}{@{} c c c c c c c @{}}
		\toprule
		\textbf{$n$} & \textbf{$p$} & \textbf{$\eta$} & \textbf{FNR} & \textbf{FPR} & \textbf{F1 Score} & \textbf{\% F1 $<$ 1} \\
		\midrule
		2500 & 5  & 0.3 & 0 (0) & 0 (0) & 1 (0) & 0.02 \\
		& 5  & 0.8 & 0 (0) & 0 (0) & 1 (0) & 0.01\\
		& 10 & 0.3 & 0 (0) & 0 (0) & 1 (0) & 0 \\
		& 10 & 0.8 & 0 (0) & 0 (0) & 1 (0) & 0.03 \\
		\midrule
		4900 & 5  & 0.3 & 0 (0) & 0 (0) & 1 (0) & 0.02 \\
		& 5  & 0.8 & 0 (0) & 0 (0) & 1 (0) & 0.03 \\
		& 10 & 0.3 & 0 (0) & 0 (0) & 1 (0) & 0 \\
		& 10 & 0.8 & 0 (0) & 0 (0) & 1 (0) & 0.02 \\
		\bottomrule
	\end{tabular}
	\label{tab_sim_results3}
\end{table}

\begin{table}[h]
	\caption{Performance of \texttt{repSpat} over 100 simulation runs for each setting with three additional noise features. For each run, 21 pairwise tests are conducted, of which six pairs correspond to true distributional differences and 15 pairs correspond to no differences. A false positive (FP) occurs when a pair of clusters sharing the same underlying distribution is incorrectly identified as different, while a false negative (FN) occurs when a pair of clusters with different distributions is incorrectly identified as similar. We report the median and interquartile range (IQR) of the proportions of false positives (FPR) and false negatives (FNR) across runs. Precision is defined as the proportion of correctly identified differences among all detected differences, and recall (sensitivity, or true positive rate) is the proportion of true differences that are correctly identified. The F1 score is computed as the harmonic mean of precision and recall. The final column reports the proportion of runs with F1 score less than 1.}
	\centering
	\renewcommand{\arraystretch}{1.4}
	\begin{tabular}{@{} c c c c c c c @{}}
		\toprule
		\textbf{$n$} & \textbf{$p$} & \textbf{$\eta$} & \textbf{FNR} & \textbf{FPR} & \textbf{F1 Score} & \textbf{\% F1 $<$ 1} \\
		\midrule
		2500 & 5  & 0.3 & 0 (0) & 0 (0) & 1 (0) & 0.01 \\
		& 5  & 0.8 & 0 (0) & 0 (0) & 1 (0) & 0.01\\
		& 10 & 0.3 & 0 (0) & 0 (0) & 1 (0) & 0 \\
		& 10 & 0.8 & 0 (0) & 0 (0) & 1 (0) & 0.05 \\
		\midrule
		4900 & 5  & 0.3 & 0 (0) & 0 (0) & 1 (0) & 0 \\
		& 5  & 0.8 & 0 (0) & 0 (0) & 1 (0) & 0 \\
		& 10 & 0.3 & 0 (0) & 0 (0) & 1 (0) & 0 \\
		& 10 & 0.8 & 0 (0) & 0 (0) & 1 (0) & 0.01 \\
		\bottomrule
	\end{tabular}
	\label{tab_sim_results4}
\end{table}

\begin{table}[h]
	\caption{Clustering performance measured by the adjusted Rand index (ARI) for \texttt{repSpat} and \texttt{Banksy} over 100 simulation runs for each setting with one noise feature. Entries report the median ARI with interquartile range (IQR) in parentheses. The final column reports the percentage of runs in which the ARI of \texttt{repSpat} is less than 1.}
	\centering
	\renewcommand{\arraystretch}{1.4}
	\begin{tabular}{@{} c c c c c c @{}} 
		\toprule
		\textbf{$n$} & \textbf{$p$} & \textbf{$\eta$} 
		& \textbf{repSpat (ARI)} 
		& \textbf{Banksy (ARI)} 
		& \textbf{\% ARI $<$ 1 (repSpat)} \\  
		\midrule
		2500 & 5  & 0.3 & 1 (0) & 0.83 (0.03) & 0.03 \\
		& 5  & 0.8 & 1 (0) & 0.82 (0.01) & 0.02 \\
		& 10 & 0.3 & 1 (0) & 0.82 (0.03) & 0 \\
		& 10 & 0.8 & 1 (0) & 0.83 (0.03) & 0.08 \\
		\midrule
		4900 & 5  & 0.3 & 1 (0) & 0.85 (0.10) & 0.01 \\
		& 5  & 0.8 & 1 (0) & 0.84 (0.01) & 0.02 \\
		& 10 & 0.3 & 1 (0) & 0.83 (0.10) & 0 \\
		& 10 & 0.8 & 1 (0) & 0.83 (0.10) & 0.02 \\
		\bottomrule
	\end{tabular}
	\label{tab_sim_comp1}
\end{table}

\begin{table}[h]
	\caption{Clustering performance measured by the adjusted Rand index (ARI) for \texttt{repSpat} and \texttt{Banksy} over 100 simulation runs for each setting with two noise features. Entries report the median ARI with interquartile range (IQR) in parentheses. The final column reports the percentage of runs in which the ARI of \texttt{repSpat} is less than 1.}
	\centering
	\renewcommand{\arraystretch}{1.4}
	\begin{tabular}{@{} c c c c c c @{}} 
		\toprule
		\textbf{$n$} & \textbf{$p$} & \textbf{$\eta$} 
		& \textbf{repSpat (ARI)} 
		& \textbf{Banksy (ARI)} 
		& \textbf{\% ARI $<$ 1 (repSpat)} \\  
		\midrule
		2500 & 5  & 0.3 & 1 (0) & 0.83 (0.01) & 0.02 \\
		& 5  & 0.8 & 1 (0) & 0.83 (0.02) & 0.01 \\
		& 10 & 0.3 & 1 (0) & 0.83 (0.02) & 0 \\
		& 10 & 0.8 & 1 (0) & 0.83 (0.04) & 0.05 \\
		\midrule
		4900 & 5  & 0.3 & 1 (0) & 0.85 (0.01) & 0.02 \\
		& 5  & 0.8 & 1 (0) & 0.85 (0.01) & 0.03 \\
		& 10 & 0.3 & 1 (0) & 0.83 (0.10) & 0 \\
		& 10 & 0.8 & 1 (0) & 0.84 (0.10) & 0.02 \\
		\bottomrule
	\end{tabular}
	\label{tab_sim_comp2}
\end{table}

\begin{table}[h]
	\caption{Clustering performance measured by the adjusted Rand index (ARI) for \texttt{repSpat} and \texttt{Banksy} over 100 simulation runs for each setting with three noise features. Entries report the median ARI with interquartile range (IQR) in parentheses. The final column reports the percentage of runs in which the ARI of \texttt{repSpat} is less than 1.}
	\centering
	\renewcommand{\arraystretch}{1.4}
	\begin{tabular}{@{} c c c c c c @{}} 
		\toprule
		\textbf{$n$} & \textbf{$p$} & \textbf{$\eta$} 
		& \textbf{repSpat (ARI)} 
		& \textbf{Banksy (ARI)} 
		& \textbf{\% ARI $<$ 1 (repSpat)} \\  
		\midrule
		2500 & 5  & 0.3 & 1 (0) & 0.83 (0.02) & 0.01 \\
		& 5  & 0.8 & 1 (0) & 0.83 (0.01) & 0.01 \\
		& 10 & 0.3 & 1 (0) & 0.83 (0.02) & 0\\
		& 10 & 0.8 & 1 (0) & 0.83 (0.03) & 0.07\\
		\midrule
		4900 & 5  & 0.3 & 1 (0) & 0.85 (0.10) & 0.02 \\
		& 5  & 0.8 & 1 (0) & 0.85 (0.01) & 0 \\
		& 10 & 0.3 & 1 (0) & 0.83 (0.10) & 0 \\
		& 10 & 0.8 & 1 (0) & 0.84 (0.10) & 0.01 \\
		\bottomrule
	\end{tabular}
	\label{tab_sim_comp3}
\end{table}

\section{Supplementary Materials}
\label{app4}
The following are the supplementary materials related to this article.

\begin{enumerate}
  \item Simulation workflow used to produce the figures and tables in Section~\ref{sim-study} and for the illustrative example.
    \begin{itemize}
      \item See \texttt{AppF-supplementary-materials-simulation-study.pdf} for details on \texttt{repSpat} performance evaluation and comparison with the \texttt{Bansky} method.
      \item See \texttt{AppF-simulation-setup-and-silhouette-scores.pdf} for selecting the number of clusters and neighborhood size using the modified silhouette score for one simulation run in the \texttt{repSpat} framework.
      \item See \texttt{AppF-illustraive-example.pdf} for an illustration of the \texttt{repSpat} method on an example dataset.
      \item See \texttt{AppF-supplementary-materials-simulation-study-noise.pdf} for details on \texttt{repSpat} performance evaluation and comparison with the \texttt{Bansky} method when some features are noisy.
    \end{itemize}
 
 \item Descriptive summares for MIBI-TOF TNBC dataset described in Section 4, inlcuding thresholds.
 	\begin{itemize}
 		\item See \texttt{AppF-descriptive-summary-MIBI-TOF-TNBC.pdf} for details. 
 	\end{itemize}
 
    \item The workflow for applying the \texttt{repSpat} method to MIBI-TOF TNBC protein intensities data \cite{keren_structured_2018} described in Section \ref{appli}.
        \begin{itemize}
            \item See \texttt{AppF-repSpat-MIBI-TOF-TNBC-intensities-10-samples.pdf} for details.
        \end{itemize}
    \item The workflow for applying the \texttt{repSpat} method to MIBI-TOF TNBC binary data \cite{keren_structured_2018} described in Section \ref{appli}.
        \begin{itemize}
            \item See \texttt{AppF-repSpat-MIBI-TOF-TNBC-binary-10-samples.pdf} for details.
        \end{itemize}
    \item R functions for \texttt{repSpat}
        \begin{itemize}
            \item The development version of the R and Python packages \texttt{repSpat} is available in a GitHub repositories and and can be provided upon request.
        \end{itemize}
    \item Simulation workflow for sampling distribution of MMD$^2$
        \begin{itemize}
            \item See \texttt{AppF-MMD-sampling-distribution-sim-data} for simulating data.
            \item See \texttt{AppF-MMD-sampling-distribution-Monte-Carlo} for creating Figures \ref{fig:mmd-prop1} - \ref{fig:mmd-prop7}
        \end{itemize}
\end{enumerate}








\renewcommand{\bibname}{References}
\clearpage
\bibliographystyle{elsarticle-harv}  
\bibliography{references}

\end{document}